\documentclass[aps,prl,twocolumn,showpacs,superscriptaddress,groupedaddress,floatfix,nofootinbib]{revtex4-1}
\pdfoutput=1

\usepackage{graphicx} % needed for pdflatex (.pdf figures)
\usepackage{bmpsize}
\usepackage{multirow}
\usepackage{dcolumn}   % needed for some tables
\usepackage{bm}        % for math
\usepackage{amsmath}
\usepackage{amssymb}   % for math
\usepackage{amsopn}
\usepackage{subfigure}
\usepackage{here}
\usepackage{amsmath}

\usepackage{lineno}
\newcommand*\conj[1]{\bar{#1}}

\def\Cs{$^{137}$Cs}
\def\Ge{$^{68}$Ge}
\def\Co{$^{60}$Co}
\def\PoBe{$^{210}$Po$^{9}$Be}
\def\Cf{$^{252}$Cf}
\def\B{$^{12}$B}
\def\N{$^{12}$N}

\def\CsS{$^{137}$Cs }

\def\CoS{$^{60}$Co }
\def\PoBeS{$^{210}$Po$^{9}$Be }
\def\CfS{$^{252}$Cf }
\def\BS{$^{12}$B }
\def\NS{$^{12}$N }
\def\LiHeS{$^{9}$Li/$^{8}$He }
\def\LiS{$^{9}$Li }

\def\He{$^{8}$He}

\def\nueb{$\overline{\nu}_{e}$}
\def\nuebS{$\overline{\nu}_{e}$ }

\newcommand{\qOneThree}{sin$^22\theta_{13}$ }
\newcommand{\dmSq}{$|\Delta m^2_{ee}|$ }

\begin{document}

%\linenumbers

\title{Spectral Measurement of the Electron Antineutrino Oscillation Amplitude and Frequency using 500 Live Days of RENO Data}

\affiliation{Institute for Universe and Elementary Particles, Chonnam National University, Gwangju 61186, Korea }

\affiliation{Department of Physics Education, Chonnam National University, Gwangju 61186, Korea }

\affiliation{Department of Physics, Chung Ang University, Seoul 06974, Korea }

\affiliation{Department of Radiology, Dongshin University, Naju 58245, Korea }

\affiliation{Department of Physics and Photon Science, Gwangju Institute of Science and Technology, Gwangju 61005, Korea }

\affiliation{Department of Physics, Gyeongsang National University, Jinju 52828, Korea }

\affiliation{Institute for Basic Science, Daejeon 34047, Korea }

\affiliation{Department of Physics, Kyungpook National University, Daegu 41566, Korea }

\affiliation{Department of Physics and Astronomy, Sejong University, Seoul 05006, Korea }

\affiliation{Department of Physics and Astronomy, Seoul National University, Seoul 08826, Korea }

\affiliation{Department of Fire Safety, Seoyeong University, Gwangju 61268, Korea }

\affiliation{Department of Physics, Sungkyunkwan University, Suwon 16419, Korea }

\author{S. H. Seo 
%\footnote{Corresponding author: 
%Tel: +82 2 880 4394 \, Fax: +82 2 884 3002 \\   
%E-mail address: shseo@phya.snu.ac.kr (S. H. Seo)}
}
\affiliation{Department of Physics and Astronomy, Seoul National University, Seoul 08826, Korea }

\author{W. Q. Choi 
\footnote{Present Address: Korea Institute of Science and Technology Information, Daejeon 34141, Korea}
}
\affiliation{Department of Physics and Astronomy, Seoul National University, Seoul 08826, Korea }

\author{H. Seo}
\affiliation{Department of Physics and Astronomy, Seoul National University, Seoul 08826, Korea }

\author{J. H. Choi}
\affiliation{Department of Radiology, Dongshin University, Naju 58245, Korea }

\author{Y. Choi}
\affiliation{Department of Physics, Sungkyunkwan University, Suwon 16419, Korea }

\author{H. I. Jang}
\affiliation{Department of Fire Safety, Seoyeong University, Gwangju 61268, Korea }

\author{J. S. Jang}
\affiliation{Department of Physics and Photon Science, Gwangju Institute of Science and Technology, Gwangju 61005, Korea }

\author{K. K. Joo}
\affiliation{Institute for Universe and Elementary Particles, Chonnam National University, Gwangju 61186, Korea }

\author{B. R. Kim}
\affiliation{Institute for Universe and Elementary Particles, Chonnam National University, Gwangju 61186, Korea }

\author{H. S. Kim}
\affiliation{Department of Physics and Astronomy, Sejong University, Seoul 05006, Korea }

\author{J. Y. Kim}
\affiliation{Institute for Universe and Elementary Particles, Chonnam National University, Gwangju 61186, Korea }

\author{S. B. Kim}
\affiliation{Department of Physics and Astronomy, Seoul National University, Seoul 08826, Korea }

\author{S. Y. Kim}
\affiliation{Department of Physics and Astronomy, Seoul National University, Seoul 08826, Korea }

\author{W. Kim}
\affiliation{Department of Physics, Kyungpook National University, Daegu 41566, Korea }

\author{E. Kwon}
\affiliation{Department of Physics and Astronomy, Seoul National University, Seoul 08826, Korea }

\author{D. H. Lee}
\affiliation{Department of Physics and Astronomy, Seoul National University, Seoul 08826, Korea }

\author{Y. C. Lee}
\affiliation{Department of Physics and Astronomy, Seoul National University, Seoul 08826, Korea }

\author{I. T. Lim}
\affiliation{Department of Physics Education, Chonnam National University, Gwangju 61186, Korea }

\author{M. Y. Pac}
\affiliation{Department of Radiology, Dongshin University, Naju 58245, Korea }

\author{I. G. Park}
\affiliation{Department of Physics, Gyeongsang National University, Jinju 52828, Korea }

\author{J. S. Park \footnote{Present Address: Institute for Basic Science, Daejeon 34047, Korea}
}
\affiliation{Department of Physics and Astronomy, Seoul National University, Seoul 08826, Korea }

\author{R. G. Park}
\affiliation{Institute for Universe and Elementary Particles, Chonnam National University, Gwangju 61186, Korea }

\author{Y. G. Seon}
\affiliation{Department of Physics, Kyungpook National University, Daegu 41566, Korea }

\author{C. D. Shin}
\affiliation{Institute for Universe and Elementary Particles, Chonnam National University, Gwangju 61186, Korea }

\author{J. H. Yang}
\affiliation{Department of Physics, Sungkyunkwan University, Suwon 16419, Korea }

\author{J. Y. Yang}
\affiliation{Department of Physics and Astronomy, Seoul National University, Seoul 08826, Korea }

\author{I. S. Yeo}
\affiliation{Institute for Universe and Elementary Particles, Chonnam National University, Gwangju 61186, Korea }

\author{I. Yu}
\affiliation{Department of Physics, Sungkyunkwan University, Suwon 16419, Korea }

\collaboration{The RENO Collaboration}

\date{\today}

\begin{abstract}
The Reactor Experiment for Neutrino Oscillation (RENO) has been taking electron antineutrino (\nueb) data from the reactors in Yonggwang, Korea, 
using two identical detectors since August 2011.
Using roughly 500 live days of data through January 2013 
we observe 290\,775 (31\,514) reactor \nuebS candidate events with 2.8\%(4.9)\% background in the near (far) detector. 
The observed visible positron spectra from the reactor \nuebS events in both detectors show discrepancy around 5~MeV with regard to the prediction from the current reactor \nuebS model. 
Based on a far-to-near ratio measurement using the spectral and rate information we have obtained 
$\sin^2 2 \theta_{13} = 0.082 \pm 0.009({\rm stat.}) \pm 0.006({\rm syst.})$ and 
$|\Delta m_{ee}^2| =[2.62_{-0.23}^{+0.21}({\rm stat.})_{-0.13}^{+0.12}({\rm syst.})]\times 10^{-3}$~eV$^2$.
\end{abstract}

\pacs{14.60.Pq, 29.40.Mc, 28.50.Hw, 13.15.+g}
\maketitle

%======================================================================
\section{\label{sec:intro} Introduction }

The historical observations of neutrino oscillations~\cite{SK,SNO,DB,RENO} have verified that neutrinos are massive. 
Existence of neutrino mass requires modification of the Standard Model and provides hints on the Grand Unification Theory.
The smallest neutrino mixing angle $\theta_{13}$ in the PMNS matrix~\cite{Ponte,MNS} is definitively
measured in 2012 by Daya Bay~\cite{DB} and RENO~\cite{RENO}.  
The leptonic CP phase $\delta_{\rm{CP}}$ and neutrino mass ordering are now
possible to be measured due to the large value of $\theta_{13}$. 
A precise measurement of $\theta_{13}$ by a reactor \nuebS experiment will greatly improve determination of the CP phase
when combined with results of accelerator neutrino experiments~\cite{T2K,NOVA}. \\
\indent Using the two identical detectors in separate locations the RENO experiment measures the reactor \nuebS survival probability,
$P_{ee} \equiv P(\overline{\nu}_{e} \rightarrow \overline{\nu}_{e})$~\cite{Petcov},

\begin{eqnarray}
\label{eq_pee}
 P_{ee} & =  & 1 - \sin^2 2 \theta_{13} ( \cos^2 \theta_{12} \sin^2 \Delta_{31} + \sin^2 \theta_{12} \sin^2 \Delta_{32} )
  \nonumber      \\
 &  &    - \cos^4 \theta_{13} \sin^2 2\theta_{12} \sin^2 \Delta_{21}
  \nonumber       \\
 &  &  \hspace*{-1.3cm} ~ \approx  1 - \sin^2 2 \theta_{13} \sin^2 \Delta_{ee}  - \cos^4 \theta_{13} \sin^2 2\theta_{12} \sin^2 \Delta_{21},
~\label{e:Pee}
\end{eqnarray}
where $\Delta_{ij} \equiv 1.267 \Delta m_{ij}^2 L/E_{\nu}$, $E_{\nu}$ is the $\overline{\nu}_e$ energy in MeV, $L$ is the distance between the reactor and detector in meters,
and $\Delta m_{ee}^2$ is the effective neutrino mass squared difference in eV$^{2}$ and defined as 
$\Delta m_{ee}^2 \equiv \cos^2 \theta_{12}\Delta m_{31}^2 + \sin^2 \theta_{12} \Delta m_{32}^2$~\cite{Parke}. 

Recently RENO has published a letter~\cite{RENO-PRL2} on the improved measurement of $\theta_{13}$ and the first measurement of $|\Delta m_{ee}^2|$ 
with a spectral shape and rate analysis using $\sim$500 live days of data. 
This paper provides more detailed description on the analysis. 

%======================================================================
\section{\label{sec:setup} Experimental Setup }

%----Experimental setup---------------------------

\begin{figure}
\begin{center}
\includegraphics[width=0.48\textwidth]{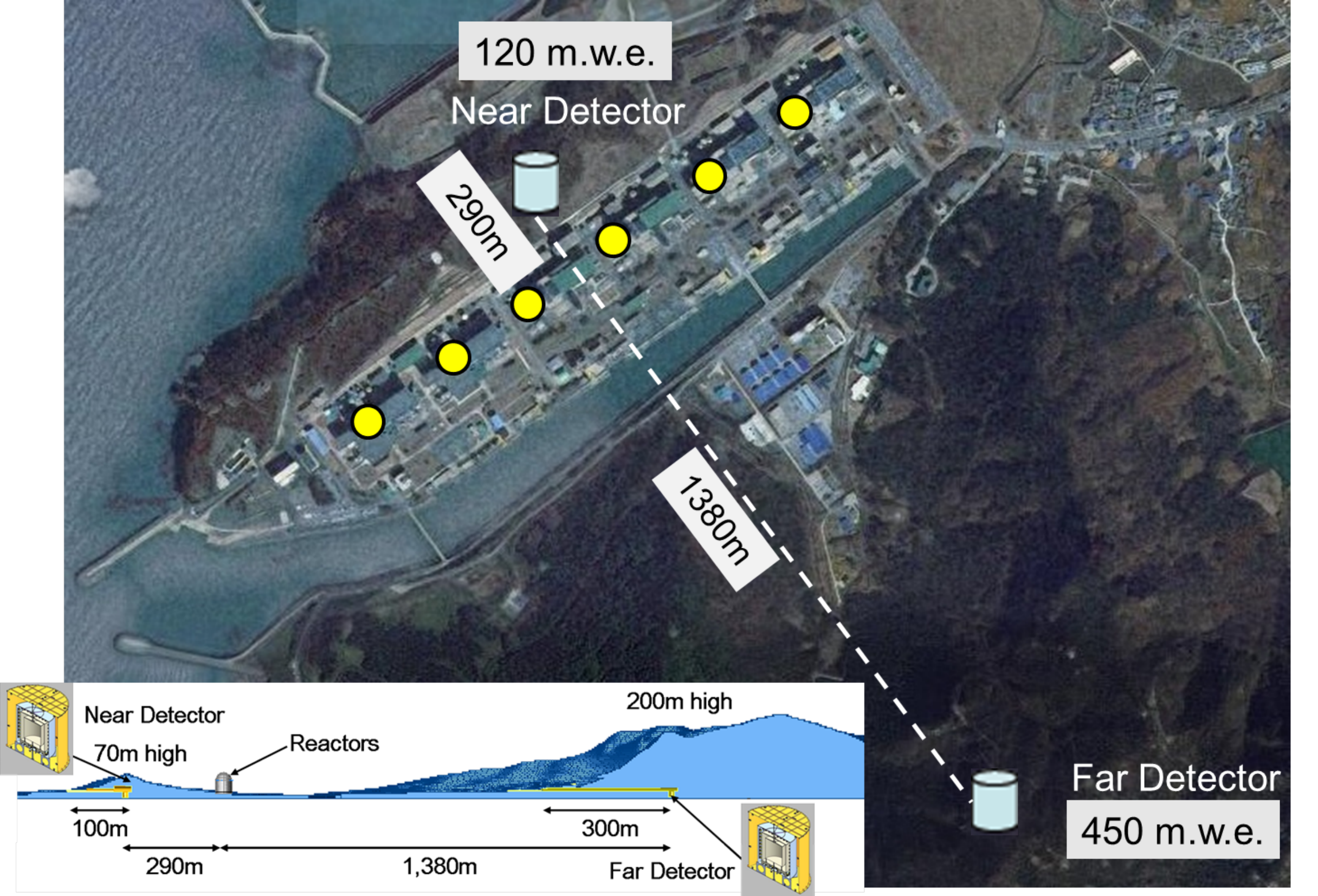}
\end{center}
\caption{(Color online) (Top) top view of the six reactors (circles) in Hanbit nuclear power plant and the location of the two detectors (cylinders).
(Bottom) side view of the RENO experimental layout.}
\label{f:site}
\end{figure}

The RENO detectors are located near the Hanbit (previously known as Yonggwang) nuclear power plant, 
operated by Korea Hydro and Nuclear Power Co., Ltd (KHNP), in Yonggwang, the southwest coast region in South Korea. 
The plant consists of six reactors linearly aligned with equal distance of $\sim$260 m
and provides total thermal output of $16.8~{\rm GW_{th}}$.
Reactors 1 and 2 each produce maximum $2.755~{\rm GW_{th}}$ and newer reactors 3--6 each produce $2.755~{\rm GW_{th}}$.\\
\indent
RENO started its civil engineering in 2007 and completed the construction of two identical detectors, ready for data taking in June 2011. 
A near (far) detector is located at 294 (1383) m from the center of the six reactors.
The near (far) detector is installed underground with an overburden of 120 (450) m.w.e.
Figure~\ref{f:site} shows a layout of the RENO experiment. 

The far-to-near ratio measurement using the two identical detectors greatly reduces the systematic uncertainties
in the measurement of $\theta_{13}$ due to the cancellation of their correlated uncertainties. 
It would be difficult to measure the mixing angle $\theta_{13}$ with a single detector 
because of the large reactor \nuebS flux uncertainty. By measuring the reactor 
neutrino flux at the near detector and predicting the expected one at far detector,
the systematic error associated with the reactor \nuebS flux uncertainty can be significantly reduced. 
The baseline distances between the detectors and reactors are measured to an accuracy of better than 10~cm using GPS and total station.

%----Detectors---------------------------
\section{\label{sec:det} The RENO Detector }

\begin{figure}
\begin{center}
\includegraphics[width=0.48\textwidth]{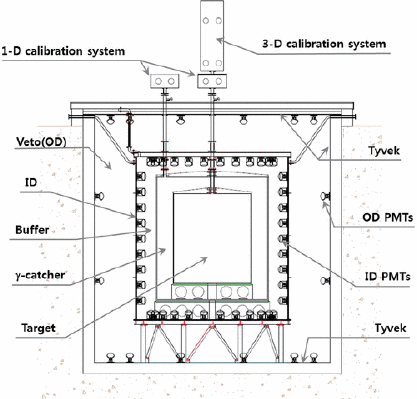}
\end{center}
\caption{The schematics of the RENO detector consisting of the ID (target, $\gamma$-catcher, and buffer) and OD (veto) detectors.
A total of 354 (67) 10~in. PMTs detect scintillation lights from the ID (OD).
}
\label{f:det}
\end{figure}

The RENO experiment detects reactor \nuebS through the inverse beta decay (IBD) reaction, $\overline{\nu}_e + p \rightarrow e^{+} + n$,
using liquid scintillator (LS) with 0.1\% gadolinium (Gd) as a target. In the IBD reaction \nuebS with energy larger than 1.81~MeV 
interacts with a free proton in hydrocarbon LS to produce a positron and a neutron. The positron carries away most kinetic energy of the incoming
\nuebS while the neutron takes only $\sim$10~keV. The positron annihilates immediately to releases 1.02~MeV as two $\gamma$-rays 
in addition to its kinetic energy. The neutron after thermalization is captured by Gd with a mean delayed time of $\sim$26~$\mu$s.\\
\indent
RENO detectors are optimized to detect reactor \nuebS and consist of four layers of nested cylindrical structures as shown in Fig.~\ref{f:det}.
They are target, $\gamma$-catcher, buffer, and veto from the innermost and contain different liquids. The main inner detector (ID) is contained in a cylindrical stainless steel vessel of 5.4~m in diameter and 5.8~m in height and houses two nested cylindrical acrylic vessels. The 1.5~m thick outer detector (OD) surrounding the ID is filled with 350 tons of highly purified water. The OD is intended to identify events coming from the outside by their Cherenkov radiation and to shield against ambient $\gamma$-rays and neutrons from the surrounding rock.\\
\indent
The innermost target vessel, a 25~mm thick acrylic vessel of 2.75~m in diameter and 3.15~m in height, holds 16~tons of 0.1\% Gd-doped LS (Gd-LS) as a neutrino target. 
It is surrounded by a 60~cm thick layer of 29~ton undoped LS in the $\gamma$-catcher, useful for recovering $\gamma$-rays escaping from the target region.
The $\gamma$-catcher liquid is contained in a 30~mm thick acrylic vessel of 4.0~m in diameter and 4.4~m in height. The acrylic vessels holding organic liquids 
are made of casting polymethyl methacrylate (PMMA (C$_5$O$_2$H$_8$)$_n$) plastic which transmits up to 92\% of visible light at 3~mm thickness and reflects 
about 4\% from the surface in LS~\cite{RENO-acrylic}. \\
\indent
Outside the $\gamma$-catcher is a 70~cm thick buffer region filled with 65 tons of mineral oil. It provides shielding against ambient $\gamma$-rays and neutrons coming from outside. 
Light signals emitted from particles interacting in ID are detected by 354 low-background 10~inch Hamamatsu R7081 photomultiplier tubes (PMTs)~\cite{RENO-PMT}
that are mounted on the inner wall of the ID. The OD is equipped with 67 10~inch R7081 water-proof PMTs mounted on the wall of the concrete veto vessel. 
The inner surface of OD is covered with Tyvek sheets to increase the light collection. \\
\indent
The LS is developed and produced as a mixture of linear alkylbenzene (LAB), 3 g/$\ell$ of PPO, and 30 mg/$\ell$ of bis-MSB. 
LAB (C$_{n}$H$_{2n+1}$-C$_{6}$H$_{5}$, n=10$-$13) is an organic solvent with a high flash point of 130$^{\circ}$C and has a good light yield and a large attenuation length. 
A Gd-carboxylate complex using Trimethylhexanoic acid was developed for the best Gd loading efficiency into LS and its long term stability~\cite{RENO-GdLS}. 
Care is taken in production of LS and Gd-LS and filling into detectors to ensure that the near and far detectors are as identical as possible.\\
\indent
The RENO detector uses cartesian coordinates of $x$, $y$, and $z$ with an origin at the center of the detector. 
The $z$ coordinate is along the cylindrical axis. 
The detectors are calibrated using radioactive sources and cosmic-ray induced backgrounds. Various radioisotopes of gamma-ray sources 
are periodically deployed in the target and gamma-catcher by a motorized pulley system in a glove box as shown in Fig.~\ref{f:det}. 
The system deploys a source along the vertical direction only. 
The relative source location can be controlled at an accuracy of a few mm by a stepper motor, but absolute vertex location accuracy is 1~cm. 
The source data are taken every one or two months to monitor the detector stability and to obtain calibration parameters. 
Also a 3-D calibration system is developed for deploying calibration sources at off cylindrical axis positions in the target. However it has not been used. 
More details on the RENO detector are found in Ref.~\cite{RENO-tdr}.

%==================================================
\section{Data Acquisition } 

The scintillation light produced in the liquid scintillator from the interaction of signal or background events are collected by the PMTs.
Analog signals are produced and sent through 25~m RG303/U single cables to the signal processing front-end boards. 
The RENO data acquisition (DAQ) system uses electronic modules developed for the Super-Kamiokande experiment~\cite{SK_DAQ} and consists of
a total of 18 front-end boards with 24 channels each, driven by a common 60~MHz master clock.
Each front-end board is equipped with eight charge-to-time conversion (QTC) chips, four time-to-digital converter (TDC) chips,
and an 100~Mbps ethernet card. The QTC chip has three inputs with different gains of 1, 7, and 49 to cover a
dynamic range from 0.2 to 2\,500~pC with a resolution of 0.1~pC at gain 1. The QTC chip measures the time and integrated charge of a PMT
analog signal and converts them to digital values. The timing information is fed into a TDC chip to be digitized and recorded. 
The signal processing time per hit is roughly 800~ns for charge sampling and digitization. 
The signal front-end boards can handle upto $\sim$100~kHz of events each with photon hits on every PMT
without dead time and does not require any hardware triggers to lower the event rate.
\\
\indent
An offline software trigger system generates buffer, veto, or buffer and veto triggers for an event if it satisfies 
an appropriate trigger condition. 
The number of PMT hits ($N_{\rm hit}$ is defined as the number of PMTs that has a signal more larger than 0.3~p.e. in a 50~ns time window.
A buffer trigger requires ID $N_{\rm{hit}} > $ 90, corresponding to 0.5$-$0.6~MeV
and is well below the 1.02~MeV minimum energy of an IBD positron signal. 
Upon a trigger an event is made by collecting all the PMT hits in a time window of $-18$ to $+18$~$\mu$s. 
The time zero is defined by the first hit time when $N_{\rm{hit}}$ is greater than 90 in a time window of 50~ns.
The only PMT hits in a time window of $-$100 to $+$50~ns are used for the event energy and vertex reconstruction.
The PMT hits outside a time window of $-$100 to $+$50~ns are recorded to monitor dark currents. 
If a trigger is issued within $18~\mu$s of the previous trigger, the PMT hits in the overlapping time windows are shared between two events.
%The only PMT hits in a time window of $-$100 to $+$50~ns are used for the event energy and vertex reconstruction. 
A veto trigger is issued for a cosmic-ray muon event and requires OD $N_{\rm{hit}} > $ 10 out of total 67 OD PMTs.
A buffer and veto trigger is issued if an event satisfies the two conditions simultaneously. \\
\indent
The average total trigger rates of the 500 day data sample are $\sim$590~Hz in the near detector and
$\sim$140~Hz in the far detector. The trigger types and rates are summarized in Table~\ref{t:trig_rate}. 
The buffer-only trigger is required for an IBD candidate and the rate is $\sim$60 ($\sim$77)~Hz for
the near (far) detector. The veto-only trigger rate is higher in the near detector having less overburden than the far detector.  
\begin{table}
\begin{center}
\caption{\label{t:trig_rate} Average trigger rates of the $\sim$500 live days of data in the RENO detectors.
The rates for the buffer-only trigger required for IBD event are $\sim$60~Hz (near) and $\sim$77~Hz (far).}
\begin{tabular*}{0.48\textwidth}{@{\extracolsep{\fill}} c r r}
\hline
\hline
Trigger type & Near & Far \\
 & (Hz) & (Hz) \\
\hline
Buffer  & 269 &  100  \\
Veto  & 529 & 61 \\
Buffer \& veto  &  209 & 23 \\
\hline
Total  & 590 & 138 \\
\hline
\hline
\end{tabular*}
\end{center}
\end{table}

\iffalse
The trigger efficiency is determined by the IBD signal loss due to the requirement of ID $N_{\rm{hit}} >$ 90. 
The RENO Monte Carlo simulation (MC), which is described later, does not reproduce the data $N_{\rm hit}$ well 
due to lack of realistic individual-channel simulation for the p.e. threshold and dark or noise hits. 
According to comparison of $N_{\rm{hit}}$ distribution between
data and MC, we find a MC equivalent requirement of $N_{\rm{hit}} >$ 84 to accept a buffer-only trigger.
Using the MC equivalent hit requirement, the trigger efficiency for the IBD signal excluding spill-in
events in the near (far) detector is estimated as 99.77$\pm$0.05\% (99.78$\pm$0.13\%) where spill-in events are events that occur outside the target 
and produce a neutron capture on Gd in the target. The position dependent DAQ inefficiency contributes to the inefficiency near the trigger threshold below $\sim$0.8~MeV. 
Our measured trigger efficiency using a \CsS source (E = 0.63~MeV) is roughly 50\% at the threshold energy of 0.5$\sim$0.6~MeV and almost 100\% at 0.8~MeV. 
The uncorrelated systematic uncertainty of the trigger efficiency is estimated as 0.01\% from the difference between near and far efficiencies.
The correlated uncertainty of the trigger efficiency is estimated as 0.01\% from the ambiguity in finding a MC equivalent $N_{\rm{hit}}$ threshold. \\
\fi
\indent
Real-time online monitoring for PMT hit rates, trigger rates, High Voltage (HV) and other interesting vaiables is performed 
to find possible data-taking troubles. Various environmental parameters including the water level and temperature are also monitored online~\cite{SCM}.
More checks are performed offline on a weekly basis for trigger rates, muon rates, flashing PMT rates,
IBD prompt and delayed candidate rates, and charge stability.    

%---Data collection & quality------------------------
\section{\label{sec:data} Data Sample }

RENO has started taking data in August 2011 and has been operating continuously so far with an accumulated average DAQ efficiency of better than 95\%
for both detectors. 
%As of August 2016 RENO has reached about 1\,700 live days of data taking and collected about 1.5 (0.15) million \nuebS events in the near (far) detector.
In this analysis 489.93 (458.49) live day data with negligible uncertainties in the far (near) detector taken from August 2011 to January 2013 is used 
to extract the neutrino mixing parameters, 
$\theta_{13}$ and $|\Delta m^2_{ee}|$. Each reactor is periodically turned off for a month every 1.5~years to replenish nuclear fuel. 
Besides these periodic turn-off there are sporadic unscheduled down-times. All of these information are provided by KHNP.
Table~\ref{t:reac-off} summarizes the reactor-off periods during the $\sim$500 live days.
\begin{table}
\caption{\label{t:reac-off} Reactor-off periods during the $\sim$500 live days.
}
\begin{center}
\begin{tabular*}{0.48\textwidth}{@{\extracolsep{\fill}} c c}
\hline
\hline
Periods  & Off reactor number \\
\hline
2011.08.30 - 2011.09.29  &  2  \\
2012.02.24 - 2012.03.21  &  1 \\
2012.05.01 - 2012.05.30  &  5 \\
2012.06.07 - 2012.07.17  &  4 \\
2012.10.19 - 2012.11.07  &  4 \\
2012.11.08 - 2012.12.30  &  3, 4, 5 \\
2012.12.31 - 2013.01.21  &  3 \\
\hline
\hline
\end{tabular*}
\end{center}
\end{table}
%=========================================================================

%---Data collection & quality------------------------
\section{\label{sec:Simulation} Detector Simulation }

The primary software tool for modelling the RENO detector response is {\sc GLG4SIM}~\cite{glg4sim}, a {\sc GEANT4}~\footnote{V4.7.1 patch01} based simulation package 
%[Ref: G.A. Horton-Smith, Generic liquid scintillator Genat4 simulation (2005) http://neutrino.phys.ksu.edu/~GLG4sim/ ]
for liquid scintillator detectors derived from {\sc KLG4SIM} of the KamLAND Collaboration. The {\sc GLG4SIM} is designed for simulation of the detailed detector response 
to particles moving through and interacting with a large volume of liquid scintillator detector. This generic program has been customized for the RENO detector. 
The {\sc GEANT4} toolkits are used for simulating the physics processes involving particles with energies above a few keV propagating through the detector materials. 
However, the optical photon production and propagation through liquid scintillator, including processes like absorption, re-emission, and elastic collisions, 
are handled by specifically written codes in {\sc GLG4SIM}, using measured optical properties of the RENO LS. The simulation includes the measured quenching 
effect of the $\gamma$-ray at low energies using a pure Ge detector.

{\sc glg4sim} has a detailed modeling of PMTs and takes into account transmission, absorption, and reflection of optical photons at the photocathode. The photocathode thickness and wavelength dependent photocathode efficiency are implemented in the PMT model.

Each photon generated in the simulation is tracked in the detector until it either reaches a PMT or is lost. The simulation takes into account 
several light propagation phenomena while tracking the photons. In the scintillator, photons can be absorbed or elastically scattered 
(Rayleigh scattering) by solvent and fluor molecules.

The absorption of photons within the acrylic vessel medium is simulated according to the absorption probability calculated with medium's attenuation length. 
Also, the reflection and refraction of photons at the surface of the acrylic vessel are simulated using the Fresnel's law. The refractive indices of 
all dielectric materials in the detector are measured at different wavelengths and implemented in the simulation.

For the simulation of neutron capture on Gd, the {\sc GLG4SIM} is used to provide a proper modeling of discrete lines of high-energy gammas and 
the continuous gamma spectrum arising from the neutron capture on Gd. Both Cherenkov radiation and scintillation light emission are simulated. 

The {\sc GEANT} used in the MC simulation is outdated due to time evolution since the start of RENO and therefore needs to be updated. 
However, this measurement is largely data-driven and thus expects to be hardly affected by the update. Systematic uncertainties may be improved 
by better understanding of detailed physics processes with an updated MC simulation.

The dead PMT fraction during the data taking reported here is less than 1\% for both
near and far detectors. 
However the dead PMTs are not accounted for in RENO MC as the time dependent charge correction in data compensates the effects of dead PMTs.
More details on RENO detector simulation is found in \cite{RENO-tdr}. 

%======================================================================
\section{Event Reconstruction}

Reconstructed energy and vertex are essential for selecting IBD candidate events against various backgrounds.
In the following subsections we describe energy and vertex reconstructions of the triggered events. 

\subsection{Energy reconstruction}

An analog signal from each PMT is amplified, integrated and then digitized by ADC in a QTC chip. 
The ADC value is then converted to a charge in pC. A charge injection board is used 
to determine an ADC-to-pC conversion factor for an individual channel of a front-end board.
Using a \CsS source, a fit to one photoelectron response 
finds a corresponding charge value of $\sim$1.6~pC. A PMT charge is measured in p.e. based on the conversion factor. 

The event energy is determined by the total charge ($Q_{\rm{tot}}$) that is defined as sum of hit PMT charges greater than 0.3 p.e. 
in a time window of $-100$ to $+50$~ns.
%The time zero is defined by the first hit time when $N_{\rm{hit}}$ is greater than 90 in a time window of 50~ns.
The event time window is determined by taking into account the size of the RENO detector and to minimize the contributions of dark hits, 
flashing PMT hits, and negative charges caused by the unsettled pedestal after a large pulse height due to a highly energetic muon. 

\begin{figure}
\begin{center}
\includegraphics[width=0.48\textwidth]{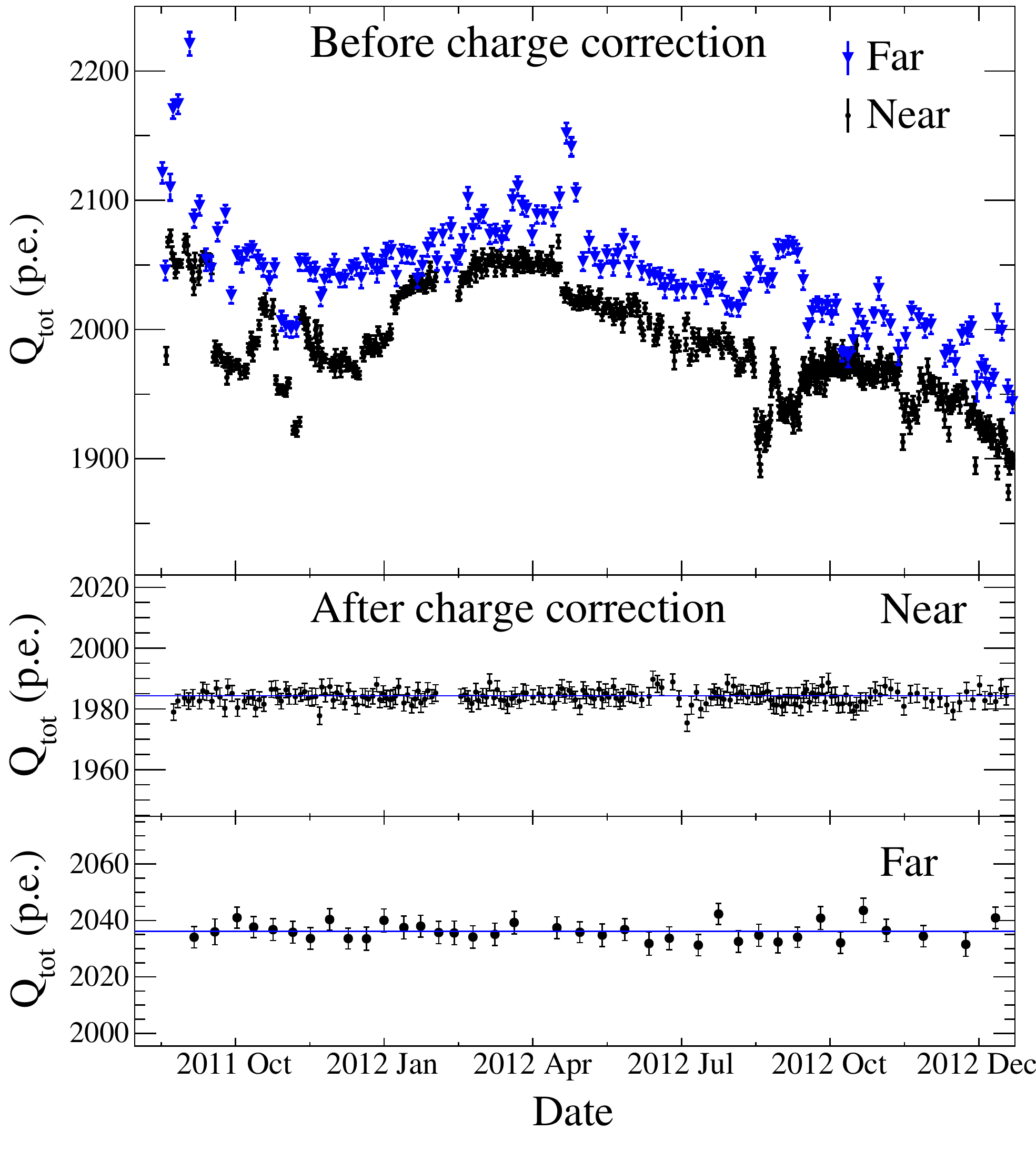}
\end{center}
\caption{
(Color online) 
Time variation of raw charges observed by IBD delayed events i.e., n-Gd capture event peaks, (top panel) 
and time stability of their corrected charges (bottom panels).
}
\label{f:s2_stability}
\end{figure}

\begin{figure}
\begin{center}
\includegraphics[width=0.48\textwidth]{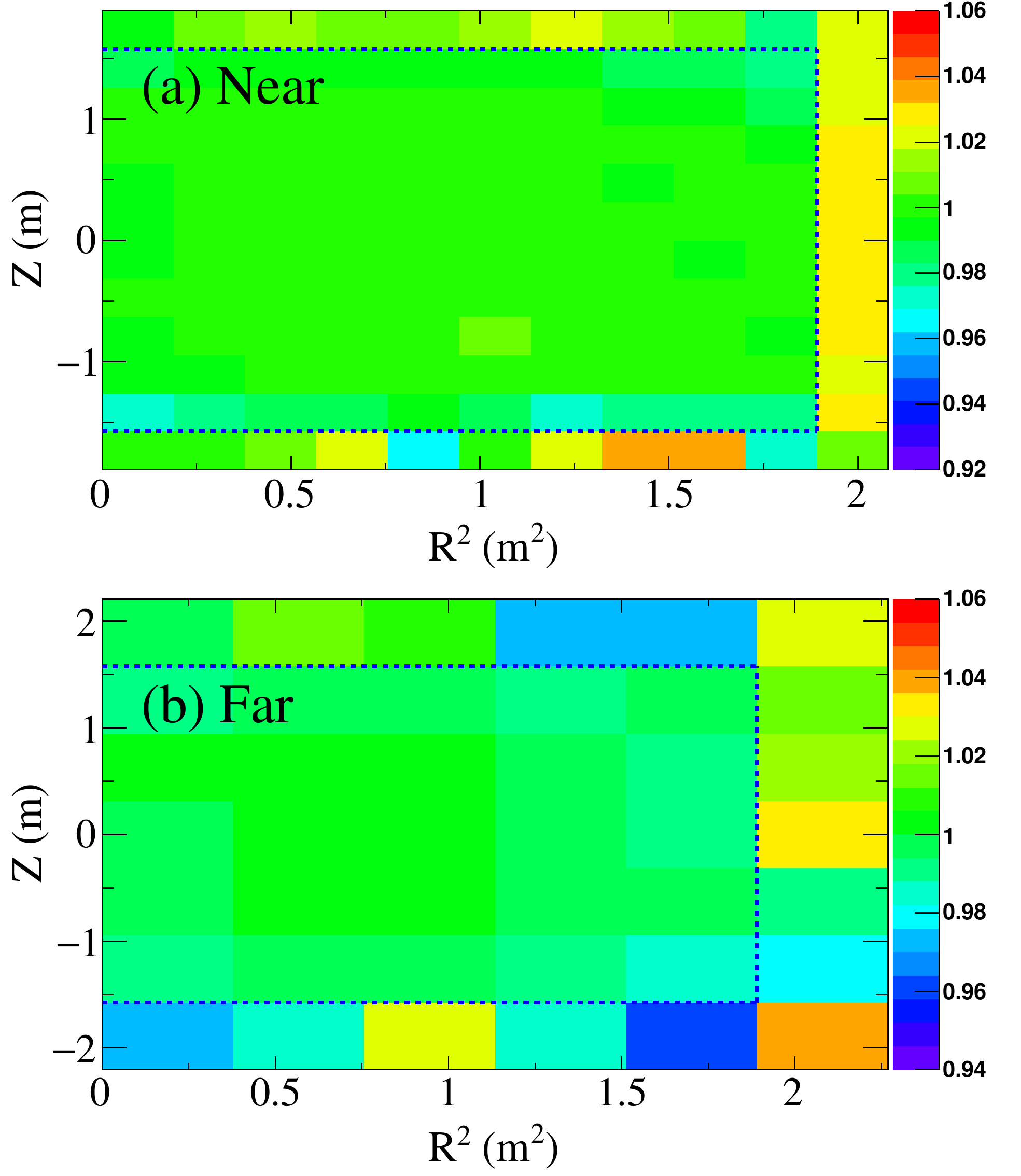}
\end{center}
\caption{
(Color online) Charge uniformity map seen with n-Gd delayed energy peaks in the (a) near and (b) far detectors.
The radial coordinate $R$ is defined by $\sqrt{x^{2}+y^{2}}$. 
The blue dotted line is a target boundary. 
The color code represents the ratio of the fitted corrected-charges in a bin to a reference value.
}
\label{f:q_unif}
\end{figure}

The raw $Q_{\rm{tot}}$ of IBD delayed signals shows a time variation as shown in the Fig.~\ref{f:s2_stability} upper panel. 
This is caused by PMT gain change, removal of flashing PMTs, and the decrease of the LS attenuation length~\cite{RENO_LS_lambda}. 
The raw charge time-variation is corrected using temporal charge correction factors obtained from the IBD delayed signal peaks
with respect to a reference value. 
Figure~\ref{f:s2_stability} lower panel shows an excellent stability of the reconstructed energies of IBD delayed signals after the temporal charge correction.
According to the charge uniformity map shown in Fig.~\ref{f:q_unif}, there is no need of spatial charge correction
since the charge differences of less than 1\% in the entire target volume are observed.
The nonuniform energy response near the target acrylic vessel is due to loss of the energy in the acrylic and due to larger scintillation of spill-in events in $\gamma$-catcher.
The energy loss effect is a bit pronounced at bottom due to the acrylic structure supporting the target and $\gamma$-catcher vessels.. 
This energy loss introduces a slight modification 
of the energy spectrum of prompt events in a few percent ($<$4\%) level, but occurs identically in the near and far detectors. 
Again our far-to-near ratio measurement minimizes a possible spectral difference between the two detectors. 

\begin{figure}
\begin{center}
\includegraphics[width=0.48\textwidth]{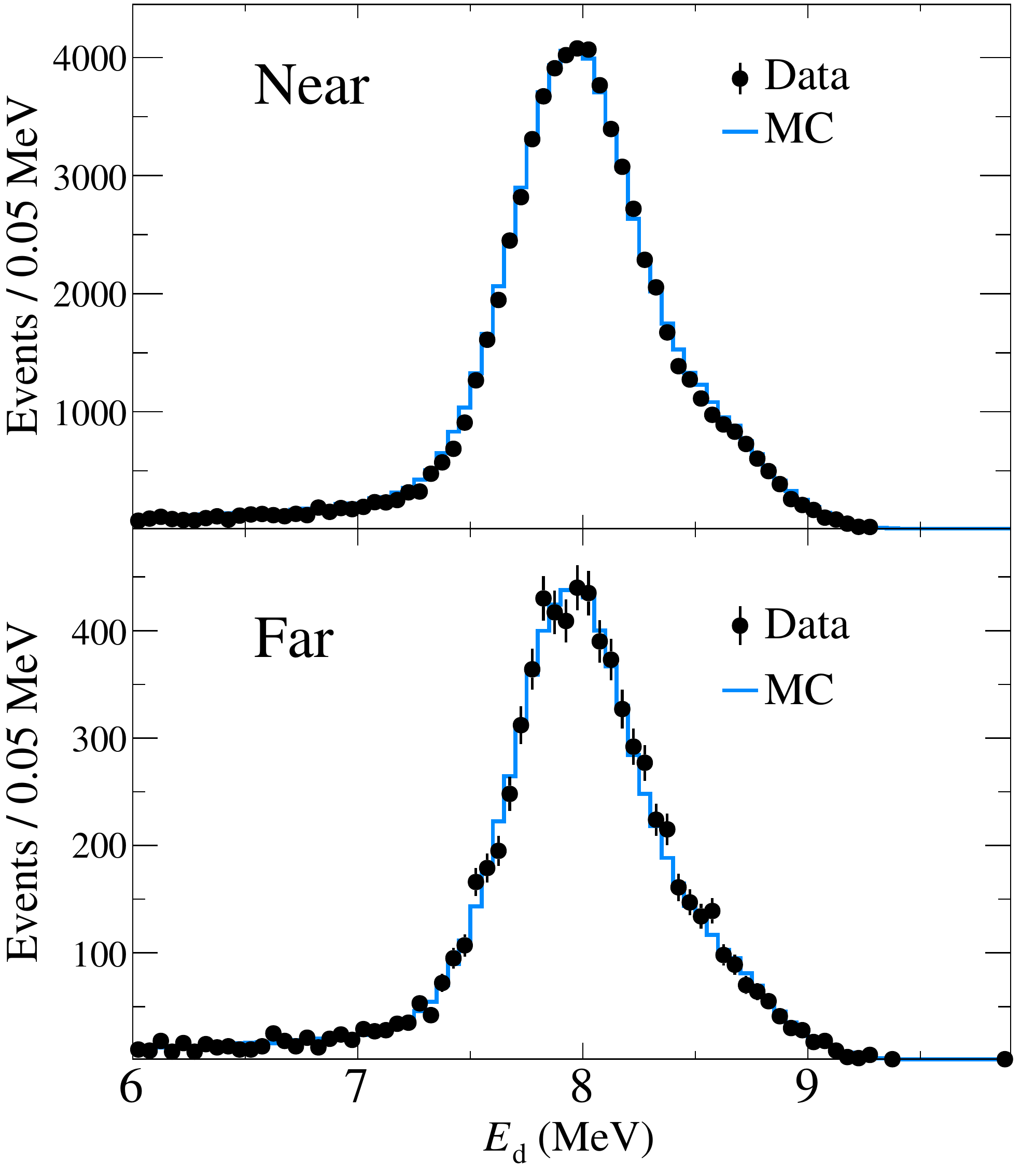}
\end{center}
\caption{(Color online) Delayed energy spectra of the IBD candidate events in the near and far detectors.
Data and MC spectra agree well. }
\label{f:s2_shape}
\end{figure}

After the raw charge correction $Q_{\rm{tot}}$ in p.e. is converted to energy in MeV
using an energy conversion function that will be described in the calibration section later. 
After the charge correction and conversion we obtain reconstructed energies. 
Figure~\ref{f:s2_shape} shows a good agreement between data and MC in the delayed signal spectrum of IBD candidate events.

%+++++++++++++++++++++++++++++++++++++++++++++++++++++++++++++++++++++++
\subsection{Muon energy estimation}

\begin{figure}
\begin{center}
\includegraphics[width=0.48\textwidth]{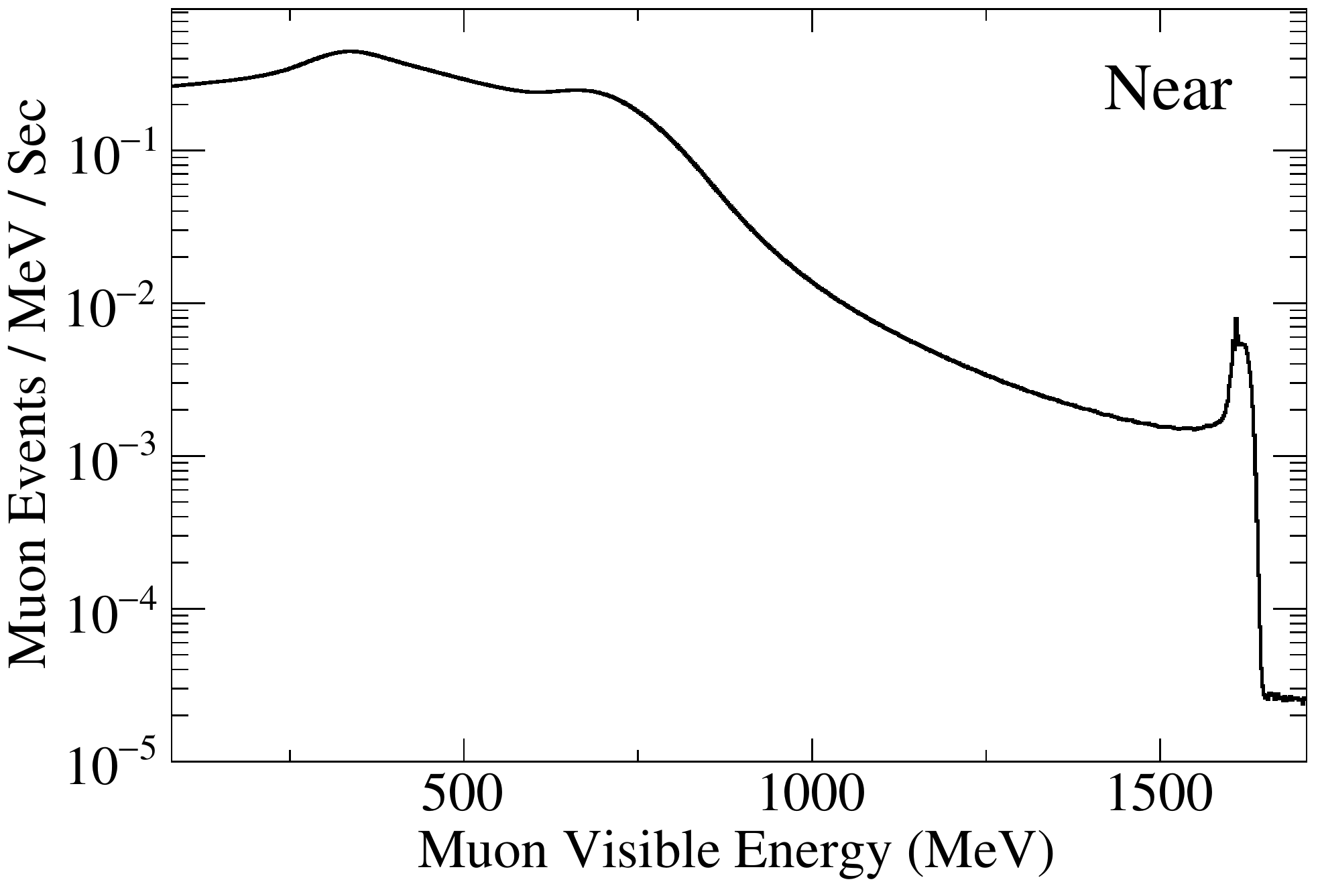}
\end{center}
\caption{
%Muon deposit energy distribution for the near detector.
%A maximum visible energy of $\sim$1700 MeV is found due to a finite size of the detector. 
Muon deposit energy distribution for the near detector.
The maximum energy of $\sim$1700~MeV is due to the saturation of the DAQ electronics. 
}
\label{f:mu_en}
\end{figure}

Cosmogenic muons introduce a main background in the IBD candidates. 
The intrinsic muon energy cannot be reconstructed, but its deposited energy inside the detector
can be reasonably measured as a visible energy proportional to its path length. 
The muon deposit energy ($E_{\mu}$) is reconstructed by the observed $Q_{\rm{tot}}$ with a conversion factor
of 250~{p.e.} per MeV. A muon is identified by an event with the deposit energy greater than 70~MeV. 
%Due to the finite size of the RENO detector the muon deposit energy can not exceed a maximum value
%corresponding to its maximum travel length as shown in Fig.~\ref{f:mu_en}.
Due to the saturation of the DAQ electronics the muon deposit energy cannot exceed a maximum value $\sim$1700~MeV 
as shown in Fig.~\ref{f:mu_en}. 
The muon charge correction is obtained from the change of the maximum deposit energy with respect to a reference value.

%+++++++++++++++++++++++++++++++++++++++++++++++++++++++++++++++++++++++
\subsection{Vertex reconstruction}

The event vertex information is useful for removing accidental backgrounds because of their uncorrelated distances between  prompt and delayed candidates. 
A simple and fast method is adopted to reconstruct an event vertex using an individual PMT charge as a weighting factor to the position of a hit PMT. 
A reconstructed vertex, $\vec{r}_{\rm{vtx}}$, is obtained as a charge weighted average of locations of all the hit PMTs, 
\begin{equation}
\vec{r}_{\rm{vtx}} = \frac{\sum_{i}(Q_{i}\cdot\vec{r}_{i})}{\sum_{i}Q_{i}}, 
\end{equation}
where $Q_{i}$ is the charge collected by the $i^{th}$ PMT, and $\vec{r}_{i}$ is a position vector of the PMT from the center of the RENO detector~\cite{RENO_vertex}. 
This method results in $\vec{r}_{\rm vtx}$ with a position dependent offset from the true vertex position mainly due to geometrical effects. 
A correction factor that depends on $\vec{r}_{vtx}$ is obtained using a simple numerical calculations that account a simple geometrical shape of detector 
and the effective attenuation length of ID materials.

\begin{figure}
\begin{center}
\includegraphics[width=0.48\textwidth]{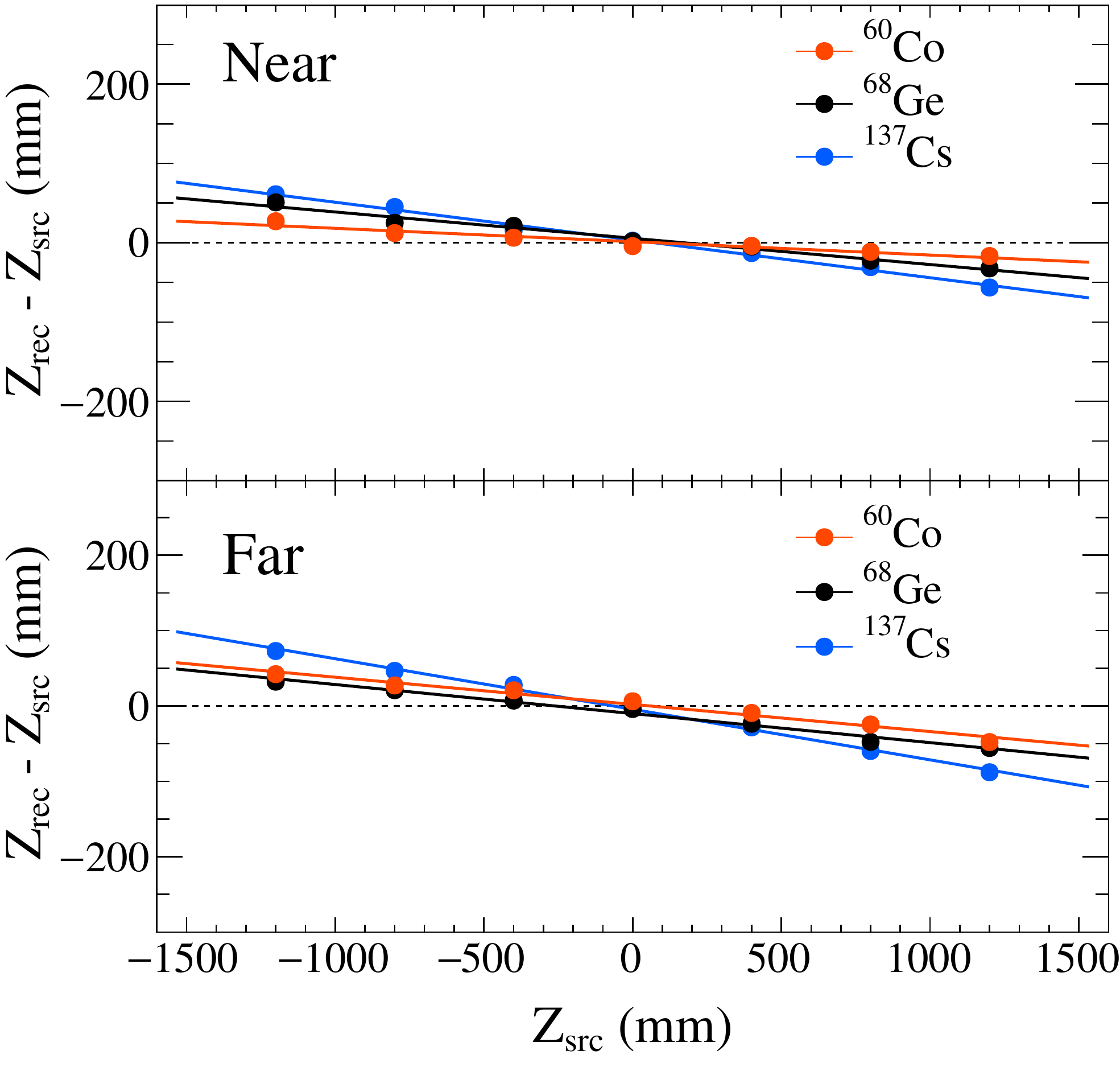}
\end{center}
\caption{(Color online) 
Difference between reconstructed vertices ($Z_{\rm{rec}}$) and actual positions ($Z_{\rm{src}}$) of \Cs, \Ge, and \CoS sources.
The reconstructed vertices show systematic deviations from the true positions at the source locations away from the center. The systematic shifts 
become lessened as the source energy is larger.
}
\label{f:vert_reco}
\end{figure}

The performance of the vertex reconstruction was checked with three calibration source data: \Cs, \Ge, and \Co. 
%Figure~\ref{f:vert_resi} shows the vertex residual distributions of the three source data taken along the z-direction of the far detector.
The vertex resolution is about 20~cm at 1~MeV, and improves at higher energies. 
\iffalse
\begin{figure}
\begin{center}
\includegraphics[width=0.48\textwidth]{figure-07.pdf}
\end{center}
\caption{
Vertex residuals for \Cs, \Ge, and \CoS source data taken in the far detector.
See text for more details.
}
\label{f:vert_resi}
\end{figure}
\fi
Figure~\ref{f:vert_reco} shows a reasonable agreement between the reconstructed and actual source positions. The difference is as large as $\sim$7\% 
for \CsS and less than $\sim$5\% for the other two sources with gamma-ray energies larger than 1~MeV. 
However, such a bias is not really problematic because the requirement of a delayed signal from neutron capture on Gd naturally selects 
the target events without the event vertex information.

%======================================================================
\section{\label{sec:calib} Energy Calibration}

\begin{figure}
\begin{center}
\includegraphics[width=0.48\textwidth]{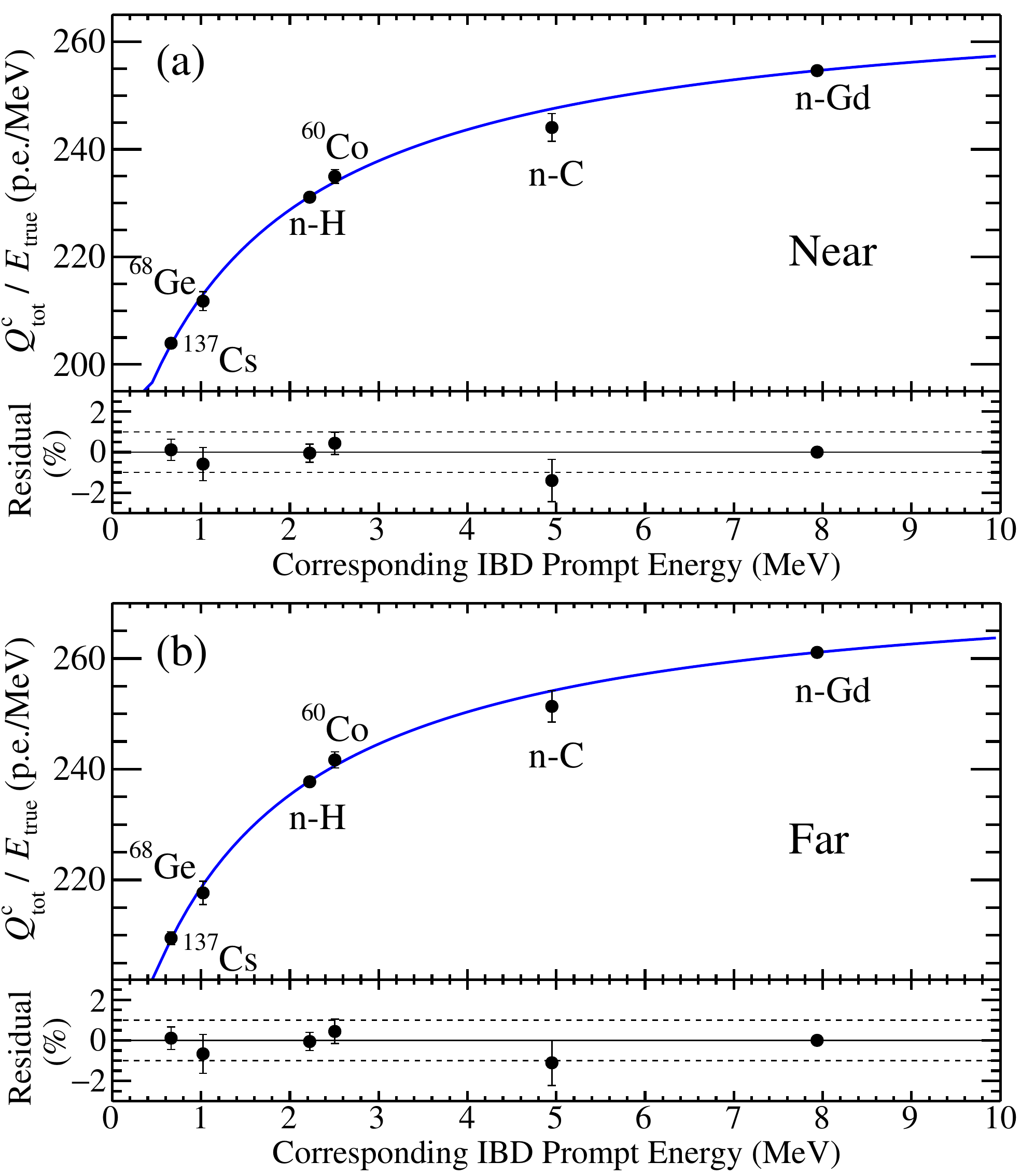}
\end{center}
\caption{
Nonlinear response of scintillating energy obtained from the visible energies of $\gamma$-rays
coming from several radioactive sources and IBD delayed signals in the near and far detectors. 
%The corresponding IBD prompt energy for each source is given by its peak energy of gamma-ray(s).
The curves are the best fits to the data points and the charge-to-energy conversion functions. Note that the n-C sample 
is obtained from the \PoBeS source and the n-H sample from the \CfS source. 
The lower panels show fractional residuals of all calibration data points from the best fit.
}
\label{energyconv}
\end{figure}

An energy measurement is essential for measuring \dmSq and $\theta_{13}$ to a lesser extent. 
To calibrate the energy scale we used the following radioactive sources with a $\mu$Ci level or lower activities: 
\Cs, \Ge, \Co, \PoBe, and \CfS. The source is enclosed in an acrylic container when taking the source data. 
Source data are taken regularly, and their observed charges are corrected for variations of gain, charge collection, and LS attenuation length
using the neutron capture peak energies.
The corrected charges are averaged and used to represent $Q_{\rm{tot}}$ for the peak energy of a $\gamma$-ray source.
%The energy loss due to the source container is estimated with MC calculation and accounted for accordingly.
The total charge $Q_{\rm tot}$, given in p.e., is converted to the corresponding absolute energy in MeV ausing a charge-to-energy conversion function 
obtained through various source calibration samples and neutron capture samples.
The conversion function from $Q_{\rm{tot}}$ to corresponding energy deposited by a positron is generated from the peak energies of these $\gamma$-ray sources.\\
\indent
The observed charges of the source data, taken at the detector center, are also corrected for different charge-response of uniformly distributed events.
The center-to-uniform corrections are $\sim$0.7\% and $\sim$0.5\% for the near and far detectors. 
The energy loss due to the source wrapper and container is estimated with MC calculation and accounted for accordingly.
The RENO MC includes measured optical properties of the LS and quenching effect of the $\gamma$-ray at low energies~\cite{RENO-GdLS}.
The quenching effect depends on the energy and the multiplicity of $\gamma$-ray released from the calibration sources. The MC simulated $Q_{\rm{tot}}$
well reproduces that of $\gamma$-ray source including the quenching effect. \\
\indent
Since a positron loses its kinetic energy via scintillating processes and annihilates with an electron and emitting two $\gamma$-rays, 
its total energy is taken as the true energy ($E_{\rm{true}}$) of positron. The observed $Q_{\rm{tot}}$ of $\gamma$-ray source is converted 
to the corresponding Qtot of a positron ($Q_{\rm{tot}}$) after all the necessary corrections using the {\sc GEANT4}. 
The $Q_{\rm{tot}}$ correction from $\gamma$-ray to positron is performed by taking the $\gamma$-ray source energy as the positron $E_{\rm{true}}$ 
or corresponding IBD prompt energy ($E_{\rm{p}}$). The converted $Q_{\rm{tot}}^{c}$ of IBD prompt energy is estimated by taking into account 
the difference in the visible energies of the $\gamma$-ray and positron through the MC simulation. The uncertainty in $Q_{\rm{tot}}^{c}$ 
due to the correction is largely correlated among data points 
and negligible compared to the source data errors including the time variation of corrected charges. 
The upper panels of Fig.~\ref{energyconv} show the nonlinear response of scintillating energy for the IBD prompt signal 
that is well described by a fitted parametrization and consistent with the MC prediction. The nonlinear response at lower energies is mainly due to the quenching effect
in the scintillator and Cherenkov radiation. 
The following empirical formula is used for the fit function, 
\begin{equation}
\label{eq_pe2MeV}
Q_{\rm{tot}}^{c}/E_{\rm{true}} = P_0 - P_1/[1 - \exp(-P_2 \cdot E_{\rm{true}} - P_3)],
\end{equation}
where $E_{\rm true}$ is in $\rm MeV$. The fit parameters $P_0$ determines a saturation level
$P_{1}$ corresponds to the magnitude of nonlinearity, and $P_{2}$ and $P_{3}$ are related to the shape of the nonlinearity.
The fitted values of the parameters are presented in Table ~\ref{t:econv}.
%The excellent fit value demonstrates the validity of the empirical fit function of the nonlinearity.
Deviation of all calibration data points with respect to the best fit is within 1\% as shown in Fig.~\ref{energyconv} lower panels.
According to the energy calibration, the observed charge $Q_{\rm{tot}}$ at the far detector is $\sim$220 p.e. per MeV at 1~MeV, and $\sim$250 p.e. per MeV at 5~MeV.
\begin{table}[b]
\begin{center}
\caption{\label{t:econv}
The fit parameter values of the charge-to-energy conversion function given in Eq.~\ref{eq_pe2MeV}}. 
\begin{tabular*}{0.48\textwidth}{@{\extracolsep{\fill}} c c c}
\hline
\hline
Parameter & Far & Near\tabularnewline
\hline
$P_0$ & 275.9$\pm$1.0 &  270.1$\pm$1.3\tabularnewline
$P_1$ & (1.698$\pm$0.151)$\times 10^{-2}$ & (1.701$\pm$0.247)$\times 10^{-2}$ \tabularnewline
$P_2$ & (1.228$\pm$0.123)$\times 10^{-4}$ & (1.161$\pm$0.117)$\times 10^{-4}$ \tabularnewline
$P_3$ & (1.735$\pm$0.176)$\times 10^{-4}$ & (1.794$\pm$0.299)$\times 10^{-4}$ \tabularnewline
\hline
\hline
\end{tabular*}
\end{center}
\end{table}

The effective attenuation lengths of the near and far detectors differ by 1.4\% at 430~nm wavelength
that is estimated by PMT charge response to the radioactive source at detector center.
The LS light yields of the two detectors differ by 2.7\% at $\sim$1~MeV. The dead PMT fraction during the data taking reported here is less than 1\% 
for both near and far detectors, and the difference between them is less than 0.5\%. This difference is compensated when the charge-to-energy 
conversion is performed using the conversion function obtained for each detector. 

\begin{figure}
\begin{center}
\includegraphics[width=0.48\textwidth]{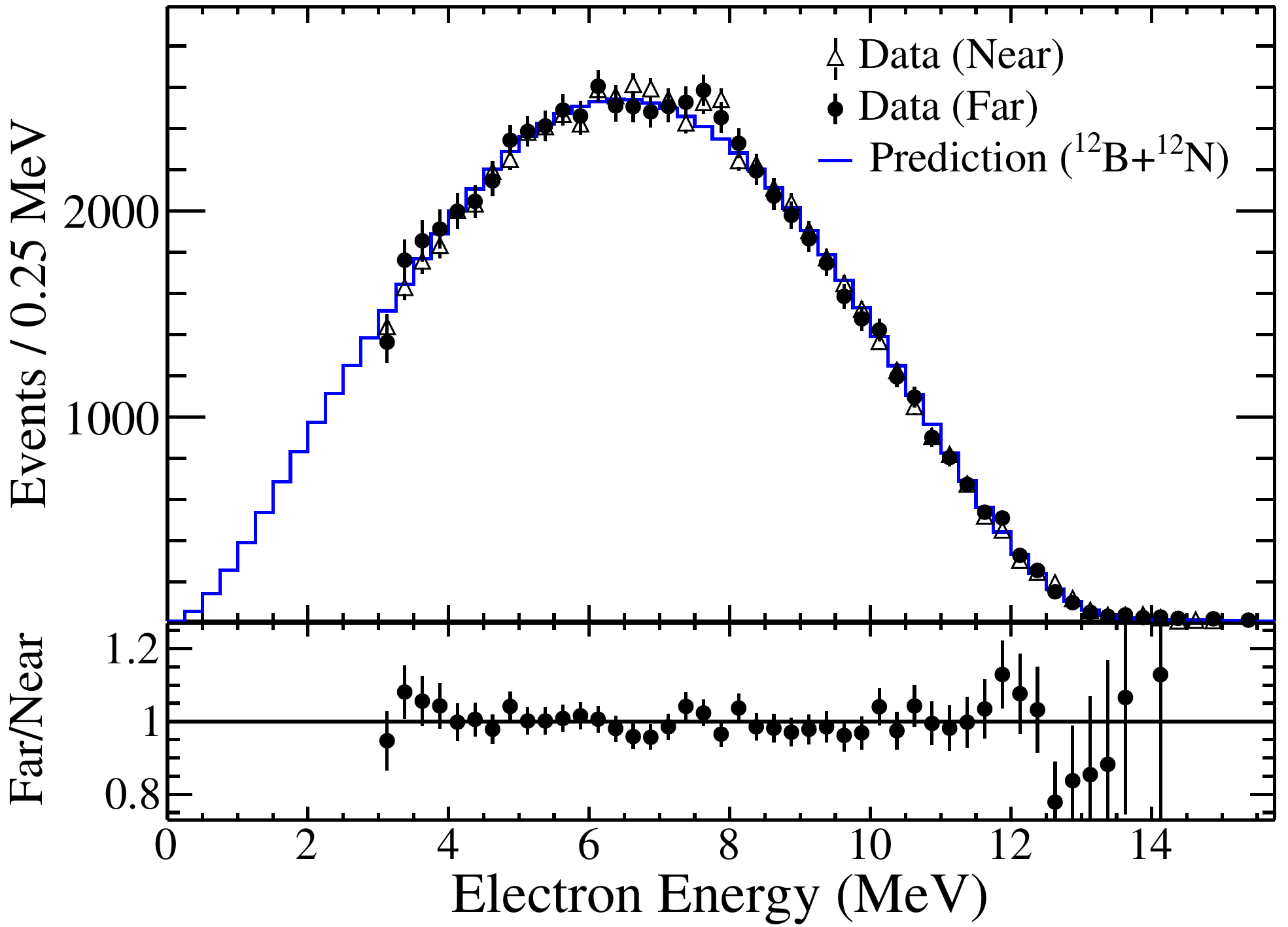}
\end{center}
\caption{
Comparison of measured and simulated energy spectra of the electrons from $\beta$-decay of unstable isotope \B,
with minute contribution from \N, produced by cosmic muons. 
The spectra are overlaid after scaling the total number of events in the near detector to that in the far detector.
The far-to-near ratio of the spectra is shown in the lower panel.
A small excess near 8~MeV is seen in both near and far detectors and may be remaining background events coming from neutron capture by Gd.
}
\label{f:boron}
\end{figure}

Cosmogenic \BS and \NS samples are used to check the validity of the charge-to-energy conversion functions. These isotopes are generated by cosmic muons 
interacting with carbons in the scintillator. The positron charge-to-energy conversion functions are modified to convert the charge in the $\beta$-decay events 
by subtracting a charge value corresponding to the positron annihilation.
Figure~\ref{f:boron} shows good agreements in the energy distributions between near and far data as well as data and MC. 
This demonstrates the obtained parametrization for the nonlinear response of electron scintillating energy works well for energies of 3 to 14~MeV within the statistical 
fluctuation of the data sample. Thus it indicates the positron energy conversion function is valid not only for the IBD energy region up to 8~MeV but also 
for the extended energy region up to 14~MeV.

\begin{figure}
\begin{center}
\includegraphics[width=0.48\textwidth]{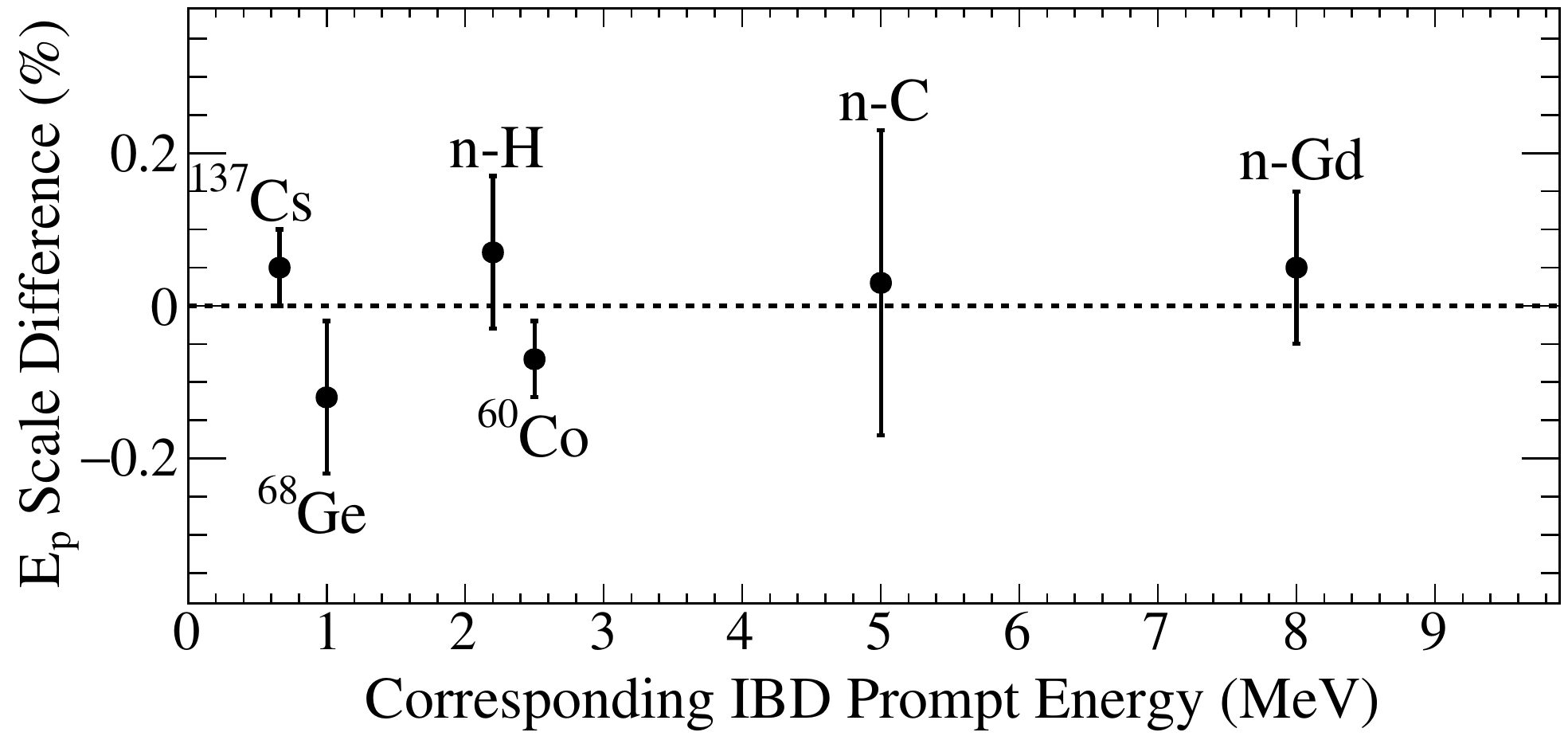}
\end{center}
\caption{
%Energy scale difference of the near and far detectors. All calibration data show the difference less than 0.15\%.
Energy-scale difference between the near and far detectors. The prompt energy difference between the two detectors is estimated by
comparing the energy spectra of $\gamma$-ray sources obtained using the charge-to-energy conversion functions.
All calibration data show the values of the difference less than 0.15\%
}
\label{f:e_uncert}
\end{figure}

The energy-scale difference between the near and far detectors contributes to the uncorrelated systematic uncertainties associated with a relative 
measurement of spectra at two detectors, whereas the correlated uncertainties to the absolute energy scale does not. 
The energy-scale difference is estimated 
by comparing of near and far spectra of calibration data and found to be less than 0.15\% as shown in Fig.~\ref{f:e_uncert}.

\begin{figure}
\begin{center}
\includegraphics[width=0.48\textwidth]{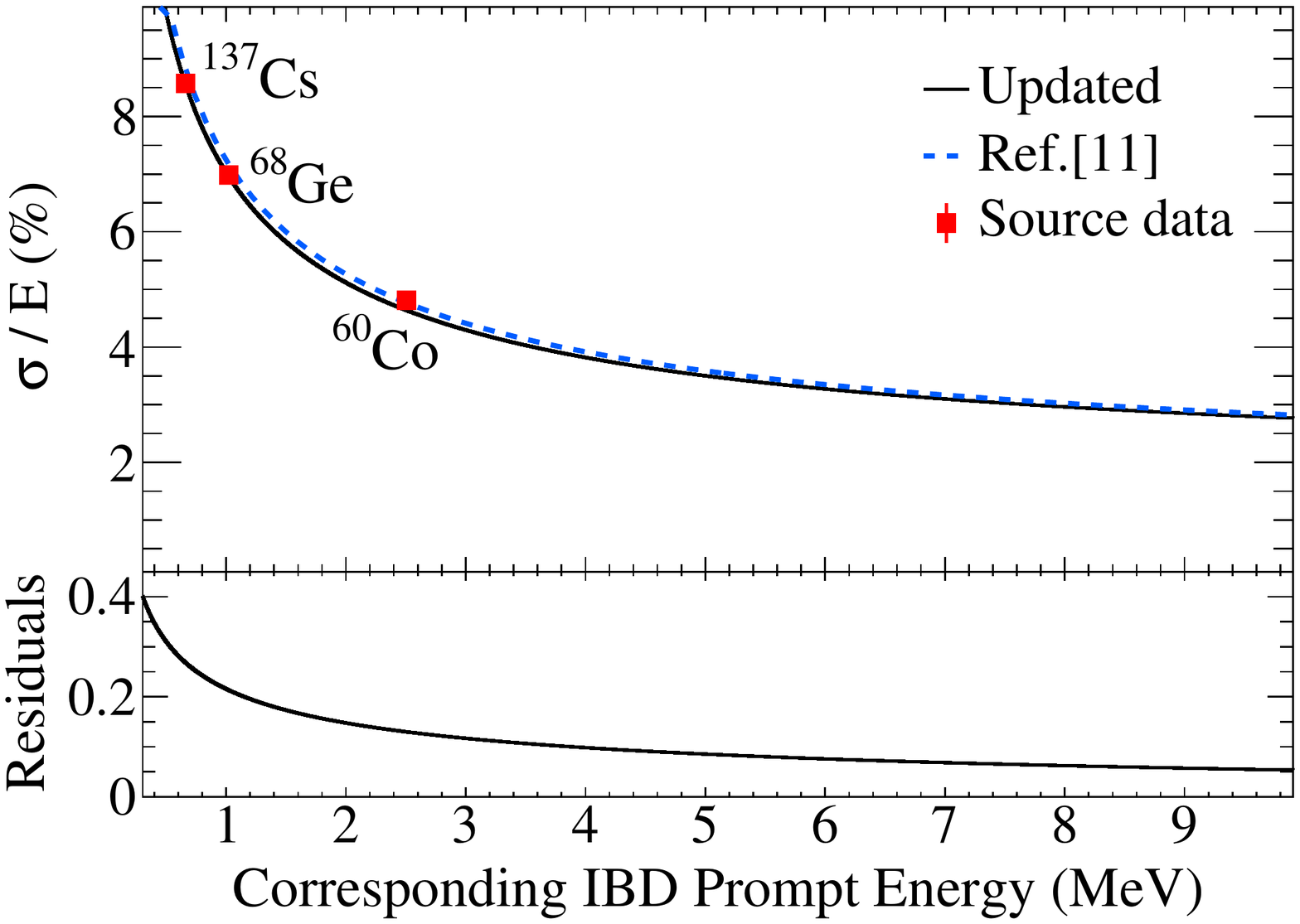}
\end{center}
\caption{
Energy resolution for the far detector as a function of prompt energy. 
The solid curve corresponds to the estimated energy resolution using a MC simulation.
The dotted curve represents the energy resolution used in Ref.~\cite{RENO-PRL2}. 
Their difference is shown as a residual in the lower panel. 
Note that each source data point is given at true energy of a gamma-ray(s).
}
\label{f:e_resol}
\end{figure}

%According to the energy calibration, the observed charge $Q_{\rm{tot}}$ at the far detector is $\sim$220 p.e. per MeV at 1~MeV, and $\sim$250 p.e. per MeV at 5~MeV.
The energy resolution is measured with the calibration data with the radioactive sources placed at the center of the detector. The obtained energy resolution 
is $\sigma /E= 7.9\%/\sqrt{E(\rm{MeV}) + 0.3}$ for the far detector with a comparable energy resolution for the near detector. The discrepancy between data and MC 
is taken into account in MC. The energy resolution is worse by a small amount due to IBD events being uniformly distributed in the target region. 
The difference is estimated to be less than 0.2\%.
The dotted curve is the energy resolution used for the results in Ref.~\cite{RENO-PRL2}.
An updated resolution is obtained to be more appropriate for the uniform IBD events.
The difference between the two energy resolution functions is minimal as shown by their residual distribution in the lower panel of Fig.~\ref{f:e_resol}. 
The measurement of \qOneThree and \dmSq is repeated with the updated energy resolution,
and the obtained values are essentially unchanged except for an increase of $0.01 \times 10^{-3}~{\rm eV}^2$ in the $|\Delta m^2_{ee}|$ value. 
Therefore, the energy resolution function used in Ref.~\cite{RENO-PRL2} is taken for the results in this paper.

%-----------------------------------------------------------------
\section{Backgrounds}

There are several background contributions to the prompt and delayed like events. They are ambient $\gamma$-rays from the surrounding rock and the detector materials, 
neutrons entering into the detector, spallation products produced by the cosmic muons, flashing lights from PMTs, electronic noise, and others. 
Two main background components for the IBD candidates are uncorrelated and correlated pairs of prompt and delayed like events.
Because of a much shallower overburden for the near detector than the far detector, the near detector suffers higher rate of cosmogenic backgrounds.

The uncorrelated IBD background is due to accidental coincidences from the random association of a prompt like event due to radioactivity and a delayed like neutron capture. 
The prompt like events are mostly ambient $\gamma$-rays from the radioactivity in the PMT glasses, LS, and surrounding rock. Most of the ambient radioactivities 
generate $\gamma$-rays of energies below 3~MeV. The delayed like events come from neutrons produced by cosmic muons in the surrounding rocks or in the detector. 

The correlated IBD backgrounds are due to fast neutrons, $\beta$-n emitters from cosmogenic \LiHeS isotopes, and \CfS contamination in the target. 
The fast neutrons are produced by cosmic muons traversing the surrounding rock and the detector. An energetic neutron entering the ID can interact in the LS 
to produce a recoil proton before being captured on Gd. The recoil proton generates scintillation lights mimicking a prompt like event. 
The \LiHeS $\beta$-n emitters are produced mostly by energetic cosmic muons because their production cross sections in carbon increase with muon energy.

The \CfS contamination background comes from the contamination of Gd-LS by a small amount of \CfS that was accidentally introduced into both detectors 
during detector calibrations in October 2012. It is found that the source container did not have a tight seal due to a loose O-ring. 
When the source was submerged in Gd-LS during source calibrations, Gd-LS seeped into the source container and a small amount of dissolved \CfS leaked into Gd-LS.
Among the $\sim$500 day data sample the last 105 (79) days of data in the far (near) detector are contaminated by the \Cf. 
Thus the \CfS background removal criteria to be described later are applied to data taken during these periods. 
It is known that a \CfS decay emits 3.7 neutrons per fission on average with a mean energy of 2.1~MeV per neutron, 
via $\alpha$-emission (96.9\%) and spontaneous fission (3.1\%).

%-----------------------------------------------------------------
\section{Event Selections}

Event selection criteria are applied to obtain IBD candidate events without distorting spectral shape of IBD signal events.
Because an IBD candidate requires a delayed signal from a neutron capture by Gd in Gd-LS, a fiducial volume naturally becomes the entire target region 
without a vertex position requirement.
As a result, the detection efficiency is enhanced by some spill-in of IBD events.

\begin{figure}
\begin{center}
\includegraphics[width=0.48\textwidth]{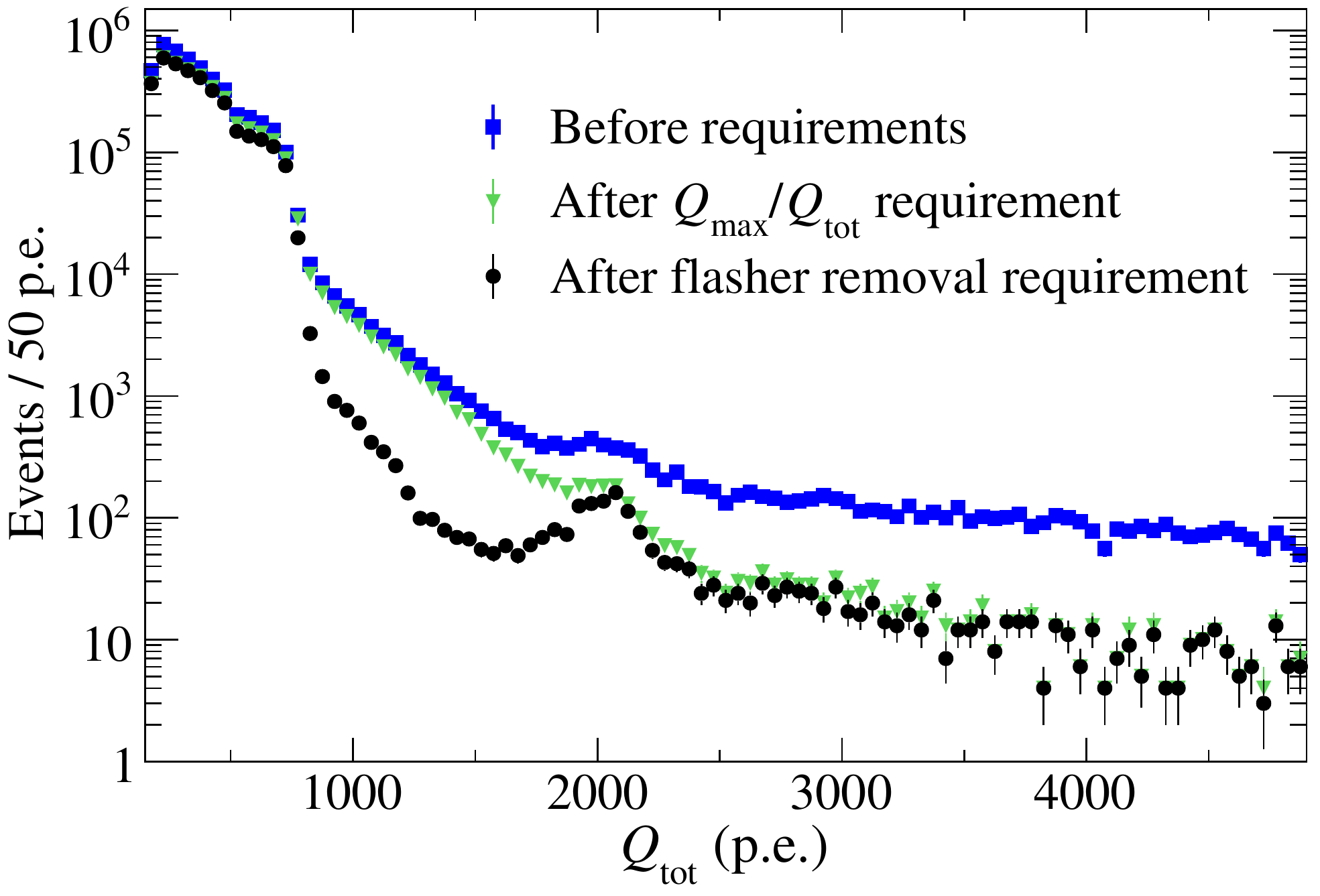}
\end{center}
\caption{(Color online) $Q_{\rm{tot}}$ distributions of events before applying any selection criterion (blue), 
applying $Q_{\rm{max}}/Q_{\rm{tot}} < 0.07$ requirement (green), and flashing PMT removal condition (black). 
}
\label{f:Qtot}
\end{figure}

\begin{figure}
\begin{center}
\includegraphics[width=0.48\textwidth]{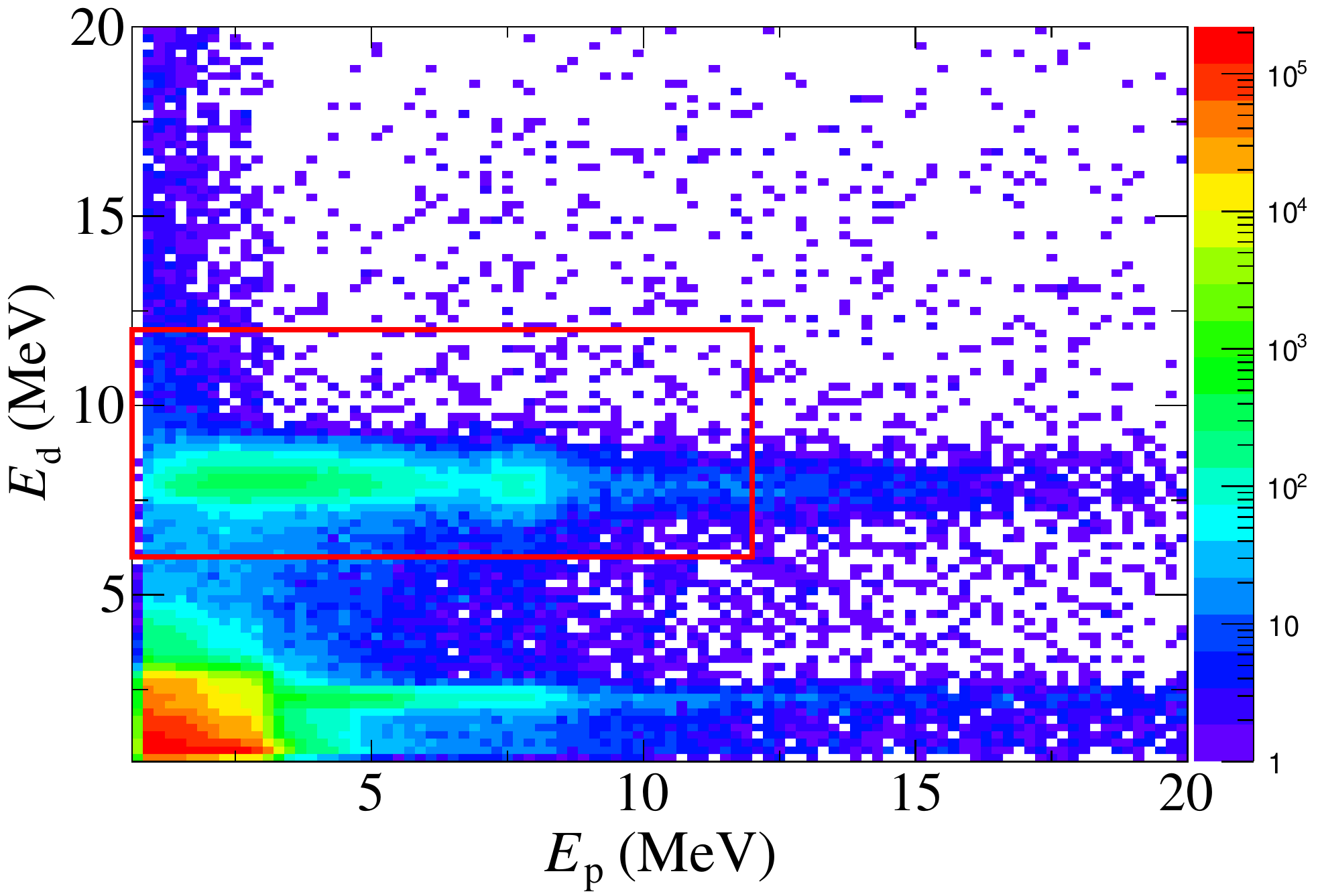}
\end{center}
\caption{(Color online) Prompt versus delayed energies of IBD candidates after the pre-selection criteria using the 400 day data prior to the \CfS contamination. 
Neutron captures on Gd (red box) and H are clearly seen in the delayed energy distribution before full selection criteria. 
Accidental background events highly populate the region below prompt energy of 3~MeV.
}
\label{f:s2_s1}
\end{figure}

Before applying prompt and delayed coincidence criteria, the following three pre-selection criteria are applied to all buffer-only triggered events:
(i) $Q_{\rm{max}}/Q_{\rm{tot}} < 0.07$ where $Q_{\rm{max}}$ is the maximum charge of any single ID PMT, to eliminate external $\gamma$-ray events and flashing PMT events;
(ii) an additional PMT hit timing and charge requirement of $Q_{\rm{max}}/Q_{\rm{tot}} < 0.07$ where an extended timing window of -400 to +800~ns is imposed to calculate
$Q_{\rm{tot}}$ and $Q_{\rm{max}}$ for this criterion, to eliminate events coming from remaining flashing PMTs effectively;
(iii) timing veto criteria to reject events associated with the cosmic muons (a) if they are within a 1~ms window following a cosmic muon of 
$E_{\mu} > $ 70~MeV, or of 20 $< E_{\mu} < $ 70~MeV for OD $N_{\rm{hit}} > $ 50,
%(b) if they are within a 1 ms window following a cosmic muon of 70 MeV $< E_{\mu} < $ 1.4 GeV, 
or (b) if they are within a 700~ms (400~ms, 200~ms) window following a cosmic muon of $E_{\mu} > $ 1.6~GeV (1.5$-$1.6~{\rm GeV}, 1.4$-$1.5~{\rm GeV}) 
for the near detector, 
or within a 700~ms (500~ms, 200~ms) window following a cosmic muon of $E_{\mu} > $ 1.5~GeV (1.2$-$1.5~GeV, 1.0$-$1.2~GeV) for the far detector. 
As shown in Fig.~\ref{f:Qtot}, the selection criteria based on $Q_{\rm{max}}/Q_{\rm{tot}}$ are efficient to eliminate external $\gamma$-ray events and flashing PMT events.
Figure~\ref{f:s2_s1} shows a clean delayed-signal of $\sim$8~MeV $\gamma$-rays from neutron captures on Gd after the pre-selection criteria and a large radioactive background
against 2.2~MeV $\gamma$-rays from neutron captures on hydrogen.

The following criteria are applied to select IBD candidates: 
(iv) a prompt energy requirement of 0.7 $< E_{\rm{p}} <$ 12~MeV; 
(v) a delayed energy requirement of 6 $< E_{\rm{d}} <$ 12~MeV where $E_{\rm{d}}$ is the energy of a delayed like event;
(vi) a time coincidence requirement of 2 $< \Delta t_{e^{+}n} <$ 100~$\mu$s where $\Delta t_{e^{+}n}$ is the time difference between the prompt like and delayed like events;
(vii) a spatial coincidence requirement of $\Delta R <$ 2.5 m where $\Delta R$ is the distance between vertices of the prompt like and delayed like events,
to eliminate remaining accidental backgrounds. The coincidence requirements of a delayed candidate are quite efficient for removing accidental backgrounds 
mostly in the low energy region of $E_{\rm{p}} <$ 3~MeV.

The following multiplicity requirements are applied to remove events of fast neutron, multiple neutrons, and the \CfS contamination background:
(viii) a timing veto requirement for rejecting coincidence pairs (a) if they are accompanied by any preceding ID or OD trigger within a 100~$\mu$s 
window before their prompt candidate, (b) if they are followed by any subsequent ID-only trigger other than those associated with the delayed candidate within a 200~$\mu$s
window from their prompt candidates, (c) if they are followed by any subsequent ID and OD trigger within a 200~{\rm{$\mu$}s} 
window from their prompt candidates, (d) if there are other subsequent pairs within  the 500~$\mu$s 
interval, (e) if they are accompanied by any prompt candidate of $E_{\rm{p}} >$ 0.7~MeV within a 300~$\mu$s 
preceding window or a 1~ms subsequent window, or (f) if they are accompanied by a prompt candidate of 
$E_{\rm{p}} >$ 3~MeV within a 10~s window and a distance of 40~cm; (ix) a spatial veto requirement for rejecting coincidence pairs in the far detector only if the vertices of their prompt candidates are located in a cylindrical volume of 30~cm in radius, centered at $x$ = +12.5~cm and $y$ = +12.5~cm and -170 $< z <$ -120~cm.

The criteria of (viii) (a), (b), (c), (d), and (e) eliminate events due to multiple neutrons or multiple interactions of a neutron with protons in the ID. 
They also eliminate the \CfS contamination background. The criteria (viii) (f) and (ix) are applied to eliminate the \CfS contamination background. 
The criterion (viii) (f) is useful for removing multiple neutron events from the \CfS decays. 
The criterion (ix) removes events from a region highly populated by events from decays of $^{252}$Cf that is thought to have settled down 
at the bottom of the target of the far detector. 

Applying the IBD selection criteria yields 31\,541 (290\,775) candidate events with $E_{\rm{p}}$ between 1.2 and 8.0~MeV for a live time of 489.93 (458.49)~days in the far (near) detector, in the period between August of 2011 and January of 2013. IBD events with $E_{\rm{p}} <$ 1.2~MeV include 
IBD events occuring in or near the tarrget vessel wall that deposit positron kinetic energy in the wall without producing scintillation lights.
These events are reconstructed to have visible energy near the positron annihilation energy of 1.02~MeV and are not well reproduced by the MC prediction. The IBD signal loss by $E_{\rm{p}} >$ 1.2~MeV requirement is roughly 2\% in both detectors.
The prompt events occuring near the target vessel wall could lose some of their energy to nonscintillating target wall and lead to slight modifications of their prompt energies.
However, the energy mismeasurement affects both the near and far detectors in the identical way and thus has a negligible effect on the results. 

The magnitudes and spectral shapes of the remaining backgrounds are estimated using background enriched samples and subtracted from the final IBD candidate samples.

%-----------------------------------------------------------------
\section{Detection efficiency}

\begin{figure}
\begin{center}
\includegraphics[width=0.48\textwidth]{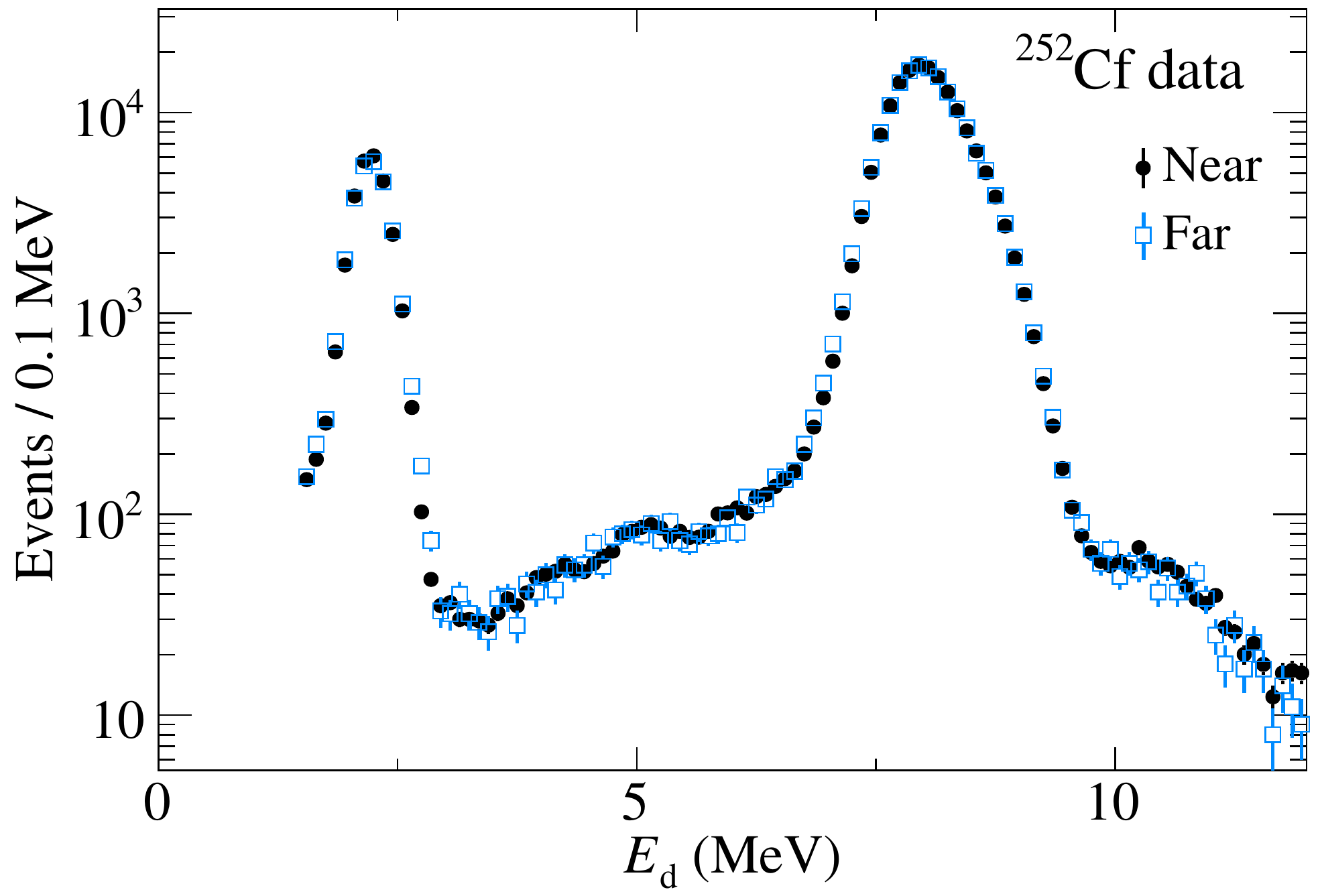}
\end{center}
\caption{\label{f:Cf} 
Comparison of delayed energy distribution of neutron captures on H or Gd using $^{252}$Cf source data.
To obtain the accurate delayed energy spectra the $^{252}$Cf source data were taken for 8 (2) hours on April 2014 (January 2013) for the near (far) detector.
}
\end{figure}

\begin{figure}
\begin{center}
\includegraphics[width=0.47\textwidth]{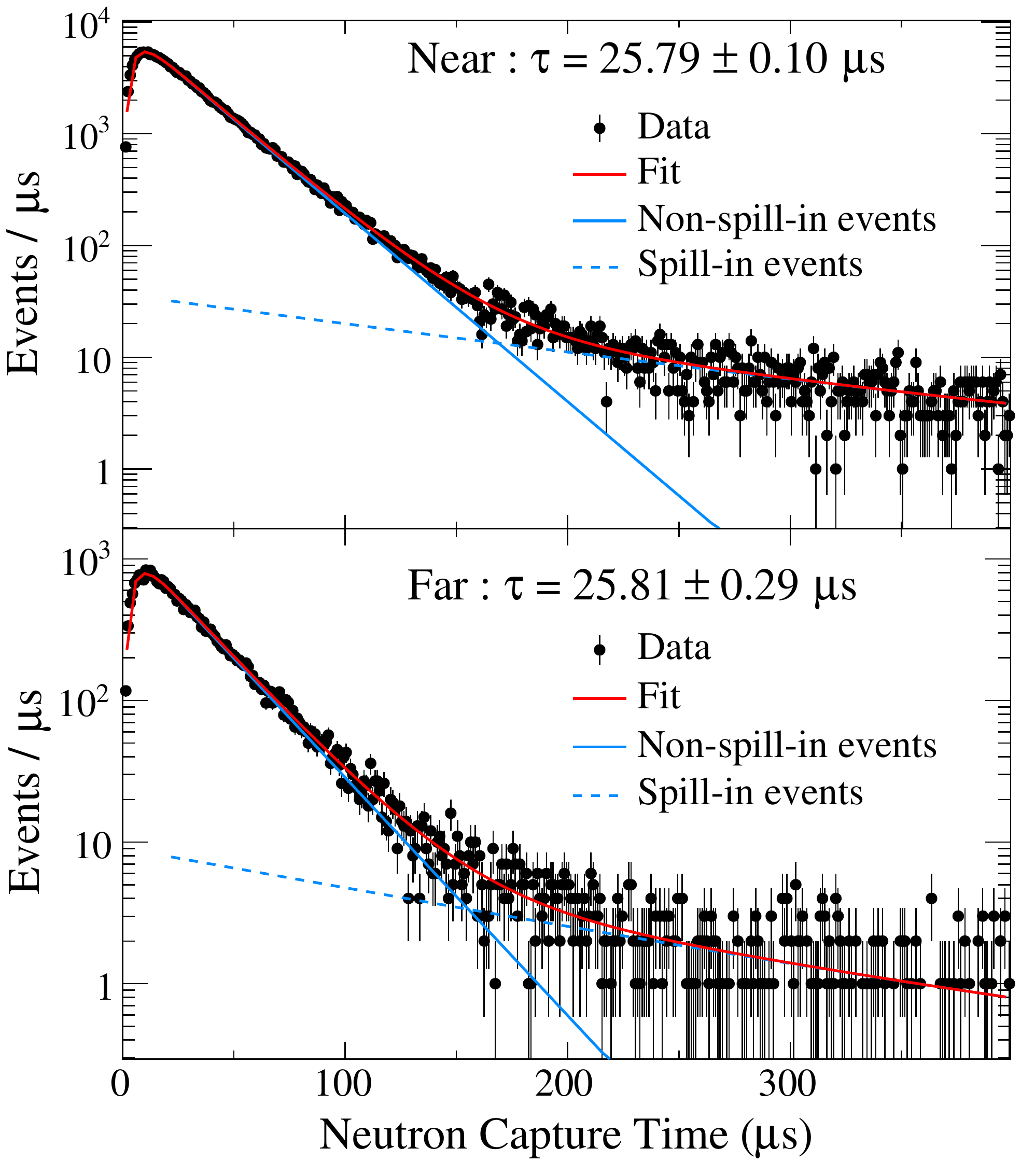}
\end{center}
\caption{
(Color online) Measured time distributions of neutron capture on Gd.
The red solid curves are the fits to the data, and the solid (dotted) blue lines
are the fitted capture time distributions of the IBD events in the central (outer)
region. 
The mean capture times in the central region of the near and far detectors are found to be consistent with each other.
The outer delayed events originating from
the vicinity of the target vessel walls show longer capture time values. A total of 1\,400
live days of data is used in these plots for more precise measurements.
}
\label{f:dT}
\end{figure}

%Detection efficiency is the fraction of the observed to the produced IBD signal events in the detector. A signal %loss comes from the imperfect response of the detector 
%and IBD selection criteria.
The detection efficiency uncertainties are categorized into correlated and uncorrelated uncertainties between the near and far detectors.
The correlated uncertainty is common to both near and far detectors and thus cancelled out for the far-to-near relative measurement
while the uncorrelated uncertainty remains with no cancellation. 
An individual detector efficiency is measured from an IBD signal enriched sample, and its uncertainty is given by a statistical uncertainty 
and uncorrelated and correlated systematic uncertainties. 
The detection efficiencies for common event selection criteria (i) to (viii) for both near and far detectors are assumed to be the same 
since the both detectors are believed to have identical performances. Therefore, the weighted mean of near and far efficiencies for each selection criterion 
is taken to be the efficiency. The systematic error of the average efficiency is estimated from data and MC.
The IBD signal enriched samples are not large enough to find all of the uncorrelated systematic uncertainties by the difference of the measured detection efficiencies. 
Some of the uncorrelated systematic uncertainties are estimated from the possible difference in properties and performances between the two detectors 
if the IBD signal enriched sample is small.
In this section, we present detection efficiencies and their uncertainties for the IBD signal events at 1.2 $< E_{\rm{p}} <$ 8.0~MeV. \\
\indent An expected number of IBD interactions is determined by reactor fluxes, an IBD cross section, and a total number of free protons in the target. 
The uncertainty of the IBD cross section from a theoretical calculation~\cite{ibd_XS} is 0.13\% and can be ignored by the relative measurement. The number of 
free protons in the target is estimated as
$(1.189\pm0.003) \times 10^{30}$, based on the measurements of LAB density (0.856$\pm$0.001 g/cm$^3$) and target volume~\cite{RENO-GdLS}. 
The uncorrelated systematic uncertainty of the number of free protons is 0.03\%, estimated from the measured volume difference of four liters 
between the near and far target vessels~\cite{RENO-acrylic}. The correlated uncertainty is 0.1\%, 
estimated from the resolution of a densitometer.\\
\indent
% trigger effi.
The trigger efficiency is determined by the IBD signal loss due to the requirement of ID $N_{\rm{hit}} >$ 90.
The RENO Monte Carlo simulation (MC), which is described later, does not reproduce the data $N_{\rm hit}$ well
due to lack of realistic individual-channel simulation for the p.e. threshold and dark or noise hits.
According to comparison of $N_{\rm{hit}}$ distribution between data and MC, 
a MC equivalent requirement of $N_{\rm{hit}} >$ 84 is found to accept a buffer-only trigger.
Using the MC equivalent hit requirement, the trigger efficiency for the IBD signal excluding spill-in
events in the near (far) detector is estimated as 99.77$\pm$0.05\% (99.78$\pm$0.13\%) where spill-in events are events that occur outside the target
and produce a neutron capture on Gd in the target. 
The trigger efficiency is also measured for the events at detector center using radioactive sources and consistent with the MC result within the uncertainty.
The position dependent DAQ inefficiency contributes to the inefficiency near the trigger threshold below $\sim$0.8~MeV.
Our measured trigger efficiency using a \CsS source (E = 0.63~MeV) is roughly 50\% at the threshold energy of 0.5$\sim$0.6~MeV and almost 100\% at 0.8~MeV.
The uncorrelated systematic uncertainty of the trigger efficiency is estimated as 0.01\% from the difference between near and far efficiencies.
The correlated uncertainty of the trigger efficiency is estimated as 0.01\% from the ambiguity in finding a MC equivalent $N_{\rm{hit}}$ threshold. \\
\iffalse
A main trigger for an IBD candidate event requires ID $N_{\rm{hit}} >$ 90 in a 50 ns time window. As described earlier, the trigger efficiency for the IBD signal 
excluding spill-in events is estimated as 99.77\% using the MC measured values of 99.77$\pm$0.05\% and 99.78$\pm$0.13\% for the near and far detector, respectively. 
Both uncorrelated and correlated systematic uncertainties are estimated as 0.01\%. 
The uncorrelated systematic uncertainty is obtained by the MC estimated efficiency difference between the near and far detectors.
The correlated systematic uncertainty is estimated from the ambiguity of the MC trigger hit threshold.\\
\fi
\indent
% Qm/Qt cut
The efficiency of the $Q_{\rm{max}}/Q_{\rm{tot}} <$ 0.07 criterion is obtained using an IBD candidate sample of almost no accidental background events that are selected 
by a stringent spatial-correlation requirement of $\Delta R <$ 0.3 m. The $Q_{\rm{max}}/Q_{\rm{tot}}$ distribution of this sample predicts an expected IBD signal loss 
in the region of $Q_{\rm{max}}/Q_{\rm{tot}} >$ 0.07, by extrapolating from the region of $Q_{\rm{max}}/Q_{\rm{tot}} <$ 0.07 using an expected shape of MC. 
The efficiency is estimated as 99.99\% using the measured values of 99.996$\pm$0.003(stat.)\% and 99.98$\pm$0.01(stat.)\% for the near and far detectors, respectively.
%The efficiency is estimated to be $>$99.98\% (99.98$\pm$0.01\%) from the near (far) data.
The correlated uncertainty is estimated from the ambiguity of the extrapolation and found to be 0.01\%.
The uncorrelated systematic uncertainty is estimated from the obtained efficiency difference between the near and far detectors and found to be 0.02\%.\\
\indent
% prompt energy cut
The efficiency of the prompt energy requirement is obtained from the fraction of events in the region of 1.2 $< E_{\rm{p}} <$ 8.0~MeV relative to total IBD events 
and estimated as 98.77\% using the measured values of 98.78$\pm$0.03(stat.)\% and 98.66$\pm$0.09(stat.)\% for the near and far detectors, respectively.
The uncorrelated systematic uncertainty is estimated to be 0.01\% by varying the energy threshold according to the energy-scale difference of 0.15\%
between the near and far detectors. The correlated uncertainty is estimated to be 0.09\% 
by varying the energy threshold according to the energy-scale uncertainty of 1.0\%. \\
\indent
% delayed energy cut
The efficiency of the delayed energy requirement is determined by the fraction of delayed events in the region of 
$6 < E_{\rm{d}} < 12$~MeV out of total delayed events of neutron capture on Gd. 
An IBD event enriched sample is used for the efficiency estimation and obtained by requiring IBD candidates to have 4 $< E_{\rm{p}} <$ 8~MeV 
to eliminate accidental and fast neutron backgrounds and 3.5 $< E_{\rm{d}} <$ 12~MeV to accept lower energy delayed events. 
According to a MC simulation, 1.16\% of the total delayed events are found at $E_{\rm{d}} <$ 3.5~MeV.
With this correction, the efficiency is estimated as 92.14\% using the measured values of 92.15$\pm$0.08(stat.)\% and 92.05$\pm$0.26(stat.)\% 
from the near and far IBD event enriched samples, respectively. 
The correlated uncertainty is estimated to be 0.50\% by considering the MC corr MC correction uncertainty below 3.5~MeV 
and varying the energy scale by its uncertainty of 1.0\%. 
The uncorrelated systematic uncertainty is estimated to be 0.05\% by changing the delayed energy requirement by 
$\pm$0.15\%, the energy scale difference between the near and far detectors.\\
\indent
% Gd capture ratio % Figure
The Gd capture fraction is measured by the ratio of neutron captures on Gd to total neutron captures on Gd or H using \CfS source data that are taken at the detector center. 
The effects of spill-in/out events at the target boundary are treated separately and described later. 
A $^{252}$Cf source sample including H capture delayed events is obtained by requiring prompt and delayed event pairs satisfying $4<E_{\rm p}<12$~MeV 
and $1.5<E_{\rm d}<12$~MeV, respectively.
An additional neutron candidate of 1.5 $< E_{\rm{d}} <$ 3~MeV or 6 $< E_{\rm{d}} <$ 10~MeV within 200~$\mu$s 
from the prompt event of a coincidece pair is required to ensure 
the delayed events are neutron capture events originating from $^{252}$Cf decay.
The obtained delayed-energy distributions show a good agreement between near and far detectors as shown in Fig.~\ref{f:Cf}. 
We obtain the Gd capture fraction by the ratio of the n-Gd events with $E_{\rm{d}} >$ 3.5~MeV 
to the total neutron capture events with $E_{\rm{d}} >$ 1.5~MeV. 
A MC simulation finds contributions of neutron captures on Gd below 3.5~MeV and of neutron captures on H below 1.5~MeV. 
With these contributions the Gd capture fraction is estimated as 85.45\% using the measured values of 85.49$\pm$0.03(stat.)\% and 85.40$\pm$0.07(stat.)\% 
from the near and far data, respectively, while it is obtained as 88.41\% from the MC. 
The measured values of the Gd capture fractions are constant in time within their uncertainties. 
The correlated uncertainty is estimated as 0.47\% mostly due to the uncertainty of the n-Gd capture cross section~\cite{nGd_XS}.
The uncorrelated systematic uncertainty is estimated as 0.1\% due to the difference of Gd concentration between the near and far detectors. 
The difference is estimated to be less than 0.1\% from the precision of dividing the Gd-LS equally for the two detectors. \\
\indent
% dT cut: 
The efficiency of the time coincidence requirement is determined by the fraction of IBD events with 2 $< \Delta t_{e^{+}n} <$ 100~{\rm $\mu$s} 
out of total IBD events. An IBD signal enriched sample is obtained by requiring IBD candidate events with 4 $< E_{\rm{p}} <$ 8~MeV in order to eliminate accidental backgrounds. 
Figure~\ref{f:dT} shows $\Delta t_{e^{+}n}$ distributions of the neutron capture on Gd for the near and far IBD signal enriched samples.
The fits to data are made by two exponential functions plus a constant that are multiplied by one minus an exponential function. 
The distributions are well described by
\begin{equation}
\begin{aligned}
N(t) = [p_0 \exp(-t/p_1) + p_2 \exp(-t/p_3) + C] \\
\cdot [1 - p_4 \exp(-t/p_5)]
\end{aligned}
\end{equation}
where the parameters of $p_0$, $p_1$, $p_4$, $p_5$ and $C$ are determined by a fit to the data.
Note that the parameters $p_2$ and $p_3$ of the second exponential function are estimated using the MC.
The first exponential function represents the capture time distribution of the IBD events without the spill-in events in the target region. 
The second exponential function is necessary to extract the contribution of the delayed events originating from the vicinity of the target vessel wall.
The delayed signal of a spill-in event tends to have a longer capture time because of its drift from the $\gamma$-catcher to the target. 
The third exponential function describes the rising capture time behavior below $\sim$10~$\mu$s where the IBD neutron is thermalized before capture.
The efficiency in the central region is obtained by the fraction of IBD events with 2 $< \Delta t_{e^{+}n} <$ 100~$\mu$s 
out of the total IBD events that are estimated from the fitted mean value of capture time using the first exponential function. 
The measured capture time values for non-spill-in events are consistent between near and far detectors. 
To obtain the efficiency of non-spill-in events, a MC simulation is used to estimate the contribution of spill-in events inside the target. 
The efficiency is estimated as 96.59\% using the measured values of 96.60$\pm$0.04(stat.)\% and 96.57$\pm$0.10(stat.)\% from the near and far data, respectively. 
The correlated uncertainty is estimated to be 0.26\% from the uncertainty associated
with a rising capture time of a delayed signal.
%The correlated uncertainty is estimated to be 0.45\% from the difference of this efficiency between data and MC at $\Delta t_{e^{+}n} <$ 2 $\mu$s. 
The uncorrelated systematic uncertainty is estimated as 0.01\% from the uncertainty of Gd concentration difference, $\sim$0.1\%, between the near and far detectors.\\
% dR cut: 
\indent
The efficiency of the spatial coincidence requirement, $\Delta R$ $< 2.5$ m is obtained from IBD candidates with $Q_{\rm{max}}/Q_{\rm{tot}} <$ 0.02.
The efficiency is estimated as 100.00\% using the measured values of 99.99$\pm$0.01(stat.)\% and 100.00$\pm$0.01(stat.)\% from the near and far data, respectively, 
assuming 100\% at $\Delta R$ $< 5$ m. 
%The efficiency is estimated as 99.99$\pm$0.01\% ($>$99.99\%) from the near (far) data assuming 100\% at $\Delta R$ $< 5$ m.
The correlated uncertainty is estimated as 0.02\% based on changing the $\Delta R$ requirement by the resolution of reconstructed vertex, 0.3 m. 
The uncorrelated systematic uncertainty is estimated as 0.02\% from the efficiency difference between the near and far detectors.\\
\indent
% spill-in efficiency
The spill-in events enhance the detection efficiency of IBD signals in the target because of additional IBD signals occurring outside the target 
but with its neutron capture by Gd in the target. 
On the other hand the reactor \nuebS interaction occurring in the target edge may be lost because of a neutron capture in the $\gamma$-catcher region by H. 
Such an event loss is accounted in the delayed energy requirement efficiency. 
The enhanced detection efficiency due to the spill-in events is estimated as 102.00\% using the measured values of 102.02\% and 101.98\% using near and far MC simulation, 
respectively. 
The uncorrelated systematic uncertainty is estimated as 0.04\% due to differences of the Gd concentration and the acrylic wall thickness of the target vessel 
between the near and far detectors. 
The correlated uncertainty is estimated as 0.61\% based on the delayed time distribution of spill-in events at $\Delta t_{e^{+}n} >$ 200~$\mu$s 
deviating from that of IBD events in the target.\\
\indent
The detection efficiencies of selection criteria that are applied to both near and far detectors are summarized in Table~\ref{t:deteff}.
\begin{table}
\caption{
Average detection efficiencies and their uncertainties of common selection criteria that are applied to both near and far detectors for the IBD candidates.
The average detection efficiency is a statistical error weighted mean of the near and far measured values.
%The error weight includes the uncorrelated systematic uncertainties in addition to the statistical errors.
}
\label{t:deteff}
\begin{center}
%\begin{tabular*}{0.96\textwidth}{@{\extracolsep{\fill}} l c c}
\begin{tabular*}{0.48\textwidth}{@{\extracolsep{\fill}} l c c c}\hline\hline
 & Efficiency & Uncorrelated & Correlated \\
 & (\%) & (\%)  & (\%) \\
\hline
IBD cross section & --  & --  & 0.13 \\ 
Target protons & -- & 0.03 & 0.1 \\
\hline
Trigger efficiency & 99.77 & 0.01 & 0.01 \\
$Q_{\rm{max}}/Q_{\rm{tot}}$, anti-flasher & 99.99 &  0.02  & 0.01 \\
Prompt energy & 98.77 & 0.01  & 0.09 \\
Delayed energy & 92.14 & 0.05 & 0.50 \\
Gd capture fraction & 85.45 & 0.10 & 0.47 \\
Time coincidence & 96.59 & 0.01 & 0.26 \\
Spatial correlation & 100.00  & 0.02 & 0.02 \\
Spill-in & 102.00 & 0.04 & 0.61 \\
\hline
Total detection & 76.47 & 0.13 & 0.97 \\
efficiency & & & \\
\hline
\hline
\end{tabular*}
\end{center}
\end{table}
Their identical performances minimize the uncorrelated systematic uncertainties and allow cancellation of the correlated systematic uncertainties 
for the ratio measurement. 
The measured efficiencies in total are 76.51$\pm$0.10(stat.)\% and 76.20$\pm$0.30(stat.)\% for the near and far detectors, respectively, 
with common uncorrelated (0.13\%) and correlated (0.97\%) uncertainties. 
The average efficiency for each selection criterion is calculated as an error weighted mean of the near and far measured values. 
The error weighting is done using a statistical error. 
The average efficiency in total is obtained as 76.47$\pm$0.16\% where the error is calculated by adding all the selection-efficiency errors in quadrature.
The near and far detection efficiencies differ from the total average efficiency by 0.07\% (near) and 0.24\% (far), respectively, and the differences are reasonably 
within the statistical errors. 
IBD signal enriched samples for some selection criteria, 
due to their small sizes, do not allow direct checks of the estimated uncorrelated uncertainties by the difference of the measured near and far efficiencies. 
With larger IBD signal enriched samples, especially in the far detector, the uncorrelated systematic uncertainties are expected to be improved in the future. 
In the rate and spectral fit the uncertainty of the far-to-near detection ratio is taken into account for one of pull parameter uncertainties. 
We obtain the uncorrelated uncertainty of the efficiency ratio as 0.21\% from combining the uncorrelated uncertainty and the weighted statistical errors 
of the measured values. We take 0.20\%
%~\footnote{The final results are almost identical using either 0.20\% or 0.21\%.
%The difference is change of the $|\Delta m_{ee}^2|$ mean value from $2.62\times 10^{-3}$~eV$^2$ to $2.63\times 10^{-3}$~eV$^2$.}
as the value of the efficiency ratio uncertainty, the same as our published result~\cite{RENO-PRL2}
because both values give essentially identical systematic errors.

\indent
Among the IBD selection criteria, the muon and multiplicity timing veto requirements are applied differently to the near and far detectors, 
and thus introduce no correlation at all between the detectors. 
The IBD signal loss due to the muon veto requirements are 21.558\% and 11.133\% for the near and far detectors, respectively,
with both of their uncertainties less than 0.03\%. 
The total IBD signal loss due to the timing veto efficiency is 27.364$\pm$0.007\% (14.691$\pm$0.021\%) for the near (far) detector as summarized in Table~\ref{t:sel_effi}.
\begin{table*}
\caption{\label{t:sel_effi} Summary of the IBD signal loss due to timing veto criteria.
The criterion with ($*$) is applied only to the $^{252}$Cf contaminated data of $\sim$100 days. 
Note that the uncertainties are treated as being fully uncorrelated between the near and far detectors.
}
\begin{center}
\begin{tabular*}{0.96\textwidth}{@{\extracolsep{\fill}} l c c}
\hline
\hline
& Signal loss (\%) & Signal loss(\%) \\
%\hline
Timing veto criteria & Near & Far \\
\hline
$\bullet$ Timing criteria associated with muon & 21.558$\pm$0.003 & 11.133$\pm$0.003 \\
$\bullet$ Adjacent IBD pair within 500~$\mu$s &  0 & 0 \\
$\bullet$ IBD candidate accompanied by any trigger &  4.672$\pm$0.001 & 1.309$\pm$0.001 \\
within 100~$\mu$s preceding time window &  & \\
$\bullet$ IBD candidate accompanied by ID-only trigger &  1.134$\pm$0.001 & 1.424$\pm$0.001 \\
within 200~$\mu$s subsequent time window &  & \\
$\bullet$ IBD candidate accompanied by prompt candidate & 0.605$\pm$0.001 & 0.163$\pm$0.001  \\
within 300~$\mu$s preceding time window & & \\
$\bullet$ IBD candidate accompanied by prompt candidate & 0.258$\pm$0.001 & 0.638$\pm$0.001 \\
within 1~ms subsequent time window & & \\
$\bullet$ IBD candidate accompanied by ID and OD triggers & 0.408$\pm$0.001 & 0.069$\pm$0.001 \\
within 200~$\mu$s subsequent time window & & \\
$\bullet$ IBD candidate accompanied by prompt candidate & 0.491$\pm$0.006 & 0.388$\pm$0.020 \\
($>$ 3~MeV) occurring within 10 sec and 40~cm ($*$) & & \\
%Hot spot removal & N/A &  0.073\pm0.004 \\
\hline
Combined IBD signal loss & 27.364$\pm$0.007 &  14.691$\pm$0.021 \\
\hline
\hline
\end{tabular*}
\end{center}
\end{table*}
%\clearpage

%===============================================================================
\section{\label{sec:bkg} Remaining Background Estimations \& Their Uncertainties}

%In the final data samples, there are some ``random coincidence'' and ``delayed coincidence'' background events which survive the IBD selection requirements. 
%We have not seen any significant evidence of other possible delayed coincidence backgrounds.
%Removing these remaining background by applying further selection criteria would require a big loss of signal. 
The remaining backgrounds after event selection requirements are subtracted from the final IBD candidate sample. 
%Correct subtraction of the backgrounds is necessary in order to obtain the prompt spectra and the rates of IBD signals for the spectral neutrino-oscillation analysis. 
The following subsections describe how to obtain the spectral shapes and rates of the remaining backgrounds. 
Since the rates and shapes of all the remaining backgrounds are measured from background enriched samples, 
their uncertainties are expected to be further reduced with more data.

%-----------------------------------------------------------------
\subsection{Accidental background}

\begin{figure}
\begin{center}
\includegraphics[width=0.48\textwidth]{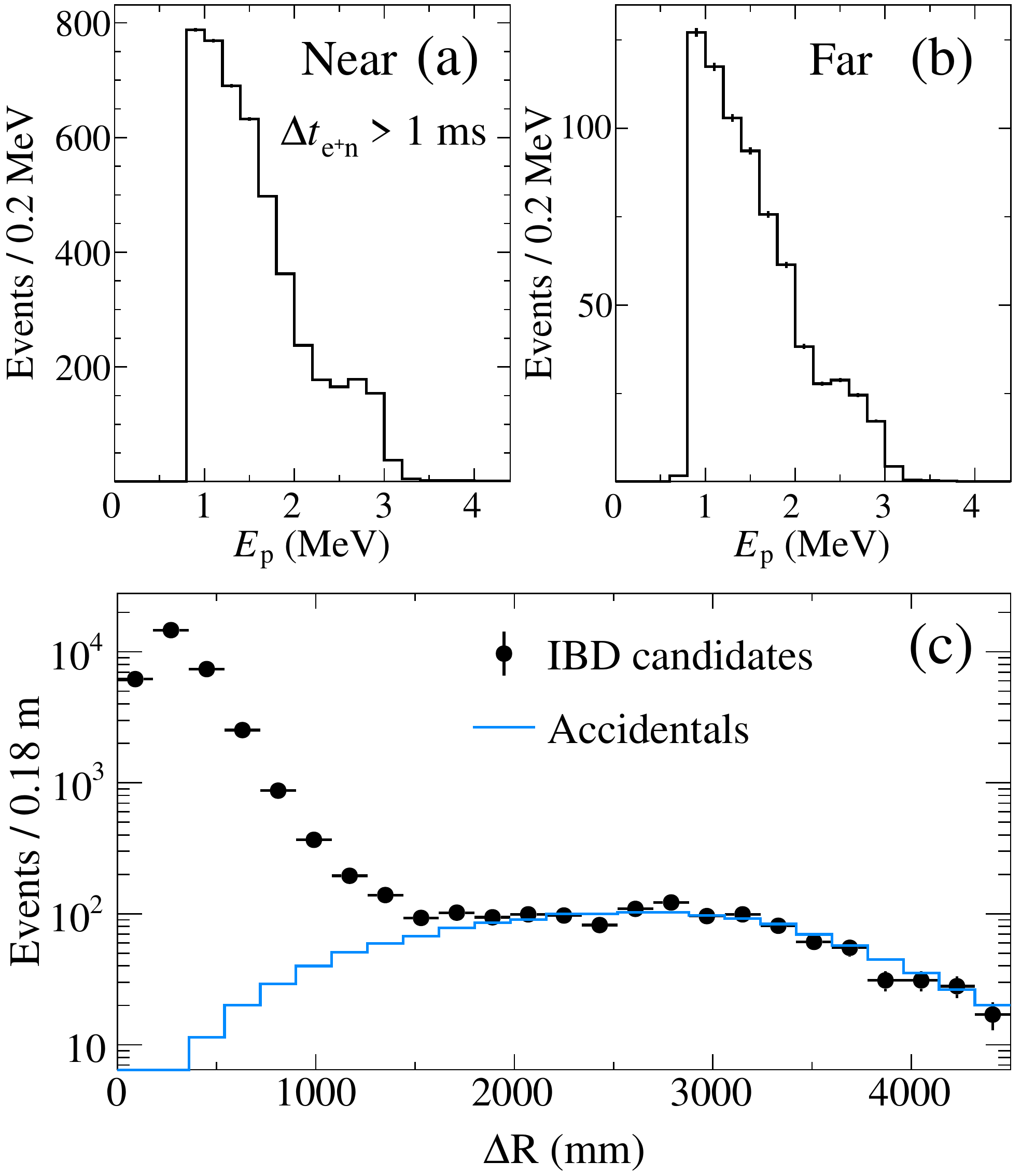}
\end{center}
\caption{\label{f:acci} 
(a) and (b) Prompt energy spectra of accidental backgrounds obtained from accidental background enriched samples that are selected by temporal association larger than 1~ms. 
They are normalized to the remaining background.
The error bars represent statistical and spectral shape uncertainties.
(c) Spatial correlation ($\Delta R$) distribution of IBD candidates with no $\Delta R$ requirement.
The amount of accidental background is obtained by a fit to data using the $\Delta R$ distribution from the accidental background enriched sample.
}
\end{figure}

Most of accidental background events are eliminated by requiring timing and spatial coincidence between the prompt like and delayed like events. 
An accidental background sample is obtained by requiring temporal dissociation between prompt and delayed like events, i.e., $\Delta t_{e^{+}n} >$ 1~ms
for the IBD sample with no $\Delta R$ requirement. 
The prompt energy spectra of the accidental backgrounds of the near and far detectors are shown in Fig.~\ref{f:acci} (a) and (b)
The energy-bin-uncorrelated uncertainty in the accidental background spectrum is obtained from the statistical error of the background enriched sample 
and estimated as 0.02 (0.01) events per day for the near (far) detector. \\
\indent
The remaining rate in the final sample is estimated by measuring the rate of random spatial associations in the IBD signal region of $\Delta R <$ 2.5~m, extrapolated 
from the background dominant region of $\Delta R >$ 1.75~m using the $\Delta R$ distribution of the accidental background spectrum as shown in Fig.~\ref{f:acci} (c).
The energy-bin-correlated uncertainty is obtained from the  fitting error and estimated as 0.08 (0.03) events per day for the near (far) detector. 
The obtained accidental-background rates are 6.89$\pm$0.09 (near) and 0.97$\pm$0.03 (far) events per day.

%-----------------------------------------------------------------
\subsection{Fast neutron background}

\begin{figure}
\begin{center}
\includegraphics[width=0.48\textwidth]{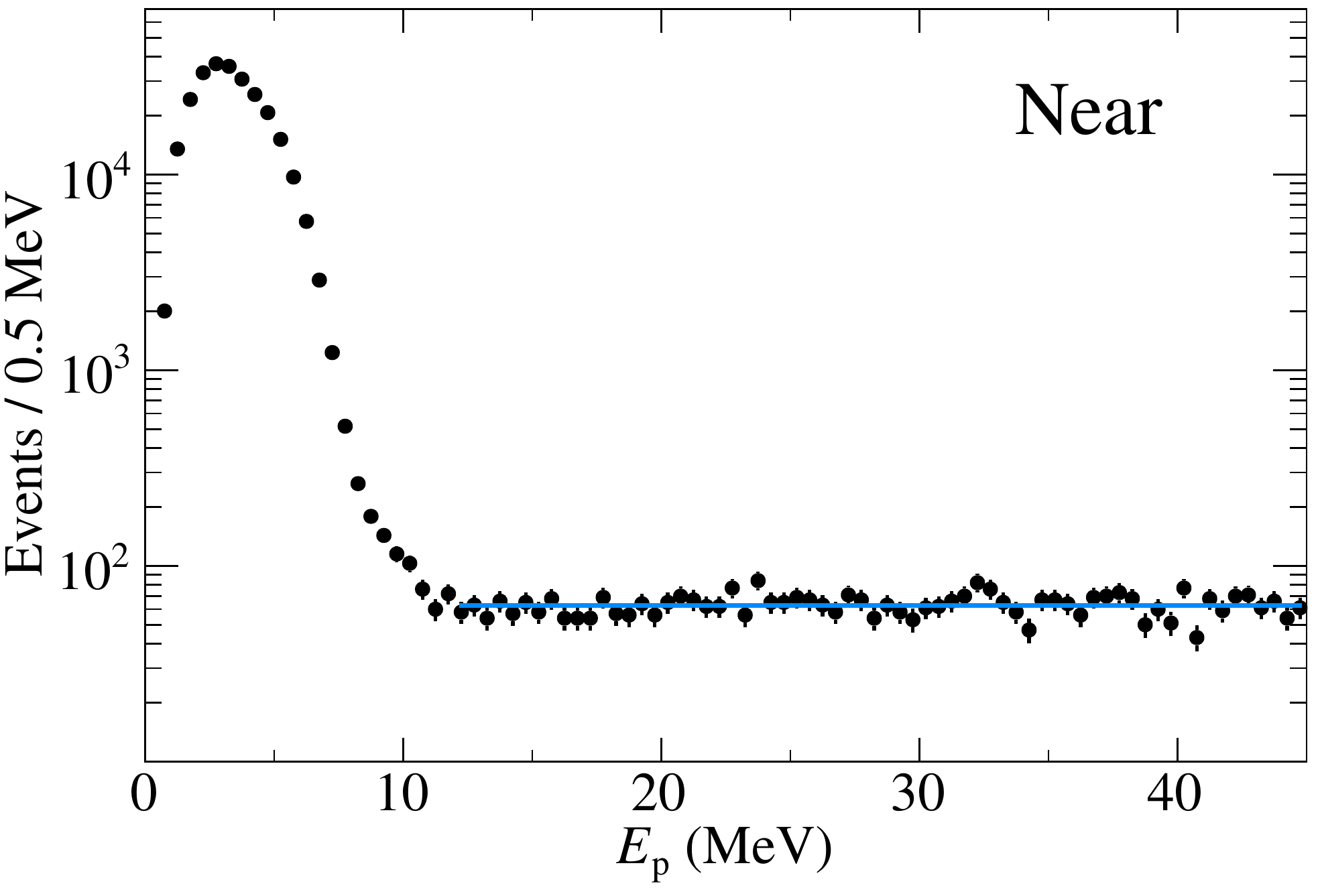}
\end{center}
\caption{\label{f:fn}
Prompt energy spectrum of IBD candidates including a flat fast neutron spectrum at $E_{\rm{p}} >$ 12~MeV.
The fast neutron background rate in the IBD candidates is estimated by extrapolating from the background dominant region assuming a flat spectrum of the background.
}
\end{figure}

\begin{figure}
\begin{center}
\includegraphics[width=0.48\textwidth]{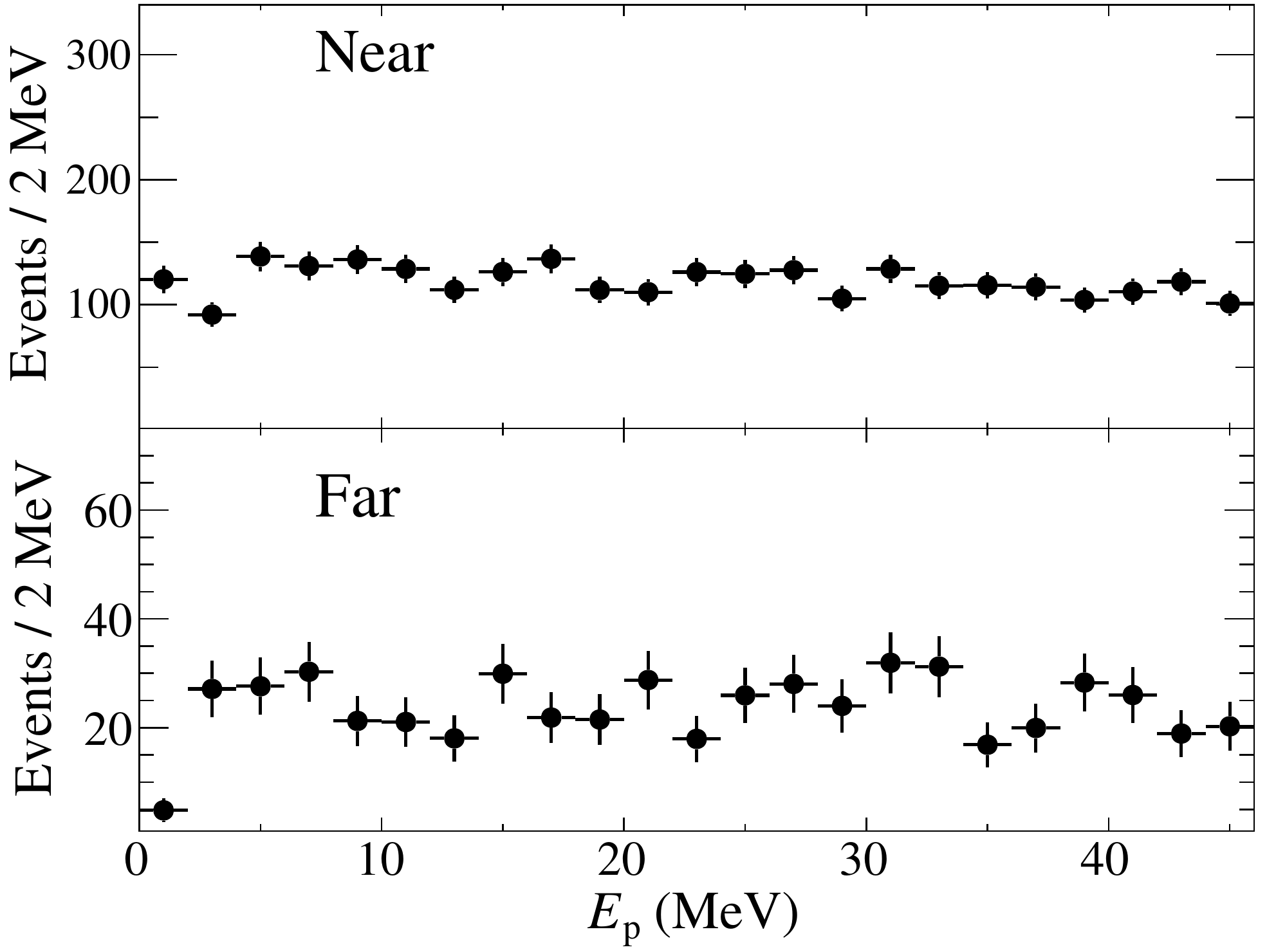}
\end{center}
\caption{\label{f:fn_shape}
Prompt energy spectra of fast neutron enriched samples in the near and far detectors.
}
\end{figure}

The fast neutron background rate in the final IBD candidate sample is estimated by being extrapolated from the background dominant energy region of 
12 $< E_{\rm{p}} <$ 40~MeV to the IBD signal region of 1.2 $< E_{\rm{p}} <$ 8.0~MeV, assuming a flat spectrum of the background as shown in Fig.~\ref{f:fn}.
A fast neutron enriched sample is obtained by selecting IBD candidates which are accompanied by any prompt candidates of
$E_{\rm{p}} >$ 0.7~MeV within a 1~ms subsequent window. The prompt events of this sample show a distribution consistent with a flat spectrum
in the IBD signal region as shown in Fig.~\ref{f:fn_shape}.
%The estimated background rate is 0.48 (2.28) events per day for the far (near) detector. 
The background rate uncertainty is obtained from the fitting error of the flat spectrum and estimated as 0.03 (0.02) events per day for the near (far) detector. 
The assumption of the flat background spectrum in the signal region is checked and validated by a fast neutron background enriched sample. 
The spectral shape uncertainty of the fast neutron background includes a possible deviation from the flat spectrum and is estimated as 0.02 (0.01) events per day 
for the near (far) detector. 
In order to estimate the deviation the background dominant region in Fig~\ref{f:fn_shape} is fitted with a first order polynomial as an alternative model.
The remaining fast neutron background rates are 2.28$\pm$0.04 (near) and 0.48$\pm$0.02 (far) events per day. \\
\indent
Some of fast neutrons lose most of their kinetic energy before reaching the target or $\gamma$-catcher regions and produce neutron capture events.
These neutron capture events are easily paired with a prompt like event to contribute to accidental backgrounds.
Those backgrounds are eliminated if any buffer and veto trigger occurs in a 100~$\mu$s window following a prompt candidate.

%-----------------------------------------------------------------
\subsection{Cosmogenic $^{9}$Li/$^{8}$He background}

\begin{figure}
\begin{center}
\includegraphics[width=0.48\textwidth]{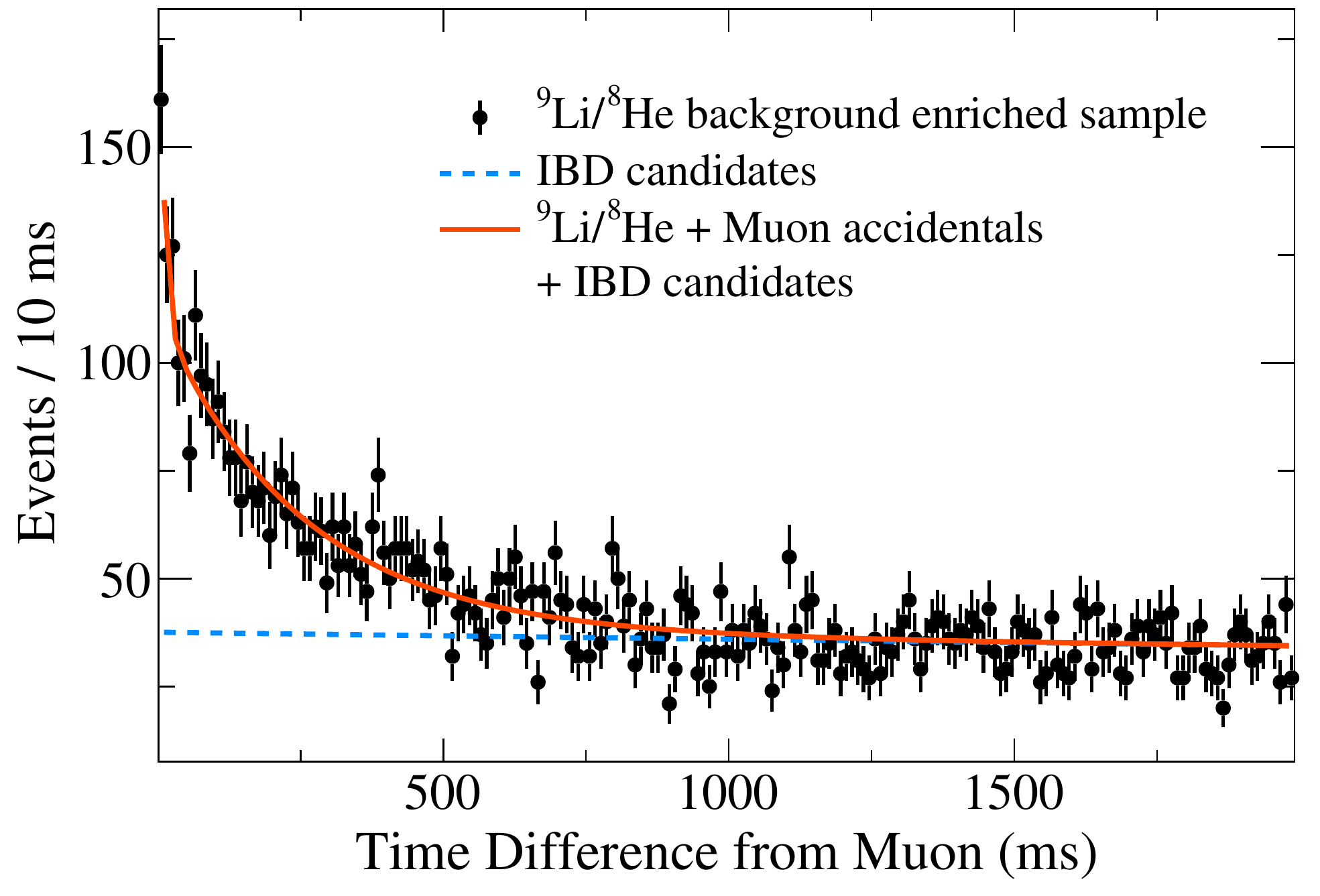}
\end{center}
\caption{\label{f:mutimediff}
(Color online) Decay time distribution of the IBD like pairs from their preceding energetic muons from a total of 1100 live days of \LiHeS background enriched sample 
in the far detector. 
The \LiHeS background is clearly seen with a measured mean decay time of $\sim$250~ms 
while muon-induced accidental background events are observed right after their preceding muons. 
}
\end{figure}

The spectral shape of the \LiHeS background is measured using a sample of IBD like pairs that are produced within 500~ms (400~ms) by energetic muons 
of $E_{\rm{\mu}} >$ 1.6~{\rm GeV} ($>$ 1.5~{\rm GeV}) for the near (far) detector. 
The distribution of time difference between an energetic muon and a subsequent IBD candidate is shown in Fig.~\ref{f:mutimediff}. 
Based on their observed spectra, the shortest decay time component is found to be the muon-induced accidental background, 
and the \LiHeS background follows after it. 
The IBD signals are temporally uncorrelated with muon events and their time differences are distributed according to the IBD rate. 
The measured mean decay time of $\sim$250~ms indicates predominant production of \LiS over \He. 

\begin{figure}
\begin{center}
\includegraphics[width=0.48\textwidth]{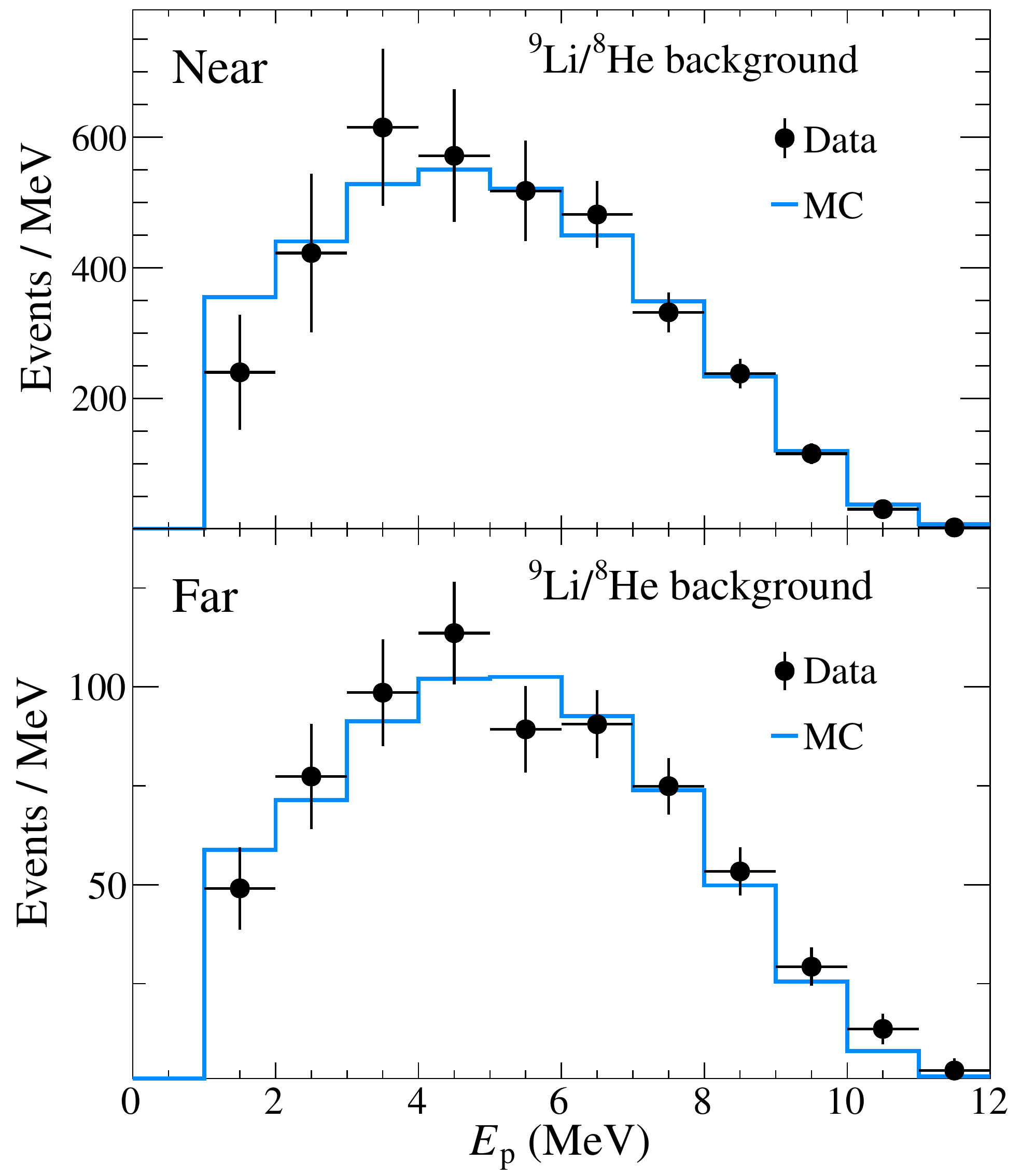}
\end{center}
\caption{\label{f:lihe}
Measured \LiHeS background spectra that are obtained from the enriched samples of 1100 live days of data after subtracting the IBD signal 
and the muon induced accidental background from those of the enriched sample. 
The MC \LiHeS background spectra (blue histograms) are overlaid with data 
where the relative fractions between $^{9}$Li and $^{8}$He are obtained from the fits to the data. 
}
\end{figure}

The measured \LiHeS background shapes as shown in Fig.~\ref{f:lihe} are obtained by subtracting the energy spectra of the IBD signal and the muon-induced accidental background 
from those of the \LiHeS background enriched samples. 
The size of the IBD signal and the muon induced accidental background are determined by a fit to the decay time distribution using three exponential functions. 
The spectral shape uncertainty comes from statistical uncertainty of the \LiHeS background enriched sample because of the subtraction and, therefore, 
is expected to be reduced by more data. 
The $^{9}$Li and $^{8}$He background shapes are also obtained from MC for comparison.
The relative fraction between $^{9}$Li and $^{8}$He is determined by a fit to the measured \LiHeS spectrum. 
The estimated $^{8}$He component is 13.6$\pm$3.9\% (1.1$\pm$1.6\%) for the near (far) detector.
The difference of the $^{8}$He components between the two detectors might be related to their different overburdens.

\begin{figure}
\begin{center}
\includegraphics[width=0.48\textwidth]{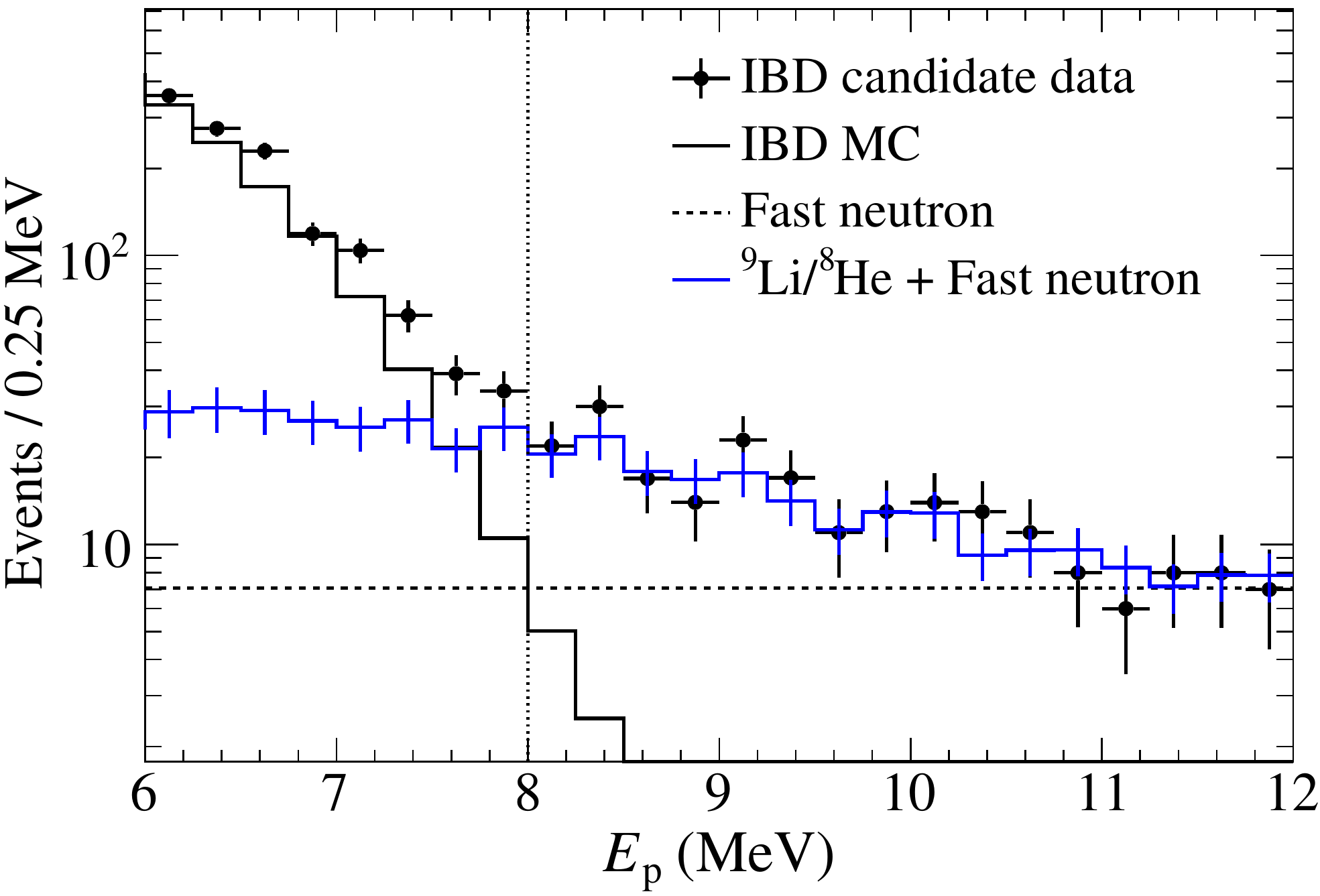}
\end{center}
\caption{
(Color online) Estimation of the remaining \LiHeS background rate in the signal region using the measured rate in the background dominant region of
$E_{\rm{p}} >$ 8~MeV in the far detector. The background rate in the signal region of $E_{\rm{p}} <$ 8~MeV
is estimated by extrapolating from the background dominant region using the measured background spectrum.
}
\label{f:lh_estim}
\end{figure}

The background rate in the IBD signal region of $E_{\rm{p}} <$ 8~{\rm MeV} is estimated by extrapolating from the background dominant region of $E_{\rm{p}} >$ 8~{\rm MeV} 
using the measured background spectrum as shown in Fig.~\ref{f:lh_estim}. 
The background rate in the region of $E_{\rm{p}} >$ 8~{\rm MeV} is estimated by a fit to the IBD candidate data using the measured \LiHeS background 
spectrum, the measured fast neutron background, and the MC IBD expectation. 
The energy-bin-uncorrelated spectral uncertainty is obtained from the measured \LiHeS spectral uncertainty and is estimated as 0.61 (0.07) events per day for the near (far) detector. 
The energy-bin-correlated uncertainty is obtained from the fit error of the background rate in the region of $E_{\rm{p}} >$ 8~MeV and estimated as 0.55 (0.22) events per day for the near (far) detector. 
The estimated \LiHeS background rates are 8.36$\pm$0.82 (near) and 1.54$\pm$0.23 (far) events per day. 

%-----------------------------------------------------------------
\subsection{$^{252}$Cf contamination background}

\begin{figure}
\begin{center}
\includegraphics[width=0.48\textwidth]{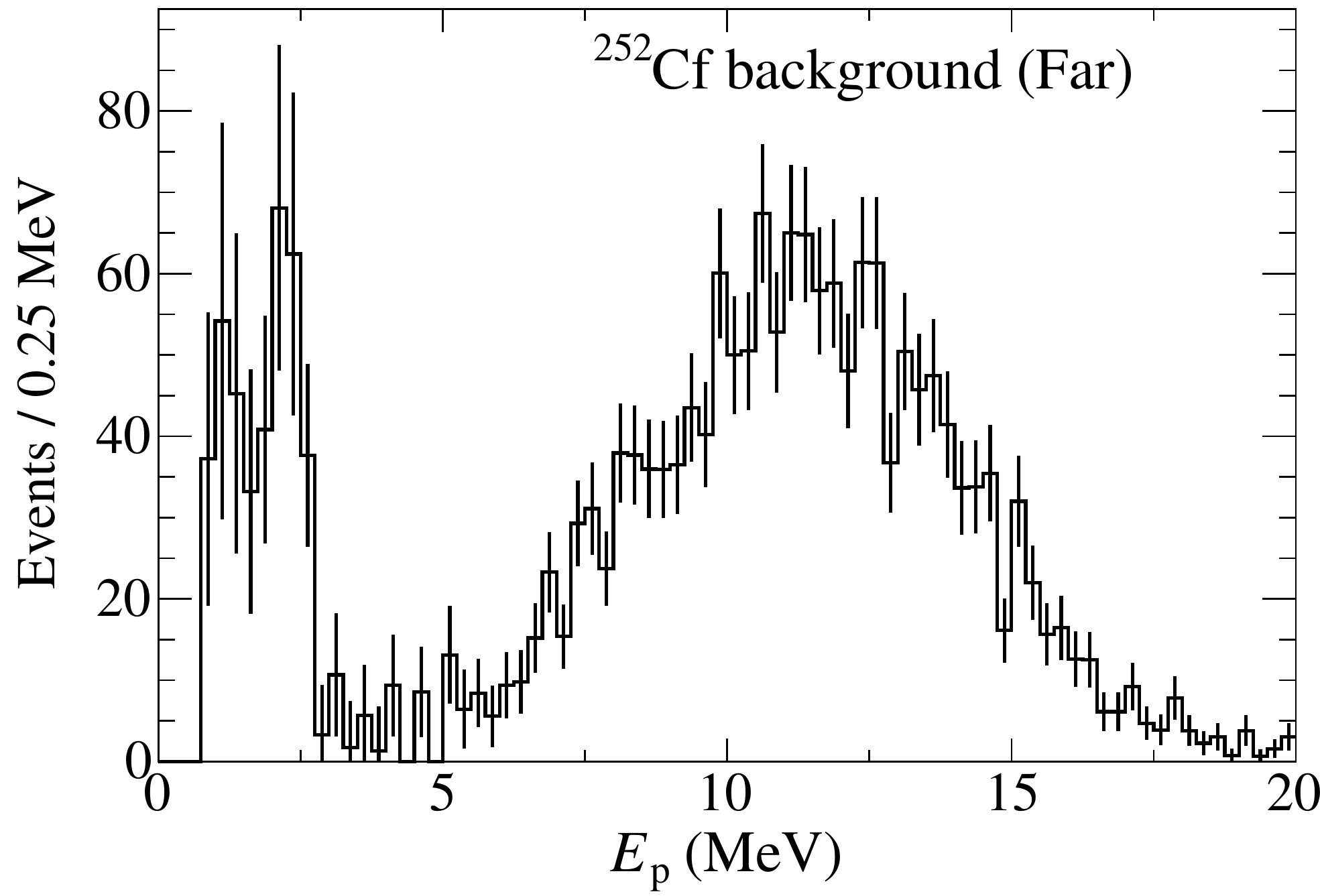}
\end{center}
\caption{\label{fig:cf_shape} \CfS background in the far detector.
The errors represent estimated shape uncertainties. 
}
\end{figure}

The amount of the initial \CfS contamination is estimated as 0.49$\pm$0.14 mBq (4.51$\pm$0.94 mBq) for the near (far) detector. 
The estimation is made based on the rejected and remaining samples after event selection requirements (viii) and (ix).
This background has a half-life of 2.7 years. Most multiple neutron events coming from the \CfS contamination
are eliminated by the stringent multiplicity requirements of no trigger or no event near an IBD event. 
After applying the requirements, 
99.9\% of the background events in the far detector is eliminated with a signal loss of 8.0$\pm$0.2\%. 
No remaining \CfS contamination background events are observed in the near detector. The remaining background rate and shape are obtained 
from the \CfS contamination candidate events that are accompanied by an additional event within a 10 s window and a distance of 40~cm from an IBD prompt event. 
Three different shape components of the background spectrum are found in this sample. 
They are two gaussian like spectral shapes peaking at 2.2 and 11.0~MeV 
and a spectral shape peaking at 1~MeV and falling rapidly up to 4~MeV as shown in Fig.~\ref{fig:cf_shape}.
It is not understood why there are three spectral shapes with different time correlation between an IBD prompt event and an associated event.
The shape of 11~MeV-- (1~MeV--) peaked component is obtained from a sample that is selected by requiring a time and spatially correlated event of $E > $ 3~MeV after (before) an IBD event. 
The associated event is likely due to the multiple neutrons or the prompt fission gammas from a \CfS decay. 
The shape of 2.2~MeV--peacked component is obtained from a sample that is selected by 
requiring a time and spatially correlated event of $E < $ 3~MeV before an IBD event. 
The remaining \CfS background spectrum in the far detector is shown in Fig.~\ref{fig:cf_shape}.\\
\indent
The rate of the 11--MeV--peaked component is estimated by fitting the $E_{\rm p}$ distribution of the prompt events with no $E_{\rm p}$ requirement 
of the IBD event candidates. The $E_{\rm p}$ distribution is fit with the 11--MeV--peaked component spectrum and a flat fast neutron spectrum above 12~MeV, 
where $^{252}$Cf background is dominant. 
%The 11~MeV background rate is estimated by a fit to the measured rate in the \CfS background dominant region of $E_{\rm{p}} > $ 12~MeV 
%using the obtained background spectrum and the measured fast neutron rate. 
The other two component rates are estimated 
from the \CfS background dominant samples that are used for obtaining their component shapes. The remaining \CfS contamination 
background rate is estimated as 0.14$\pm$0.03 events per day for the far detector. The energy-bin-uncorrelated spectral uncertainty is obtained 
from the measured  background spectral error and is estimated as 0.025 events per day for the far detector. 
The energy-bin-correlated uncertainty is obtained from the fit error of the background rate in the region of $E_{\rm{p}} > $ 8~MeV
and estimated as 0.015 events per day for the far detector. 

%------------------------------------------------------------
\subsection{Summary of the backgrounds and their uncertainties}

The total remaining background rates for 1.2 $< E_{\rm{p}} < $ 8~MeV in the final IBD candidate samples are estimated as 17.54$\pm$0.83 (near) 
and 3.14$\pm$0.23 (far) events per day. After the background subtraction, the IBD signal rates are 616.67$\pm$1.44 (near) and 61.24$\pm$0.42 (far) events per day. 
Table~\ref{t:ibd} summarizes the observed IBD and estimated background rates.
The livetime is calculated as the sum of the duration of each physics data-taking run used in the analysis and its uncertainty is estimated to be negligible.
%The livetime is calculated from the start- and end-time stamps in each run excluding calibration, non-physical or bad runs. 
%Based on the DAQ clock accuracy the livetime uncertainty is negligible.
\begin{table}
\begin{center}
\caption{\label{t:ibd} 
Observed IBD and estimated background rates per day for 1.2 $< E_{\rm{p}} < $ 8~MeV.
}
\begin{tabular*}{0.48\textwidth}{@{\extracolsep{\fill}} lcc}
\hline
\hline
Detector & Near & Far  \\
\hline
IBD rate    & 616.67$\pm$1.44 & 61.24$\pm$0.42  \\
after background subtraction  &  &   \\

Total background rate  & 17.54$\pm$0.83 & 3.14$\pm$0.23   \\

DAQ live time (days)  & 458.49 &  489.93 \\
\hline
Accidental rate  & 6.89$\pm$0.09  &  0.97$\pm$0.03  \\

\LiHeS rate    & 8.36$\pm$0.82 &  1.54$\pm$0.23  \\

Fast neutron rate    & 2.28$\pm$0.04 &  0.48$\pm$0.02 \\

\CfS contamination rate  & 0.000$\pm$0.001 &  0.14$\pm$0.03 \\
\hline
\hline
\end{tabular*}
\end{center}
\end{table}

\begin{figure}
\begin{center}
\includegraphics[width=0.48\textwidth]{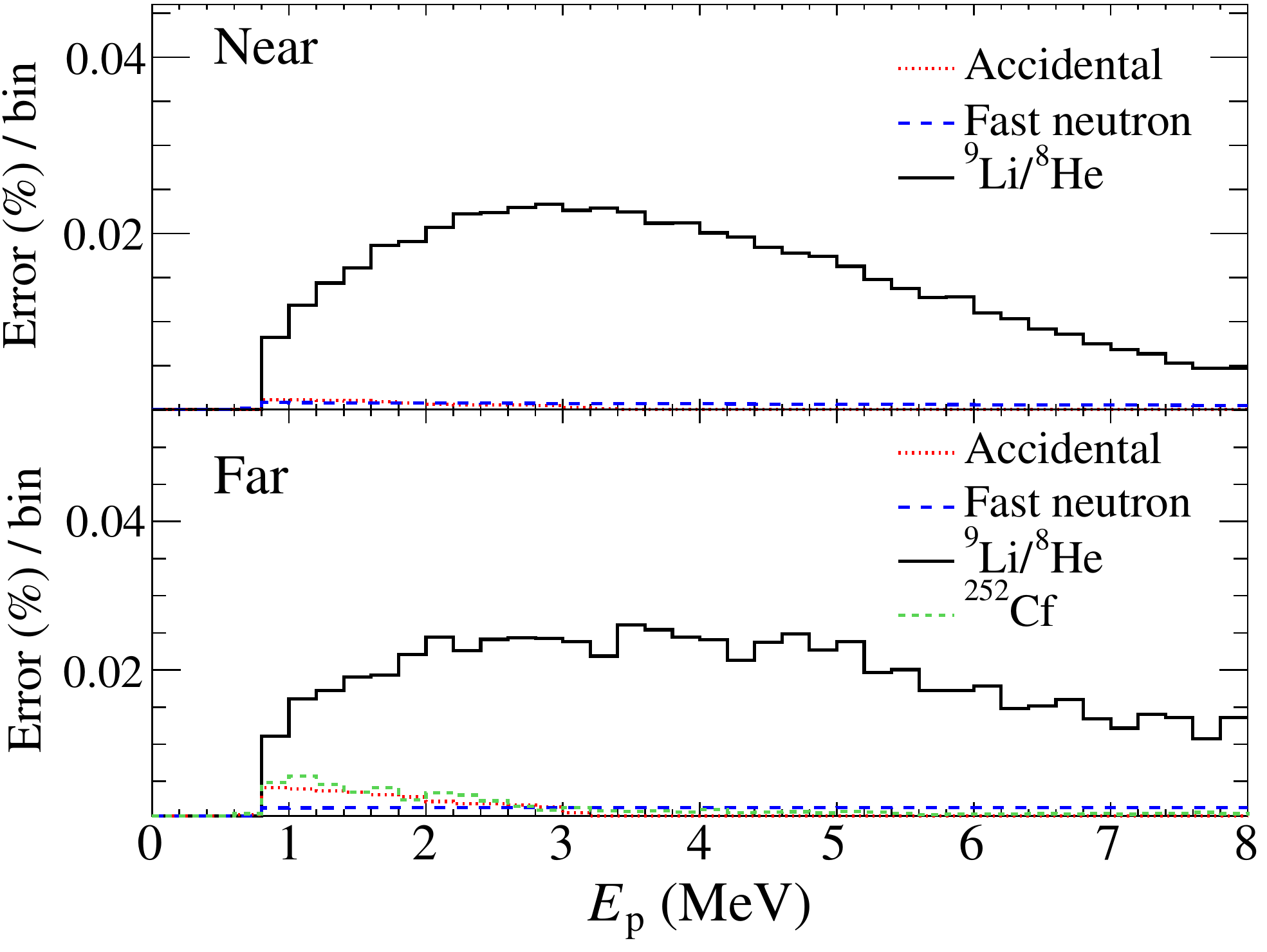}
\end{center}
\caption{\label{f:bkg_uu} 
(Color online) Bin-to-bin uncorrelated spectral uncertainties of remaining backgrounds in the final IBD candidate samples. 
The \LiHeS background is the most dominant source of the background uncertainty.
}
\end{figure}

\begin{figure}
\begin{center}
\includegraphics[width=0.48\textwidth]{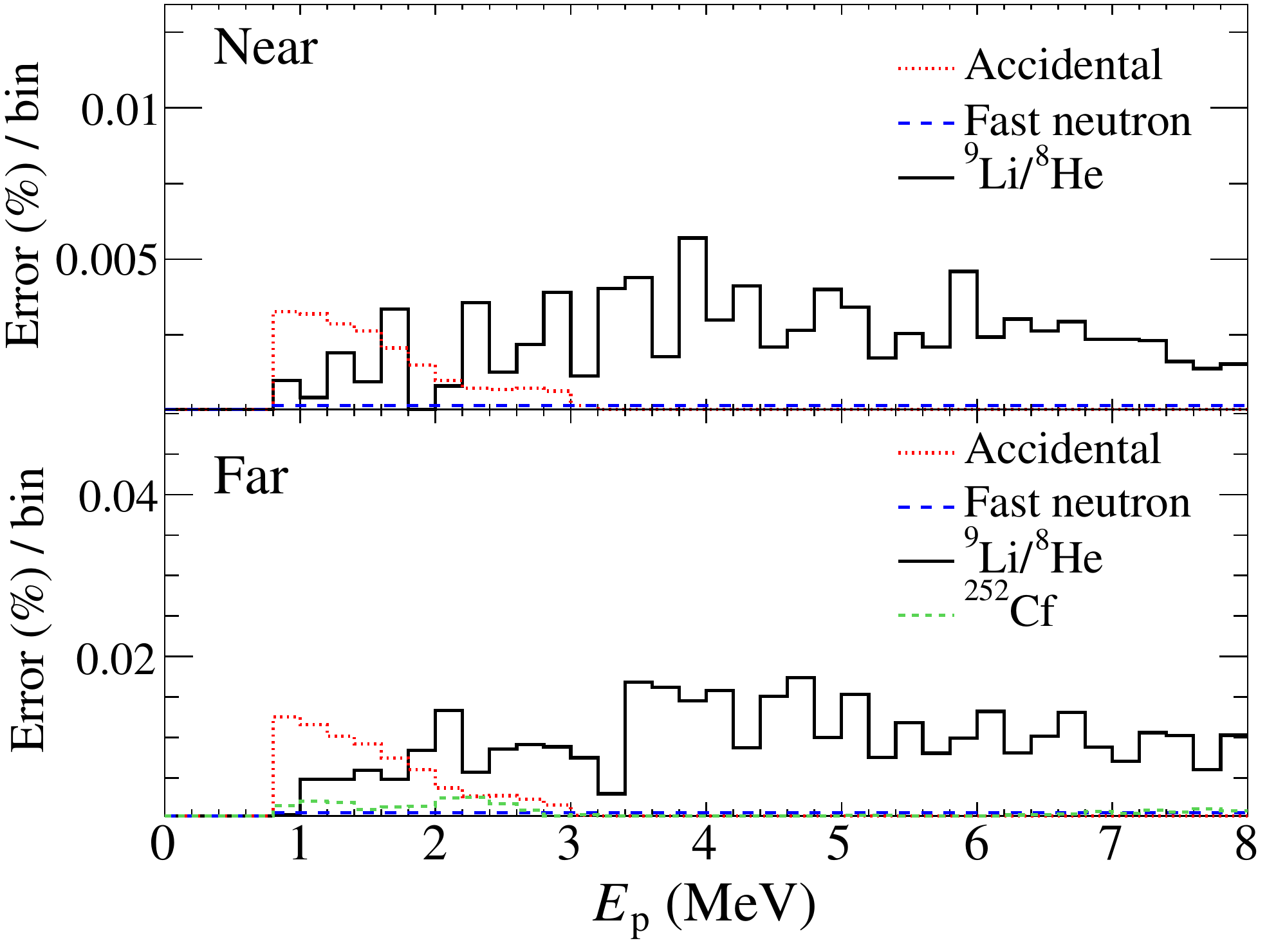}
\end{center}
\caption{\label{f:bkg_cu} 
(Color online) Bin-to-bin correlated spectral uncertainties of remaining backgrounds in the final IBD candidate samples. 
}
\end{figure}

Figures~\ref{f:bkg_uu} and ~\ref{f:bkg_cu} show bin-to-bin uncorrelated and correlated uncertainties 
of measured background spectra, respectively. The largest uncertainty comes from the \LiHeS background.
Note that the largest bin-to-bin correlated uncertainty at $E_{\rm{p}} < $ 2.0~MeV is due to the accidental background.

%======================================================================
\section{Expected Reactor Neutrino Flux}

The expected rates and spectra of reactor antineutrinos are calculated for the duration of physics data-taking by taking into account the varying thermal powers, 
fission fractions of four fuel isotopes, energy release per fission, and fission and capture cross sections. The expected number of reactor \nuebS in a detector 
is computed using the following formula~\cite{Klimov1994}, 
\begin{equation}
\begin{aligned}
\label{eq:flux}
n_{\nu} &= \frac{N_{\rm{p}}}{4\pi L^{2}}\frac{[\sum_{i}\alpha_{i}\conj{\sigma_{i}}]}{[\sum_{i}\alpha_{i}E_{i}]}P_{\rm{th}} \\ &= \frac{N_{\rm{p}}}{4\pi L^{2}} \frac{\sigma_{5}[1 + \sum_{i}\alpha_{i}(\conj{\sigma_{i}}/\sigma_{5} - 1)]}{E_{5}[1 + \sum_{i}\alpha_{i}(E_{i}/E_{5} - 1)]}P_{\rm{th}},
\end{aligned}
\end{equation} 
where $N_{\rm{p}}$ is the number of free protons in target, 
$L$ is the distance between a reactor and a detector, $P_{\rm{th}}$ is a reactor thermal power, 
$i$ is an index for each isotope of $^{235}$U, $^{238}$U, $^{239}$Pu, and $^{241}$Pu, 
$\alpha_{i}$ is the fission fraction of the $i$th isotope,
$E_{i}$($E_{5}$) is the energy released by the $i$th isotope ($^{235}$U),
$\conj{\sigma_{i}} = \int \sigma (E_{\nu})\phi_{i}(E_{\nu})dE_{\nu}$ is the average fission cross section of the $i$th isotope,
and $\sigma_{5}$ is cross-section for $^{235}$U. 
Note that $\phi_{i}(E_{\nu})$ is a \nuebS reference energy spectrum per isotope. \\
%and can be expressed as $\phi_{i}(E_{\nu}) = exp\sum(a_{n}E_{\nu}^{n})$ with known coefficients $a_{n}$~\cite{Mueller,Huber}. \\

\indent The average relative fission fractions of $^{235}$U, $^{238}$U, $^{239}$Pu, and $^{241}$Pu during the $\sim$500 live days of data-taking period are
0.569\,:\,0.073\,:\,0.302\,:\,0.056 for the near detector and 0.572\,:\,0.073\,:\,0.299\,:\,0.056 for the far detector. 
These values are obtained by taking weighted average of reactor cycles with reactor \nuebS fluxes 
according to the thermal outputs and baselines. \\

\indent 
The thermal energy release per fission is given in Ref.~\cite{Kopeikin}, 
and its uncertainty introduces 0.2\% for a correlated uncertainty. 
The daily thermal output measurement with a 0.5\% uncertainty per reactor is provided by the Hanbit power plant~\cite{Pth}. 
The uncertainty is partially correlated between reactors.
However, the uncertainty is weakly correlated between the near and far detectors because of multiple reactors and thus unknown information 
on an individual reactor \nuebS source. 

The relative fission fraction of the four main isotopes are estimated with quoted 4$-$10\% uncertainties by the Hanbit power plant, 
using the {\sc ANC} reactor simulation code~\cite{ANC}. The fission fraction uncertainties are consistent with other evaluations~\cite{FF}. 
The resultant uncertainty in the expected reactor \nuebS 
flux is estimated as 0.7\% using pseudoexperiments in which the relative isotope fractions are varied within their uncertainties. 
The fission fraction uncertainties for this analysis are assumed to be uncorrelated from reactor to reactor and from cycle to cycle although 
a large fraction of the uncertainty could be correlated among reactors according to Ref.~\cite{FF}. Thus the uncertainty uncorrelated from reactor to reactor 
may be reduced in the future work if the multi-reactor flux average is carefully treated. 
In the current analysis we have not attempted to reduce it because 
the energy-dependent variation due to the isotope fission fraction uncertainties is much smaller than the detector energy scale uncertainty. 
Based on the obtained thermal output and the relative fission fraction an expected number of reactor \nuebS 
is obtained from Eq.~\ref{eq:flux} that can be rewritten as $n_{\nu} = \gamma_{0} (1 + k(t)) P_{\rm{th}}$.
Note that $\gamma_{0}$ is determined by the experimental setup parameters and is a constant in time, 
and $k(t)$ is a time variation parameter of fuel isotopes. 
An expected number of reactor \nuebS events 
in a detector is calculated by adding all reactor contributions with individual baseline consideration and by taking into account cross section, live time, and detection efficiency.\\

\indent The systematic uncertainties associated with the reactors are listed in Table~\ref{t:r_sys}. 
The reactor \nuebS flux uncertainties uncorrelated among reactors come from baseline distance, reactor thermal power, and fission fraction. 
The positions of two detectors and six reactors are surveyed with GPS and total station to determine the baseline distances 
between the detectors and reactors to an accuracy better than 10~cm. Reactor \nuebS fluxes at the two detectors are obtained by calculating 
the flux reduction due to baseline distance to a precision better than 0.1\%. The baseline distance uncertainty is much smaller than other two. The total uncorrelated uncertainty of reactor flux is estimated as 0.9\%.
The correlated uncertainty in the fission reaction cross sections is found in Ref.~\cite{Saclay2011},
and the correlated uncertainty of reference energy spectra is given in Refs.~\cite{Mueller,Huber}. 
The total correlated uncertainty is 2.0\% and is cancelled out in the far-to-near ratio measurement.
\begin{table}[hb]
 \caption{Uncorrelated and correlated systematic uncertainties among reactors that are used in the reactor \nuebS flux estimation.}
\label{t:r_sys}
 \begin{center}
 \begin{tabular*}{0.48\textwidth}{@{\extracolsep{\fill}} l r r }

    \hline \hline
        &   Uncorrelated &    Correlated  \\
        &   (\%) & (\%)  \\
  \hline
   Baseline           &  0.03          &                \\
   Thermal power      &   0.5          &      $-$       \\
   Fission fraction   &   0.7          &      $-$       \\
   Fission reaction cross section &  $-$ &  1.9         \\
   Reference energy spectra       &   $-$  &  0.5         \\
   Energy per fission &     $-$            &  0.2         \\
  \hline
   Combined            &     0.9        &  2.0        \\
\hline
\hline
\end{tabular*}
\end{center}
\end{table}

Table~\ref{t:r_num} shows expected number of reactor \nuebS in near and far detectors from each reactor without oscillation. 

\begin{table}[hb]
 \caption{Expected number of reactor \nuebS in near and far detectors from each reactor without oscillation.}
\label{t:r_num}
 \begin{center}
 \begin{tabular*}{0.48\textwidth}{@{\extracolsep{\fill}} c r r }

    \hline \hline
Reactor  &   Near           &    Far  \\
%         &   (fraction \%)  & (fraction \%)  \\
  \hline
   1     &  4,1267 (7.6\%)    &   7,860 (15.1\%)   \\
   2     &  90,463 (16.6\%)   &   8,965 (17.2\%)   \\
   3     &  170,679 (31.3\%)  &   8,657 (16.6\%)   \\
   4     &  155,431 (28.5\%)  &   9,977 (19.2\%)   \\
   5     &  56,631 (10.4\%)   &   8,362 (16.1\%)   \\
   6     &  30,132 (5.5\%)    &   8,257 (15.9\%)    \\
%  \hline
%   Combined            &     0.9        &  2.0        \\
\hline
\hline
\end{tabular*}
\end{center}
\end{table}

%--------------------------------------------------
\section{Expected and Observed IBD Rates and Spectra}
\begin{figure}
\begin{center}
\includegraphics[width=0.48\textwidth]{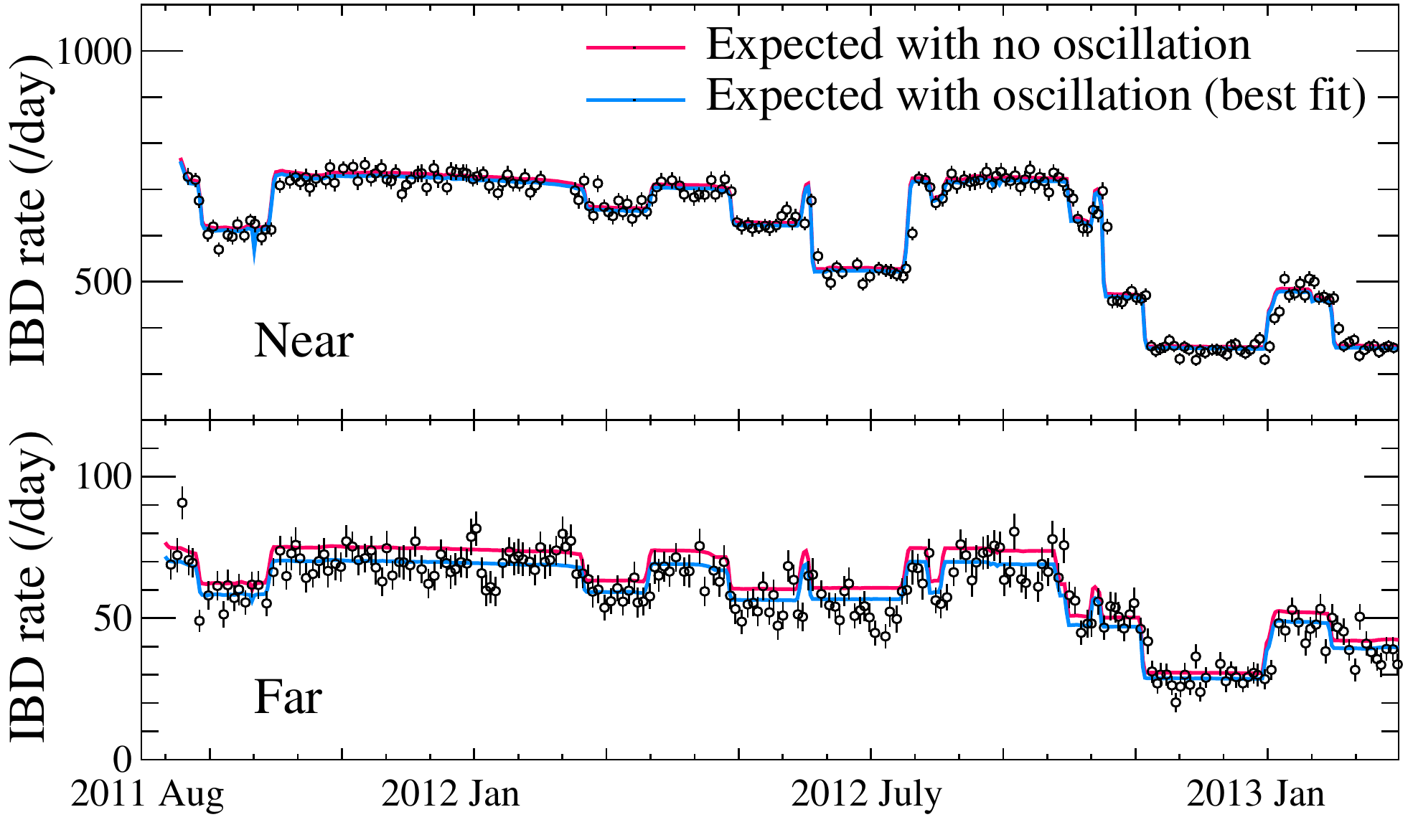}
\caption{
(Color online) Measured daily-average rates of reactor \nuebS after background subtraction in the near and far detectors as a function of running time. 
The red curves are the predicted rates for no oscillation. The blue curves are the predicted rates with the best-fit parameters 
and agree well with the measured one.
}
\label{f:daily_IBD}
\end{center}
\end{figure}

\begin{figure}
\begin{center}
\includegraphics[width=0.48\textwidth]{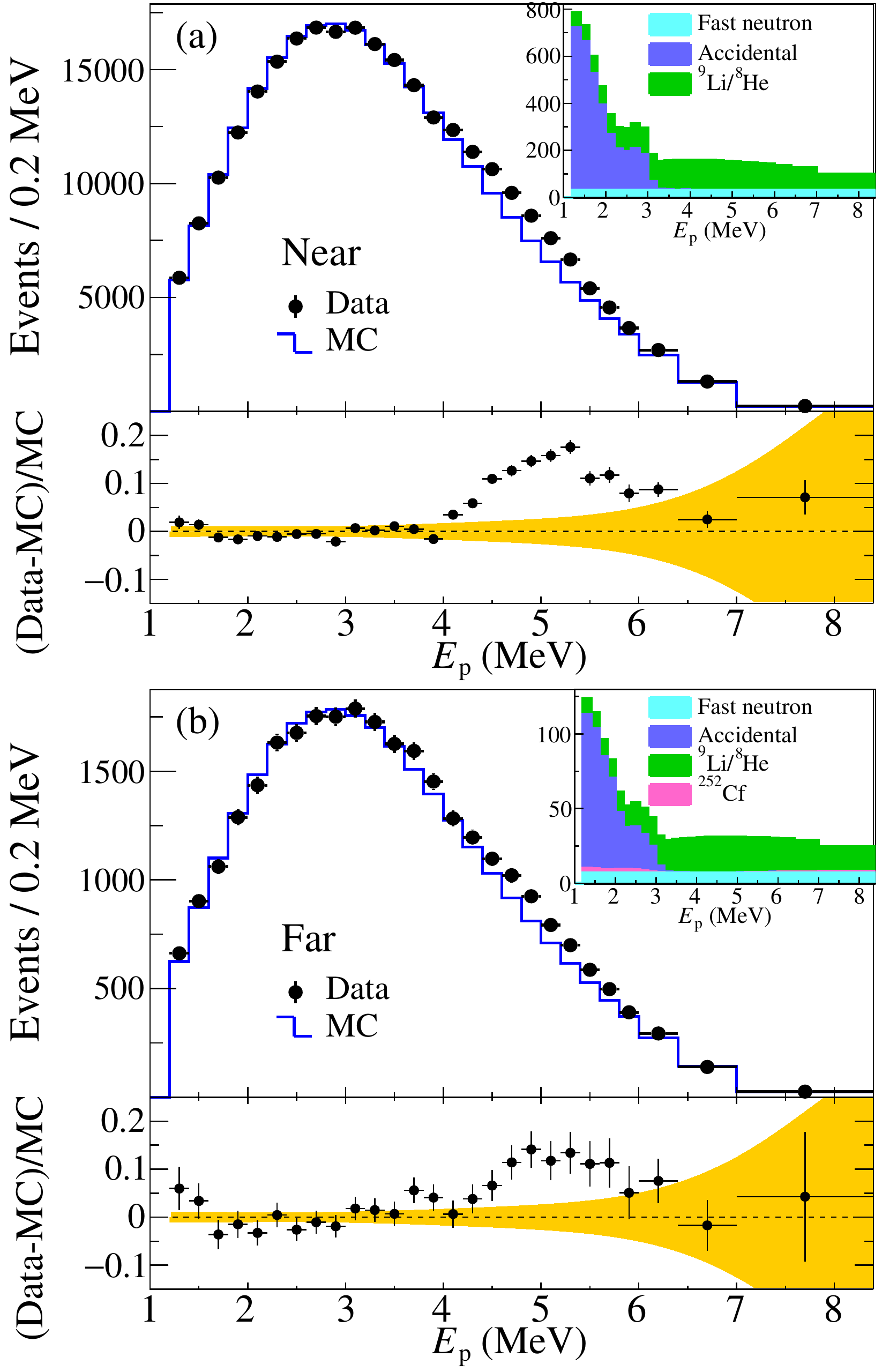}
\caption{
(Color online) Spectral shape comparison of observed and expected IBD prompt events in the (a) near and (b) far detectors. The observed spectra are obtained from subtracting 
the remaining background spectra as shown in the insets. The expected distributions are obtained from the best-fit oscillation results discussed later 
that are applied to the no-oscillation MC spectra. The expected spectra are normalized to data spectra in the region excluding 3.6 $< E_{\rm{p}} <$ 6.6~MeV. The discrepancy between data and MC prediction is clearly seen at 4--6~MeV. The observed excess 
is correlated with the reactor power, and corresponds to 3\% of the total number of IBD events. The deviation from the expectation is larger than 
the uncertainty of the expected spectrum (shaded band).
}
\label{f:5MeV}
\end{center}
\end{figure}

\begin{figure}
\begin{center}
\includegraphics[width=0.48\textwidth]{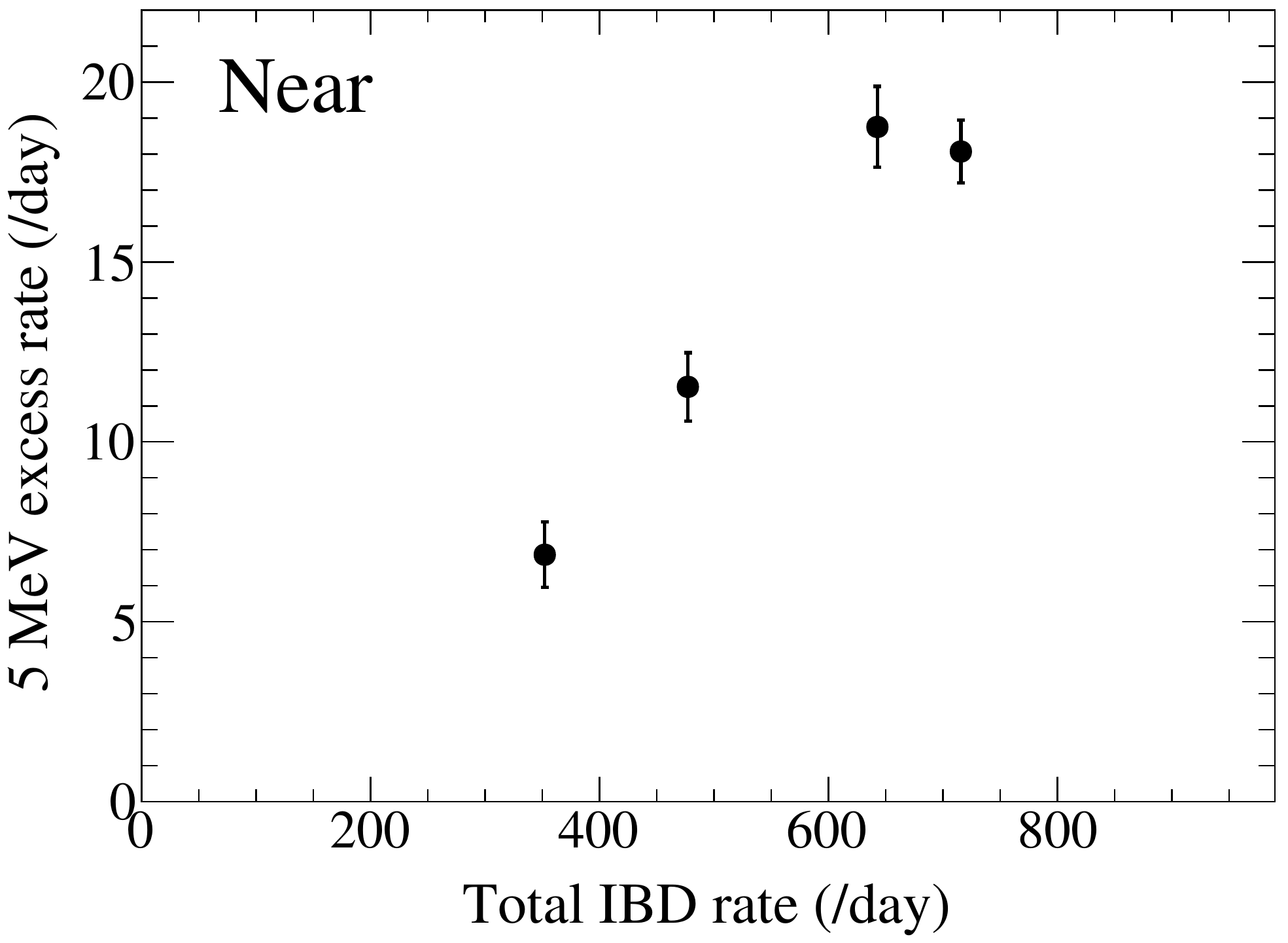}
\caption{
Correlation between the 5-MeV excess rate and the total IBD event rate in a total of 1400 live days of the near detector data. 
The error bar represents statistical uncertainty only. 
%The dotted line is a fit to data assuming their linear correlation. 
}
\label{f:5MeV_pow}
\end{center}
\end{figure}

\indent
Figure~\ref{f:daily_IBD} shows the measured daily rates of IBD candidates after background subtraction in the near and far detectors. 
The reactors were turned off for fuel replacement and maintenance.
The expected rates assuming no oscillations are shown for comparison. The measured IBD rate in the far detector is clearly lower than the expected one, 
indicating the reactor \nuebS disappearance. The expected rates with the best-fit parameters are also shown and agree well with the measured IBD rates. \\
\indent
Figure~\ref{f:5MeV} shows a spectral shape comparison between the observed IBD prompt spectrum after background subtraction and the prediction 
from a reactor \nuebS model~\cite{Mueller,Huber} using the far-to-near ratio measurement result. 
The fractional difference between data and prediction 
is also shown in the lower panel. A clear discrepancy is observed in the region of 5~MeV in both detectors. To compare the spectral shape, the MC predicted spectrum is normalized to the observed one in the region excluding 3.6 $< E_{\rm{p}} <$ 6.6~MeV. 
The excess of events is estimated as about 3\% of the total observed IBD events in both detectors. 
Furthermore, the 5-MeV excess is observed to be proportional to the reactor thermal power where the rate is calculated from the events in excess 
at 3.6 $< E_{\rm{p}} <$ 6.6~MeV relative to the nominal model prediction~\cite{Mueller,Huber}. 
Figure~\ref{f:5MeV_pow} shows a clear correlation between
the 5-MeV excess rate and the total IBD rate that corresponds to the reactor thermal power. This observation indicates that this excess indeed
arises from the reactor \nuebS and thus suggests needs for reevaluation and modification of the current reactor \nuebS model~\cite{Mueller,Huber}.

\section{\label{sec:sys}Systematic Uncertainties }
\indent
Systematic uncertainties of energy scale, backgrounds, detection efficiency, and reactor \nuebS flux are described in the earlier sections 
and summarized in Table~\ref{t:sys}. For the far-to-near ratio measurement the only uncorrelated systematic uncertainties contribute to the uncertainties of measured values. 
The energy-dependent detection efficiency is not considered in this analysis. Because of difference in overburden, we assume no correlated
uncertainties between the near and far detectors. Therefore, to be conservative, the background uncertainty of each detector is fully taken as uncorrelated systematic one 
in a $\chi^{2}$ fit described later. \\
\indent
In summary the relative energy-scale difference is 0.15\%, the uncorrelated systematic uncertainty of the detection efficiency is 0.2\%, 
the systematic uncertainties of the total backgrounds are 4.7\% (near) and 7.3\% (far), and the uncorrelated systematic uncertainty of reactor \nuebS flux is 0.9\%.
\begin{table}[b]
\caption{\label{t:sys} Summary of the systematic uncertainties. 
The detection efficiency uncertainty includes the systematic uncertainty for the signal loss due to the timing veto criteria given in Table V.
}
\begin{center}
\begin{tabular*}{0.48\textwidth}{@{\extracolsep{\fill}} l r r}
\hline
\hline
Uncertainty source &  & Uncorrelated   \\
\hline
Reactor &               & 0.9\% \\
Detection efficiency  &  &  0.2\% \\
+ timing veto &  & \\
Energy scale &          &  0.15\%  \\
\hline\hline
 & Bin-correlated  & Bin-uncorrelated\\
Total background & 3.18\% (near) & 3.48\% (near) \\
                 & 6.97\% (far) & 2.51\% (far) \\
\hline
Accidental & 1.19\% (near) & 0.30\% (near) \\
           &  2.98\% (far) & 0.82\% (far) \\
Fast neutron & 1.45\% (near) & 1.01\% (near) \\
             &  3.11\% (far) & 1.04\% (far) \\
\LiHeS &  6.58\% (near) & 7.28\% (near) \\
       &  14.0\% (far) & 4.80\% (far) \\
\CfS  & 17.4\% (far) & 20.8\% (far) \\
\hline
\hline
\end{tabular*}
\end{center}
\end{table}

%======================================================================
\section{\label{sec:results} Results }

The relative measurement makes the method insensitive 
to correlated uncertainties between the near and far detectors and reduces uncorrelated reactor uncertainties. The measurement 
results are presented based on three different analysis methods to validate their consistencies. They are rate-only, rate and spectrum, 
and spectrum-only analyses.  
The results shown here are found in Ref.~\cite{RENO-PRL2}.

%-----------------------------------------------------------------
\subsection{Rate-only results}

In the rate-only analysis the oscillation amplitude of neutrino survival probability is extracted from the information on the observed 
reactor \nuebS rates only, without using the prompt energy spectra. We observe a clear deficit of reactor \nuebS in the far detector. 
Using the deficit information, a rate-only analysis obtains the value of $\sin^2 2 \theta_{13}$ as 0.087$\pm$0.009(stat.)$\pm$0.007(syst.), 
where the world average value of $|\Delta m_{ee}^2|= (2.49\pm0.06) \times 10^{-3}$~eV$^2$ is used~\cite{PDG}. 
The $\chi^{2}$ fit for the result is described in Ref.~\cite{RENO}. The systematic error of $\sin^2 2 \theta_{13}$ is reduced from 0.019 to 0.007, 
mainly due to the reduced background rate and uncertainty, relative to the first measurement in 2012~\cite{RENO}.
In addition, the statistical error is reduced from 0.013 to 0.009. Note that the largest reduction of the background rate and uncertainty comes from the \LiHeS background.

%-----------------------------------------------------------------
\subsection{Rate and spectrum results}

\begin{figure}[h]
\begin{center}
\includegraphics[width=0.48\textwidth]{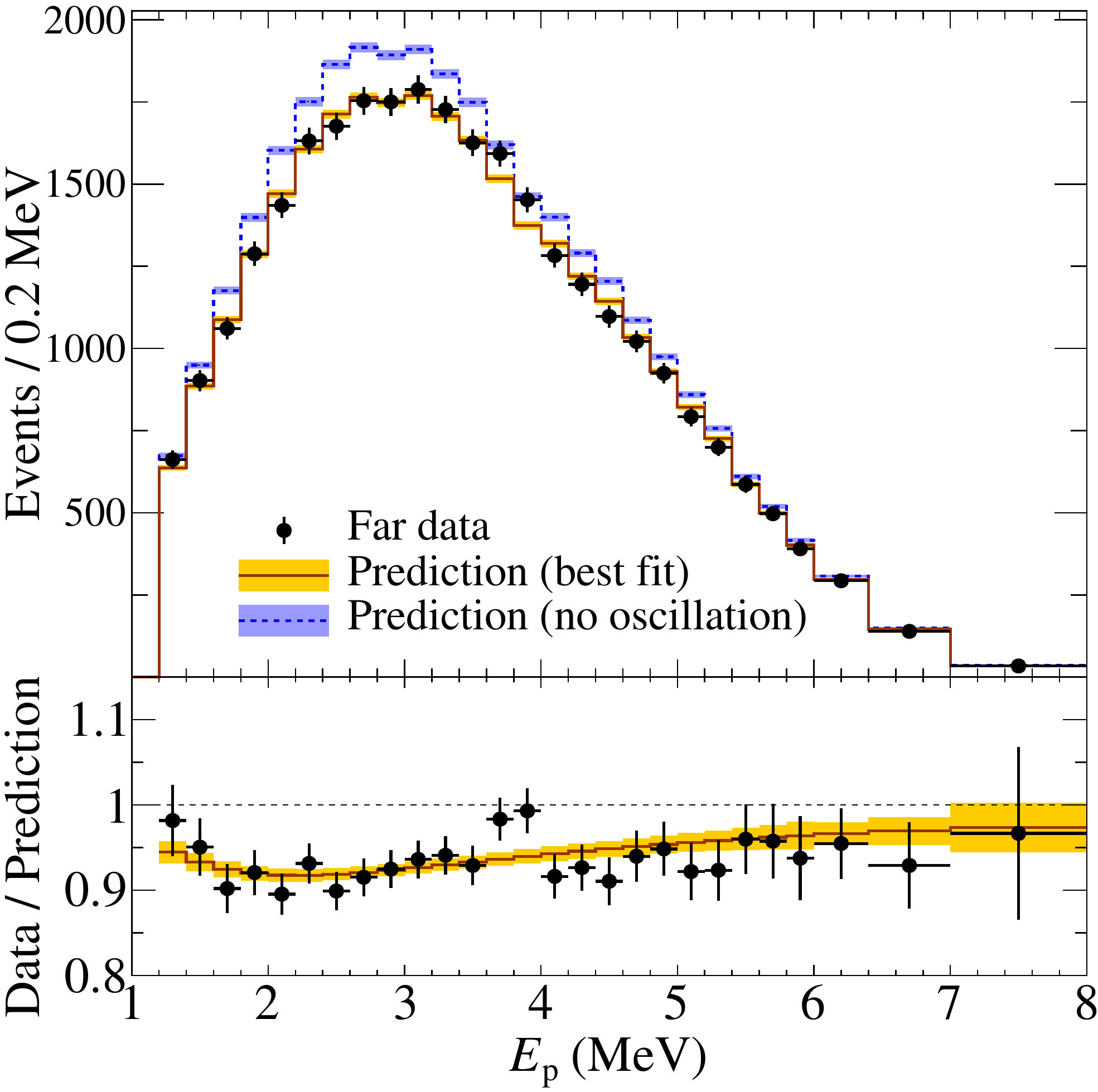}
\caption{
(Color online) Top: comparison of the observed IBD prompt spectrum in the far detector (dots) with the no-oscillation prediction (blue shaded histogram) 
obtained from the measurement in the near detector. The prediction from the best-fit oscillation parameters is also shown (yellow shaded histogram). Both blue and yellow bands represent uncertainties.
Bottom: ratio of IBD events measured in the far detector to the no-oscillation prediction (dots) and the ratio from the MC 
simulation with best-fit results folded in  (shaded band). Errors are statistical uncertainties only although both statistical and 
systematic uncertainties are included in the $\chi^{2}$ fitting.
}
\label{f:spectra-oscillation-fit}
\end{center}
\end{figure}

\begin{figure}
\begin{center}
\includegraphics[width=0.48\textwidth]{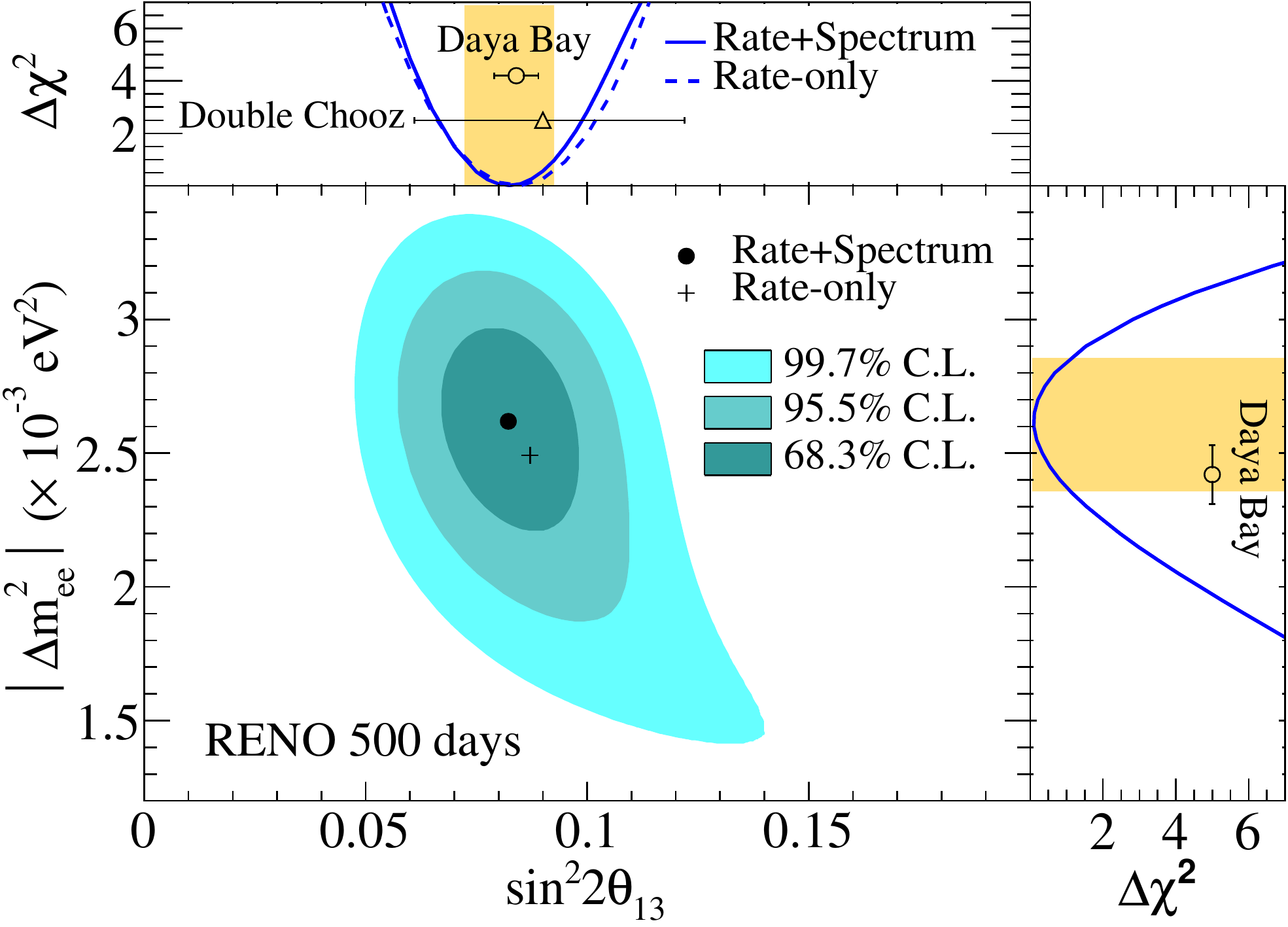}
\caption{
(Color online) Allowed regions of 68.3, 95.5, and 99.7\% C.L. in the $|\Delta m_{ee}^2|$ vs. $\sin^2 2\theta_{13}$ plane.
The best-fit values are shown as the black dot. The $\Delta \chi^{2}$ distribution for $\sin^2 2\theta_{13}$ (top)
and $|\Delta m_{ee}^2|$ (right) are also shown with an 1~$\sigma$ band. The rate-only result for $\sin^2 2\theta_{13}$
is shown as the cross. The results from Daya Bay~\cite{DB-spect} and Double Chooz~\cite{DC-recent} 
collaborations are also shown for comparison. 
}
\label{f:contour-allowed}
\end{center}
\end{figure}

In the rate and spectrum analysis the oscillation amplitude and frequency of neutrino survival probability are measured based on the information 
on the observed reactor \nuebS rates and spectra. We observe a clear energy-dependent deficit of reactor \nuebS in the far detector. 
Even with the unexpected structure around 5~MeV, the oscillation amplitude and frequency can be determined from a fit to the measured far-to-near ratio of IBD prompt spectra. 
The determination is not affected by the presence of the 5-MeV excess because of its cancellation in the ratio measurement.
For determination of $|\Delta m^2_{ee}|$ and $\sin^2 2\theta_{13}$, a $\chi^{2}$ with pull parameter terms of systematic uncertainties is constructed 
using the spectral ratio measurement and is minimized by varying the oscillation parameters and pull parameters~\cite{Anderson}.
The following $\chi^2$ function is used for the rate and shape analysis, 
\begin{eqnarray}
 \chi^2  & = & \sum_{i=1}^{\rm{N}_{\rm{bins}}} \frac{ (O_i^{\rm{F/N}} - T_i^{\rm{F/N}})^2 }{U_i^{\rm{F/N}}}
                          + \sum_{d={\rm N, F}} \left( \frac{b^{d}}{\sigma_{\rm{bkg}}^{d}} \right)^2
 \nonumber       \\
&&                        + \sum_{r=1}^{6} \left( \frac{f}{\sigma_{\rm{flux}}^r} \right)^2
                          + \left( \frac{\epsilon}{\sigma_{\rm{eff}}} \right)^2
                          + \left( \frac{\eta}{\sigma_{\rm{scale}}} \right)^2 ,
\end{eqnarray}
where $O_i^{\rm{F/N}}$ is the observed far-to-near ratio of IBD candidates in the $i$-th $E_{\rm{p}}$ bin after background subtraction,
$T_i^{\rm{F/N}} = T_i^{\rm{F/N}} (b^d, f, \epsilon, \eta; \theta_{13}, |\Delta m_{ee}^2|) $ is the expected far-to-near ratio of IBD events,
and $U_i^{\rm{F/N}}$ is the statistical uncertainty of $O_i^{\rm{F/N}}$.\\
\indent
The expected ratio $T_i^{\rm{F/N}}$ is calculated using the reactor $\overline{\nu}_e$ model, IBD cross section,
and the detection efficiency together with the signal loss due to the timing veto criteria, and folding the $\overline{\nu}_e$ survival probability and detector effects.
The systematic uncertainty sources are embedded by pull parameters ($b^d$, $f$, $\epsilon$, and $\eta$) with associated 
uncertainties ($\sigma_{\rm{bkg}}^d$, $\sigma_{\rm{flux}}^r$, $\sigma_{\rm{eff}}$, and $\sigma_{\rm{scale}}$).
The pull parameters allow variations from the expected far-to-near ratio of IBD events within their corresponding systematic uncertainties. 
The pull parameters $b^d$ and $\eta$ introduce deviations from the expected spectra accounting for the effects of the associated energy dependent 
systematic uncertainties. For the spectral deviations the energy-bin correlated and uncorrelated uncertainties are separately taken into account. 
The uncorrelated reactor-flux systematic uncertainty $\sigma_{\rm{flux}}^r$ is 0.9\%, the uncorrelated detection and timing veto systematic uncertainty $\sigma_{\rm{eff}}$ is 0.2\%,
the uncorrelated energy-scale systematic uncertainty $\sigma_{\rm{scale}}$ is 0.15\%, and the background uncertainty $\sigma_{\rm{bkg}}^d$ is 4.7\% and 7.3\%
for near and far detectors, respectively. 
The $\chi^{2}$ is constructed as a sum of two periods, before ($\sim$400 days) and after ($\sim$100 days) $^{252}$Cf contamination. 
A profile likelihood method is used to incorporate the systematic uncertainties in the fit. 
The best-fit values obtained from the rate and spectrum analysis are $\sin^2 2 \theta_{13} = 0.082\pm0.009({\rm stat.})\pm0.006({\rm syst.})$ 
and $|\Delta m_{ee}^2| =[2.62_{-0.23}^{+0.21}({\rm stat.})_{-0.13}^{+0.12}({\rm syst.})]\times 10^{-3}$~eV$^2$ with $\chi^2 /\rm{NDF} = 58.9/66$, 
where $\rm{NDF}$ is the number of degrees of freedom. 
This result is consistent with that of the rate-only analysis within their errors.
Another fit result is also obtained assuming an independent pull parameter for each energy bin to allow 
maximum variation of the background shapes within their uncertainties. The total systematic errors for both $\sin^2 2\theta_{13}$ and $|\Delta m_{ee}^2|$
remain almost unchanged by the fit.\\
\indent Table~\ref{t:sys_error} presents systematic uncertainties of $\sin^2 2\theta_{13}$ and $|\Delta m_{ee}^2|$ 
from several uncertainty sources. The uncertainties of energy-scale and backgrounds are the dominant sources of the total systematic uncertainty 
for $|\Delta m_{ee}^2|$.  The measured value of $|\Delta m_{ee}^2|$ corresponds to 
$|\Delta m_{31}^2| = (2.64_{-0.26}^{+0.24})\times 10^{-3}$~eV$^2$ ($|\Delta m_{31}^2| =[2.60_{-0.26}^{+0.24}]\times 10^{-3}$~eV$^2$)
for the normal (inverted) neutrino mass ordering, using measured oscillation parameters of
$\sin^2 2\theta_{12} = 0.846\pm0.021$ and $\Delta m_{21}^2 = (7.53\pm0.18)\times 10^{-3}$~eV$^2$~\cite{PDG}. \\
\begin{table}[h]
\caption{\label{t:sys_error} Systematic uncertainties from various uncertainty sources.
The dominant sources of the total systematic uncertainties for $|\Delta m_{ee}^2|$ are the uncertainties of energy-scale and backgrounds.
}
\begin{center}
\begin{tabular*}{0.48\textwidth}{@{\extracolsep{\fill}} l c c }
  \hline \hline
                                 & $\delta |\Delta m_{ee}^2|$ ($\times 10^{-3}$ eV$^2$)
                                 & $\delta (\sin^2 2\theta_{13})$       \\
  \hline
 Reactor                     &  $+0.018$, $-0.018$       &   $+0.0026$, $-0.0028$   \\
Detection efficiency      &  $+0.020$, $-0.022$     &   $+0.0028$, $-0.0029$  \\
Energy scale               &  $+0.081$, $-0.094$  &     $+0.0026$, $-0.0015$ \\
Backgrounds                &   $+0.084$, $-0.106$    &   $+0.0030$, $-0.0028$    \\
    \hline
 Total                        &   $+0.115$, $-0.133$     &  $+0.0055$, $-0.0052$  \\
  \hline \hline
\end{tabular*}
\end{center}
\end{table}
\indent Figure~\ref{f:spectra-oscillation-fit} shows the background-subtracted, observed spectrum at the far detector 
compared to the one expected with no oscillation and the one expected with the best-fit oscillation parameters at the far detector. 
The expected spectrum with no oscillation is obtained by weighting the spectrum at the near detector with no-oscillation assumption in order to include the 5-MeV excess. 
The expected spectrum with the best-fit oscillation parameters is obtained by applying the measured values of $\sin^2 2\theta_{13}$ and $|\Delta m_{ee}^2|$
to the one expected with no oscillation at the far detector. The observed spectrum at the far detector shows a clear energy dependent 
disappearance of reactor \nuebS events consistent with neutrino oscillations. 
A weak deviation from the expectation is observed near $E_{\rm p}$=3.8~MeV in Fig.~\ref{f:spectra-oscillation-fit} and will be kept monitored for its persistency with more data. 
%\begin{figure}[hbt]
Figure~\ref{f:contour-allowed}
shows 68.3, 95.5, and 99.7\% C.L. allowed regions for the neutrino oscillation parameters $|\Delta m_{ee}^2|$ and $\sin^2 2\theta_{13}$.
The results from other reactor experiments~\cite{DB-spect, DC-recent} are also shown in the figure. 
%-----------------------------------------------------------------
\subsection{Spectrum-only results}

The spectrum-only analysis uses only spectral shape information with a free normalization that allows variation in the expected IBD signal rates. 
This method obtains the oscillation frequency of $|\Delta m_{ee}^2|$ from the energy dependent disappearance of the reactor \nuebS 
without using the information on the total-rate deficit although it does not provide a sensitive measurement of $\sin^2 2\theta_{13}$. 
The spectrum-only analysis yields $|\Delta m_{ee}^2| = (2.62^{+0.38}_{-0.41}) \times 10^{-3}$~eV$^2$
and $\sin^2 2\theta_{13} = 0.066^{+0.042}_{-0.046} $ with $\chi^2 /\rm{NDF} = 58.8/67$. 
This result is consistent with those from the rate and spectrum analysis and the rate-only analysis within the errors.
%-----------------------------------------------------------------
\subsection{Energy and baseline dependent reactor \nuebS disappearance}

\begin{figure}[h]
\begin{center}
\includegraphics[width=0.48\textwidth]{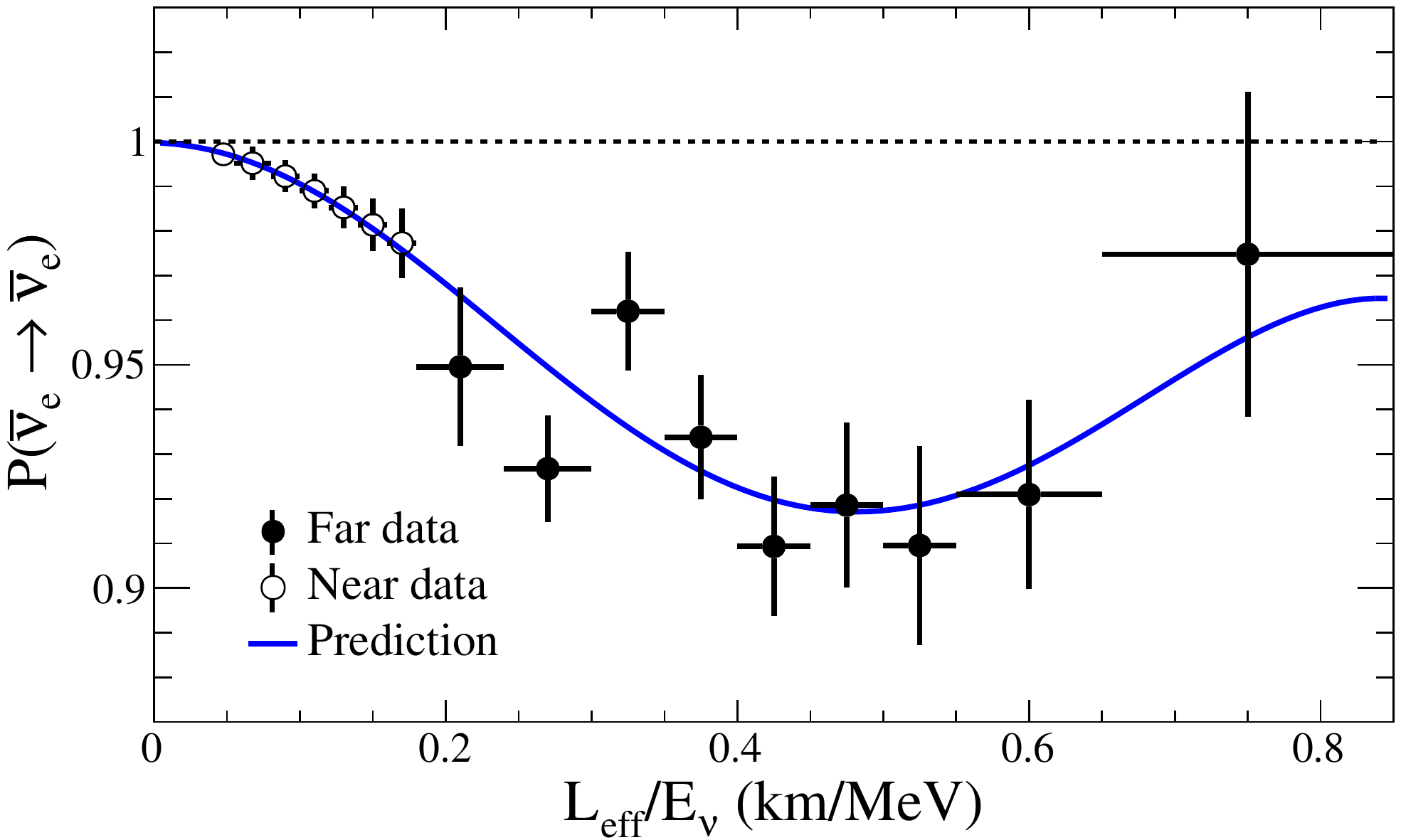}
\caption{
Measured reactor $\overline{\nu}_e$ survival probability in the far detector as a function of $L_{\rm eff}$/$E_{\nu}$. 
The curve is a predicted survival probability, obtained from the observed probability in the near detector, for the best-fit values of 
$|\Delta m_{ee}^2|$ and $\sin^2 2\theta_{13}$. The $L_{\rm eff}$/$E_{\nu}$ value of each data point
is given by the average of the counts in each bin.}
\label{f:baseline-energy}
\end{center}
\end{figure}

The survival probability of reactor \nuebS is a function of a baseline $L$ over neutrino energy $E_{\nu}$ as written in Eq.~\ref{eq_pee}.
Because of having multiple reactors as neutrino sources, an effective baseline $L_{\rm eff}$ is defined by the reactor-detector distance 
weighted by the IBD event rate from each reactor. Note that $L_{\rm eff}$ is time dependent due to the IBD event rate weighting. The neutrino energy $E_{\nu}$
is converted from the IBD prompt energy. A daily $L_{\rm eff}$/$E_{\nu}$ distribution of the IBD events is obtained from the background 
subtracted IBD event spectrum and the daily $L_{\rm eff}$. 
The observed $L_{\rm eff}$/$E_{\nu}$ distribution is obtained by summing up the daily distributions weighted by a daily IBD rate.
The measured survival probability is obtained by the ratio of the observed IBD events to the expected ones with no oscillation in each bin of 
$L_{\rm eff}$/$E_{\nu}$.
Figure~\ref{f:baseline-energy} shows the measured survival probability of reactor \nuebS in the far detector as a function of 
$L_{\rm eff}$/$E_{\nu}$. A predicted survival probability is obtained from the observed probability distribution in the near detector and the best–fit oscillation values.  
Because of the observed 5-MeV excess, the expected $L_{\rm eff}$/$E_{\nu}$
distribution is derived from the measured spectrum in the near detector instead of the IBD MC spectrum. A clear $L_{\rm eff}$/$E_{\nu}$-dependent 
disappearance of reactor \nuebS is observed and demonstrates the periodic feature of neutrino oscillation. 

%======================================================================
\section{\label{sec:conclusion} Summary and prospects }

Using about 500 live days of data RENO has observed a clear energy dependent disappearance of reactor \nuebS 
using two identical detectors and obtained 
$\sin^2 2\theta_{13} = 0.082\pm0.010$ and $|\Delta m_{ee}^2| =[2.62_{-0.26}^{+0.24}]\times 10^{-3}$~eV$^2$
based on the measured disappearance expected from neutrino oscillations. 
RENO has measured $\sin^2 2\theta_{13}$ more precisely and $|\Delta m_{ee}^2|$ for the first time with the rate and spectrum analysis. 
The systematic uncertainty of $\sin^2 2\theta_{13}$ has been significantly reduced from 0.019~\cite{RENO} 
to 0.006 due to the improvement in reducing the background uncertainties, especially the most dominant
\LiHeS background rate and its uncertainty.
A clear IBD spectral difference from the current reactor \nuebS model is observed in the region of 5~MeV in both detectors, 
with an excess corresponding to about 3\% of the total observed IBD events. The observed excess is clearly correlated 
with the reactor thermal power, indicating the excess arises from the reactor \nuebS.\\
\indent
Table~\ref{t:comp} presents comparison of the measured values of $\sin^2 2\theta_{13}$ and $|\Delta m_{ee}^2|$
between the first RENO measurement in 2012~\cite{RENO} and the current measurement. 
The precision on $\sin^2 2\theta_{13}$ is improved from 20.4\% to 13.4\%,
and the $|\Delta m_{ee}^2|$ precision is 9.9\%.
The background systematic uncertainties estimated from data are expected to be reduced with more data. 
The precision is expected to be $\sim$5\% for both oscillation parameters with ten live years of data. 
\begin{table}[h]
\begin{center}
\caption{\label{t:comp} 
Comparison of the measured $\sin^2 2\theta_{13}$ and $|\Delta m_{ee}^2|$ 
between the first measurement in 2012~\cite{RENO} and the current measurement.
}
\begin{tabular*}{0.48\textwidth}{@{\extracolsep{\fill}} l r r}
\hline
\hline
Results & 2012 & Current  \\
\hline
Live days & 220  &  500  \\
\hline
\qOneThree  & 0.113$\pm$0.023 &  0.082$\pm$0.011 \\
Precision  &  20.4\% & 13.4\% \\
\hline
\dmSq ($\times 10^{-3} \rm{eV}^{2}$) & 2.32 (PDG 2012) & 2.62$\pm$0.26 \\
Precision &  --- & 9.9\% \\
\hline
\hline
\end{tabular*}
\end{center}
\end{table}

%--Acknowledgements
\section{ACKNOWLEDGEMENTS}
The RENO experiment is supported by the National Research Foundation of Korea (NRF) grant No. 2009-0083526 funded by the Korea Ministry of Science, ICT \& Future Planning. Some of us have been supported by a fund from the BK21 of NRF.
We gratefully acknowledge the cooperation of the Hanbit Nuclear Power Site and the Korea Hydro \& Nuclear Power Co., Ltd. (KHNP).
We thank KISTI for providing computing and network resources through GSDC, and all the technical and administrative people who greatly helped in making this experiment possible.

%======================================================================

\end{document}